\input psfig.sty 
%
%
%

\ifx\mnmacrosloaded\undefined \input mn\fi

%

\newif\ifAMStwofonts

\ifCUPmtplainloaded \else
  \NewTextAlphabet{textbfit} {cmbxti10} {}
  \NewTextAlphabet{textbfss} {cmssbx10} {}
  \NewMathAlphabet{mathbfit} {cmbxti10} {} 
  \NewMathAlphabet{mathbfss} {cmssbx10} {} 
  \ifAMStwofonts
    \NewSymbolFont{upmath} {eurm10}
    \NewSymbolFont{AMSa} {msam10}
    \NewMathSymbol{\upi}     {0}{upmath}{19}
    \NewMathSymbol{\umu}     {0}{upmath}{16}
    \NewMathSymbol{\upartial}{0}{upmath}{40}
    \NewMathSymbol{\leqslant}{3}{AMSa}{36}
    \NewMathSymbol{\geqslant}{3}{AMSa}{3E}

    \let\leq=\leqslant \let\le=\leqslant
     \let\ge=\geqslant
  \else
    \def\umu{\mu}
    \def\upi{\pi}
    \def\upartial{\partial}
  \fi
\fi


\pageoffset{-2.5pc}{0pc}

\loadboldmathnames



\onecolumn        
\pagerange{}    
\pubyear{2004}
\volume{}

\begintopmatter  

\title{Bar-induced perturbation strengths of the galaxies in the Ohio State University Bright Galaxy Survey (OSUBGS) I}
\author{Eija Laurikainen and Heikki Salo}
\affiliation{Division of Astronomy, Dep. of Phys. Sci, FIN-90014, Finland}

and

\author{Ronald Buta and Sergiy Vasylyev}
\affiliation{Department of Physics and Astronomy, Box 870324, Univ.of Alabama, Tuscaloosa, AL 35487, USA}

\shortauthor{E. Laurikainen et al.}
\shorttitle{Bar strengths for OSUBGS sample}



email:eija.laurikainen@oulu.fi

\abstract {Bar-induced perturbation strengths are calculated for a well 
defined magnitude limited sample of 180 spiral galaxies, based on the
Ohio State University Bright Galaxy Survey. We use a gravitational
torque method, the ratio of the maximal tangential force to the mean
axisymmetric radial force, as a quantitative measure of the bar
strength. The gravitational potential is inferred from an $H$-band
light distribution by assuming that the $M/L$-ratio is constant
throughout the disk. Galaxies are deprojected using orientation
parameters based on $B$-band images. In order to eliminate artificial
stretching of the bulge, two-dimensional bar-bulge-disk decomposition
has been used to derive a reliable bulge model. This bulge model is
subtracted from an image, the disk is deprojected assuming it is thin,
and then the bulge is added back by assuming that its mass
distribution is spherically symmetric. We find that removing the
artificial bulge stretch is important especially for galaxies having
bars inside large bulges. We also find that the masses of the bulges
can be significantly overestimated if bars are not taken into account
in the decomposition.

Bars are identified using Fourier methods by requiring that the phases
of the main modes ($m$=2, $m$=4) are maintained nearly constant in the
bar region.  With such methods, bars are found in $65\%$ of the
galaxies in our sample, most of them being classified as SB-type
systems in the near-IR by Eskridge and coworkers. We also suggest
that as much as $\approx$$70 \%$ of the galaxies classified as
SAB-types in the near-IR might actually be non-barred systems, many of
them having central ovals.  It is also possible that a small fraction
of the SAB-type galaxies have weak non-classical bars with spiral-like
morphologies.  }


\maketitle  

\section{INTRODUCTION}

In two previous papers (Buta, Laurikainen, \& Salo 2004=BLS; Laurikainen, 
Salo, \& Buta 2004=LSB), we have used a gravitational torque parameter to investigate
two issues: the distribution of maximum relative bar torques in spiral
galaxies, and the impact of bar strength on nuclear activity type in
the same galaxies. These studies have utilized a detailed analysis of
158 galaxies from the Ohio State University Bright Galaxy Survey (OSUBGS,
Eskridge et al.  2002=EFP) and 22 galaxies from the Two Micron All-Sky
Survey (2MASS, Skrutskie et al.  1997). In the present paper, we
present the full details of our methods used to derive maximum
relative gravitational torques for bars and, in some cases,
spirals. Our method is a refinement of the approach used by Buta \&
Block (2001=BB) to derive quantitative bar strengths from
near-infrared images.

\section{The sample and data reductions}
The OSUBGS forms a sample of 205
nearby spiral galaxies, originally selected to fill the following
criteria: type index in the Third Reference Catalog of Bright 
Galaxies (de Vaucouleurs et al. 1991, RC3)  $0 < T < 10$ 
(S0/a to Sm), total magnitudes $B_T < 12.0$, standard isophotal
galaxy diameters $D_{25} < 6.5'$, and declinations in the range 
$-80^\circ < \delta  < +50^\circ$. In this study 
only OSUBGS galaxies with inclinations less than $65^\circ$ are included, which
limited the sample to 158 galaxies. This corresponds approximately
to RC3 standard isophotal axis ratio
$logR_{25}$ $\leq$ 0.38. The 22 2MASS galaxies satisfy the same criteria as 
the galaxies in the OSUBGS, except that they have 
$D_{25} >$6\rlap{.}$^{\prime}$5.
The final sample of 180 galaxies is representative of typical luminous
spiral galaxies in the nearby universe. It was shown by
BLS that the sample is  
biased mainly against late-type, low-luminosity barred spirals, based
on a comparison to
a distance-limited sample of 1264 galaxies from the catalog of Tully (1988).

We use the $B$ and $H$-band images of the OSUBGS sample, available as an early 
release by EFP \footnote{${}^{1}$}{http://www.astronomy.ohio-state.edu/\~{}survey/EDR/index.html}. 
The $B$-band images reach typically the surface brightnesses 
of $26 \ mag \ arcsec^{-2}$, whereas the limiting surface brightness
in the $H$-band is typically $20 \ mag \ arcsec^{-2}$ (EFP). Our reduction 
steps consist of cleaning the images of bad pixels and foreground 
stars, background subtraction, and for some of the galaxies also removal of
background gradients. We also checked the image scales and orientations 
of the $B$-band images using the $H$-band images as a frame of reference.
This is important, because the orientation
parameters are measured in the $B$-band, which are then used to deproject 
the $H$-band images to face-on orientation.

The pixel sizes were checked using the IRAF routine GEOMAP: the
$H$-band images were used as reference frames and magnification factors
were calculated for the $B$-band images. Generally the same
pixel sizes were obtained that were listed in the image headers of the
$B$-band images; however, for 17 of the galaxies the header value
was too small by a factor of 2 (0\rlap{.}$^{\prime\prime}$36 instead of 
0\rlap{.}$^{\prime\prime}$72). All of these 
galaxies were observed using the same
telescope, the Perkins 1.8-m reflector, and the same instrument, 
a Low NSF T1800 CCD, 
in the $B$-band. In the $H$-band the 1.8-m Perkins reflector was also used,
but with the OSIRIS instrument attached. As the same instrumentation 
was used also for 66 other galaxies, that had no 
inconsistencies in the pixel sizes between the $B$ and $H$-band images, 
we conclude that there 
must be an error in the B-band image headers of the 17 discrepant galaxies.
The corrected pixel sizes are denoted by footnotes in Table 1.
In the 2MASS sample the scale of the $H$-band images is 
1\rlap{.}$^{\prime\prime}$0 pix$^{-1}$.

The IRAF-routine GEOMAP was used also to check the relative differences
in the direction of North in the sky plane between the $B$- and $H$-band 
images. The list of star positions for GEOMAP was produced by a semiautomatic 
IDL routine. Before applying GEOMAP, 
the $B$ and $H$-band images were transposed to have the same orientations in 
the sky. The shifts in the position angles between the two bands were found 
to be $4.4^{\circ}$ at maximum ($d \phi$ in Table 1). In order to have the
correct position angle in the $H$-band image the shift given in the table must
be added to the position angle of the $B$-band image.
 
Some of the images include large numbers of foreground stars so that
point spread function (psf) fitting had to be used to automatically 
remove the stars.
Stars were identified with IRAF routine DAOFIND using the typical 
psf FWHM in each image. IRAF routines PSF and ALLSTAR were used to 
subtract stars. In some cases the model
psf was not good enough to remove all the stellar flux, probably
because it was not exactly the same for faint and bright stars, or 
because the psf varied in different parts of the image. In that case cleaning 
was completed manually using the IRAF-routine IMEDIT, which replaced
the residual in an aperture by a two-dimensional polynomial fit interpolated 
from pixels in the surrounding annulus. The sky background
in the $H$-band images was already subtracted by EFP while combining the
images, but the subtraction was fine-tuned by us, based on the 
mean values of the sky measured in different image positions. The sky gradients
were generally small, but for some of the images IMSURFIT was 
applied to calculate linear polynomial fits in the x or y-directions 
of the images.

2MASS images have approximately the same spatial resolution 
as the OSUBGS images, but they are not as deep.
Also, the field of view was generally rather 
small so that mosaics of five or more images were typically
constructed. Before combining the images the over-scan regions
were removed and the background values given in the image headers 
were subtracted. Because the background levels varied
in different parts of the images, the subtraction was fine-tuned
paying attention only to the background near to the galaxy.
Positioning of the frames was done using stars common
in the image fields. Finally the mosaics were cleaned of 
foreground stars and bad pixels.

\section{Orientation parameters}
Well defined orientation parameters are important for the measurements
of the perturbation strengths: namely it has been shown by BB
and Laurikainen \& Salo (2002=LS) that already uncertainties of 10 $\%$
in inclination can cause uncertainties of 10-15 $\%$
in the perturbation strength. All of our sample galaxies
have orientation parameters listed in RC3. However, the OSU $B$-band
images have sufficient field of view, and are of sufficient depth and
quality compared to what was used for RC3, that they can be used to 
improve the orientation parameters in each case.
They are also deeper than the OSU $H$-band images.
The radial profiles of the major-axis position angles ($\phi$) 
and minor-to-major
axis ratios of the disks ($q=b/a$) were calculated using the ELLIPSE routine 
in IRAF. A linear radial scaling was used, and for the upper and lower 
deviant pixels a 3$\sigma$ clipping criterion was applied. The 
position angles and corresponding axial ratios are listed in Table 1; 
these are means over the regions indicated in the table, and represent
the outer parts of the disks rather than a specific isophotal level
(as in RC3). For each parameter, 
the errors are standard deviations of the mean. For comparison 
the orientation parameters (at $\mu_B$ = 25.00 mag arcsec$^{-2}$)
of the same galaxies from RC3 are also listed.  

A few of the galaxies are in pairs, so that the outer parts of 
the disks are partially superimposed by the neighboring galaxies. 
For these galaxies  
masks were first created for the overlapping regions and then ellipse fits were
performed on the unmasked
parts of the images. Finally the masked areas were replaced by the mean in an
annulus at each radius covering the unmasked part of the annulus. 
For NGC 1808 the outer disk was so 
faint that the ellipse fitting failed for a large part of the
disk, and the orientation parameters were obtained by fitting an ellipse to 
the shape of the outer pseudo-ring. Because only a narrow zone could be used 
for the fit, the errors in the orientation parameters are
those given by the ellipse-fitting routine. For NGC 1300 an automatic
fit to the outer ring failed and the position angle and inclination
were estimated manually. For the galaxies NGC 289 and NGC 3893 the image
field was slightly too small for reliable measurements, but
even so the values we get are similar to those given in RC3.
In any case, we used the RC3 orientation parameters for these galaxies. 
For three of the galaxies, NGC 278, 4138 and 5248 we had no $B$-band image
so that the orientation parameters were determined using the $H$-band 
images. Comparison to the Digitized Sky Survey image showed that they
were deep enough for that purpose. The galaxies NGC 2139 and NGC 4496A
are possible mergers, and therefore the uncertainty in their orientation
parameters is large.  
For the galaxies in the 2MASS-sample we used mainly the orientation 
parameters as given in RC3, because the images were not 
very deep.

In Figure 1 we compare our measured orientation parameters, axis
ratio $q$ and major axis position angle $\phi$, with values from
RC3 and from Garcia-Gomez et al. (2004), who used the OSUBGS images
to derive orientation parameters for most of the same galaxies using
Fourier techniques. We do the comparisons by plotting the differences,
$|\Delta q|$ and $|\Delta$ $\phi |$, versus the means of $q$ and $\phi$
for each pair of sources. The results show good agreement between
our values and those of Garcia-Gomez et al. (2004), with $|\Delta q|$
averaging less than 0.05 and with $|\Delta \phi|$ averaging less than
10$^{\circ}$ for $<q>$ $\leq$ 0.65. The errors in position angle
increase towards larger values of $<q>$ in this comparison, as
would be expected since the position angle is undefined for round
galaxies. The comparisons with RC3 parameters show similar trends
with some significant disagreements even at intermediate inclinations.

The position angles and inclinations in RC3 are 
given at distances corresponding to the surface brightness level
of $25 \ mag \ arcsec^{-2}$, whereas the $B$-band images by EFP typically 
reach the surface brightness of $26 \ mag \ arcsec^{-2}$, which means that
the images we use are deeper. This explains many of the disagreements between
our measurements and those given in RC3. For some of the galaxies the
deviations from RC3 values are very large, of which examples are 
NGC 4593 and NGC 4643. In these two cases $\phi$ in RC3 deviates 
even by $43^0$ and $74^0$ from our
measurements, respectively. In the latter case $q$ also deviates 
quite significantly ($q$(RC3)=0.74, $q$(meas)=0.82). Other examples are 
NGC 1300 and NGC 1808, for which we use
outer rings/pseudo-rings to estimate the orientation parameters. 
There is no doubt that when the galaxy inclinations 
are not insignificant, large errors in the position angles
may have drastic consequences to the perturbation strengths. However,
the uncertainties in the orientation parameters are not very important
for galaxies seen in nearly face-on orientation. 

\section{2D Bulge/disk/bar decomposition}

\subsection{The decomposition method}

The structural decomposition is needed for several purposes in this 
study (see Section 5):
for calculating non-axisymmetric forces we use a method which requires
information on the vertical scale height of the disk. But as it
cannot be measured directly for low inclination galaxies,   
the radial scale length of the disk is needed to estimate it indirectly. 
Also, while
deprojecting the images a bulge correction is applied treating the
bulges as separate structural components in galaxies.

In recent years two-dimensional (2D) decomposition methods have been
widely used to separate the structural components like bulges
and disks in galaxies. 2D methods
are especially useful for separating the non-axisymmetric structures, 
like bars, ovals, and rings from the disk, but that advantage has only 
rarely been used (de Jong 1996; Peng et al. 2002; Peng 2002; de Souza, Gadotti,
\& dos Anjos 2004). The first 
2D-methods used the $R^{1/4}$ law profile for the bulge (Byun \& 
Freeman 1995; Shaw \& Gilmore 1989; de Jong 1996; Wadadekar, 
Robbason, \& Kembavi 1999), whereas later studies have shown that 
the more general S\'ersic's $R^{1/n}$ function (S\'ersic 1968) can better
account for the bulge profiles in spiral galaxies. The S\'ersic's function
has been recently used in conjunction with 2D-decomposition methods by 
M\"ollenhoff \& Heidt (2001), 
Simard et al. (2002), Peng et al. (2002), MacArthur, Courteau and Holtzman
(2003), and de Souza, Gadotti, \& dos Anjos (2004),
all developed for different purposes; for
example, the method by Simard et al. was developed for treating HST-images
with low S/N-ratios, whereas that by Peng et al. is suitable
also for detailed structural analysis of nearby galaxies.

In order to separate bars and disks from bulges we use $H$-band images and 2D 
3-component decomposition. 
For our purposes a reliable estimation of the bulge light distribution 
is important, 
because overestimating the bulge easily causes us to underestimate the 
bar strength, especially in early-type galaxies (see Section 5). 
Although bars can be easily identified 
in many galaxies, their fitting in the decompositions 
is sometimes complicated, because bars might have non-flattened 
central structures, and
it is not always clear whether boxy/peanut shaped structures
in galaxies are bars or flattened bulges.
We took the approach that both bars and ovals are fitted in the
decomposition if they are visible in the 
original images, or if they can be detected by Fourier methods
(see Section 5). For the inclination limit we use, 
possible boxy/peanut 
shaped structures are mostly invisible. 
Some of the galaxies in our sample
have bars made up of two components, a thin long bar and a thicker and
shorter component, both of which were modeled as a single component.
However, since the main purpose for including a bar
to the decomposition is to help to extract a more reliable bulge model, 
its exact treatment is not very crucial. 

The bulge and the disk are described as in M\"ollenhoff \& Heidt (2001),
using an exponential function for the disk and the $R^{1/n}$ S\'ersic's
function for the bulge: 

$$I_d(r) \ = \ I_{0d}  \exp[-(r/h_r)],$$
$$I_b(r) \ = \ I_{0b} \exp[-(r/h_b)^{\beta}],$$

\noindent where $h_r$ and $h_b$ are the scale parameters of the disk 
and bulge, 
$I_{0d}$ and $I_{0b}$ are the central surface densities, and
$\beta = 1/n$ determines the slope of the projected surface brightness 
distribution of the bulge. 
For the disk, the radius $r$ is calculated along the disk plane, 
whereas for the spherical bulge 
model, $r$ is the projected distance from the center.
Special cases of the S\'ersic function are the 
exponential function with $n$=1 and the de Vaucouleurs function with 
$n$=4. Additionally, a bar/oval component was
added, described by a function:

$$I_{bar}(r) \ = \ I_{0bar} \ * \ (1-m^2)^{n_{bar}+0.5},$$

\noindent where $m^2= (x_b/a)^2 + (y_b/b)^2 < 1$, while $a$ and $b$ are 
the bar major and 
minor axis, $I_{0bar}$ is the central surface brightness of the bar and
$n_{bar}$ is the exponent of the bar model. This corresponds to a 
projected surface 
density of a prolate Ferrer's bar, with $a > b = c$, seen along the c-axis.
Here $x_b$ and  $y_b$ are coordinates in the disk plane in a system aligned
with the bar major axis, making an angle $\phi_{bar}$ with the nodal 
line of the disk. We generally used $n_{bar}=2$. Because
bars sometimes have quite complicated structures, the Ferrer's model is only 
an approximation of the true bar intensity distribution. 

Iterative fits were done to the images in magnitude units using a
weighting function of $w_i = 1 / r_i$, where $r_i$ is the distance
from the galaxy center along the disk plane. This means that each
radial zone has the same total weight. In Laurikainen \& Salo (2000)
we showed that in one dimension unweighted magnitude fits
(corresponding most closely to the adopted 2D weighting) are most
stable, in a sense that the fit results were least affected by adding
artificial noise to the profiles.
In order to allow for seeing the model bulge profile was convolved with a 
Gaussian (psf) using the FWHM measured for each galaxy. 
The maximum radius used in the fit was taken to be the radius
at which no negative pixels appeared in the image. This is important because
the decompositions were performed with images converted to magnitude units. 
Generally, bars were fitted leaving the
parameters describing the size, orientation ($a, b, \phi_{bar}$) and the flux 
level of the bar ($I_{0bar}$) 
free. However, when the surface brightness of the bar was close to that of the
disk, the bar major axis had to be fixed, based on visual
estimation in the original image. The total number of fitted parameters
was nine, of which four describe the bar model. 
In galaxies with
bright active nuclei (AGN) the bulges were contaminated by the flux of the AGN
in a region corresponding to the size of the seeing disk. In the
$H$-band images a typical psf had a FWHM of 
$1\rlap{.}^{\prime\prime}8 \pm 0\rlap{.}^{\prime\prime}3$,
corresponding to two innermost pixels in the galaxy center.
Spiral arms were not a major problem 
for the fits, because the arms are less prominent in the 
near-IR than in the optical. 

The measured structural parameters are 
listed in Table 2, the parameters $r_{eff}$ and $\beta$ describing the bulge
and $h_r$ is the exponential scale length of the disk. The effective radius,  
$r_{eff}$, is calculated from the scale parameter $h_b$ of the bulge. 
The last column also denotes
whether a bar/oval was fitted in the decomposition. 
If no bulge model is given in the table, the galaxy had no detectable bulge. 
A galaxy was considered to have no bulge, if the azimuthally averaged 
surface density profile had no detectable bulge-like component outside the
seeing disk and no Sersic's model could be fitted 
to the assumed bulge region (unclear cases are discussed in Section 4.2).
In order to estimate the relative mass of the bulge properly, the bulge
model was always taken from the decompositions made for the H-band images.
However, when the H-band images were not deep enough for reliable estimation
of the scale length of the disk, $h_r$ was estimated from the decompositions
applied to the deeper B-band images. Also, for three of the galaxies in the
2MASS sample, $h_r$ given by MH was used.
Some examples of the decompositions are shown in Fig. 2.

\subsection{Discussion of the sample galaxies}

The main limitation of our decompositions is that not all $H$-band images were
deep enough for detecting the outermost parts of the disks,
which is the case for the galaxies NGC 1300, 1808, 2090,  
3227, 3949, 4051, 4151, 4504, 4698, 4930 and 5701. However, even for 
these galaxies a large fraction of the exponential disk was still visible  
so that the bulge could be separated from the disk quite well. The image field
was slightly smaller than the size of the galaxy for NGC 2442, which
made it difficult to derive a reliable radial scale length for this galaxy.
Fitting was generally made using an automatic procedure, but for some of the
galaxies it led to unphysical results: for example, for NGC 1058 
it was possible to obtain a reasonable fit with either a small or a large 
bulge, both having similar 
global $\chi^2$, describing the difference between the model image
and the observed image. A similar case is shown also by MacArthur, Courteau
and Holtzman (2003) 
(see their Figure 9). While MacArthur and coworkers solved the problem by 
computing separate inner and outer $\chi^2$ residuals, we
limited the number of iterations so that the physically unreasonably
large bulge would be avoided. Also, galaxies like NGC 210,
NGC 1808 and NGC 5701 having prominent outer rings complicates the
interpretation of the structural decomposition.
We found that including a bar in the decomposition can significantly 
modify the scale parameters of the bulge and the disk, so that
ignoring a bar model in the decomposition would overestimate the B/D-ratio
and the shape parameter n of the bulge. Also, if the bar is very prominent,
ignoring a bar model would overestimate the scale length of the disk.
\vskip 0.25cm
In the following we discuss some 
individual cases.
\vskip 0.25cm

{\bf NGC 1187 and NGC 1302} (see Fig. 2a): In the case of NGC 1187, in which
the bar is small, 
including a bar component to the 
decomposition decreases the shape parameter of the bulge from 
$n$ = 2.0 to 1.4, 
thus making the bulge profile appear more exponential.  
Consequently the total flux of the bulge is decreased by $30\%$. 
However, the disk model is not affected.
An example of a galaxy with a larger bar  
is NGC 1302, for which inclusion of the bar
affects not only the fitted bulge, but also the radial scale 
length of the disk, which is reduced by $~12\%$. The bar resides inside
a roundish inner ring and is prominent in the bulge dominated region of the 
galaxy. For this galaxy 
the bulge/disk decomposition without any bar model would overestimate 
the mass of the bulge by as much as $36 \%$.
The effect of the bar model in the decomposition for NGC 1302 is 
demonstrated in a different manner in Fig. 3. Two residual images are
shown for two different decompositions: in one image the bulge model is 
subtracted from the original image, whereas in the other image
the whole galaxy model is subtracted.
It is clear that if no bar model is included to the decomposition
the bulge model becomes too large, mainly because a considerable
amount of bar flux is assigned to the bulge.
\vskip 0.25cm
{\bf NGC 210:} This galaxy has a bar whose surface 
brightness is high in comparison to that of the underlying disk, which is 
dominated by an outer pseudo-ring. Because of the prominent
bar it is not possible to perform any reliable decomposition for 
this galaxy without also modeling the bar (which in blue light is a 
well-known SAB-type oval). However the 
prominence of the outer ring made decompositions problematic.
In fact, this is true for outer-ringed galaxies in general.
\vskip 0.25cm
{\bf NGC 3166, 4699, 4939, 5701, 6394:} For these galaxies 
the bar resides mostly inside the large bulge, of which NGC 3166 and 
NGC 4699 are shown as examples in Fig. 2. The bars are so strong
that, for example, for NGC 4699, according
to our decomposition model, the surface brightnesses of the
bar and the bulge are nearly similar at the edge of the bar.  
It is clear that in these galaxies ignoring the bar would
overestimate the $B/D$-ratio.
\vskip 0.25cm
{\bf NGC 6902:} In most cases a bar can be detected in the surface brightness
profile of the galaxy, but for NGC 6902
the bar is too weak to appear as a bump in the profile. However,
the bar is visible by eye in the original image (inside a prominent
inner ring at $r < 10''$; see also Crocker, Baugus,
\& Buta 1996) and is detectable also by Fourier methods. In 
this galaxy 
the prominent bump visible in the
surface brightness profile at $r$=10-40'' is due to an oval, which was modeled
by a Ferrer's function. 
\vskip 0.25cm
{\bf NGC 1084, 4698, 4962, 5962, 6753:} 
There are some non-barred galaxies in the sample whose decompositions 
were improved by modeling ovals by a Ferrer's function.  
As an example the decomposition for NGC 6753 is shown in Fig. 2.
In this case the oval has a relatively low surface brightness 
and cannot be directly distinguished in the profile. However,
it is prominent in direct images and divides the zone between
bright nuclear and outer rings (see Crocker, Baugus, \& Buta 1996).
\vskip 0.25cm
The decomposition remained unsatisfactory for the galaxies NGC 4487,
4900, 2139 and NGC 4618. Characteristic for all these galaxies is that
they have little or no bulge, or that the bulge has 
very low surface brightness
in comparison to that of the disk. As an example the decomposition of 
NGC 4487 is shown in Fig. 2. For this galaxy the bulge model
has a very large shape parameter ($n=5$), which appears 
immediately after giving the initial parameters of the fit. Such a centrally 
peaked extended bulge is the only solution for this galaxy, 
but it is not clear whether the solution is physically reasonable.
At least it is not intuitively expected in the profile, where
the bulge does not look very prominent. 
Among the sample galaxies there is also one galaxy, NGC 4900, 
having a strong Freeman 
type II profile, which makes it impossible to fit any usual global 
disk model, unless a truncated disk is assumed. In this study  
the radial scale length of the disk for this galaxy
was estimated from the outer regions of the disk, whereas the bulge model
was extracted using the inner portions of the image. The
decomposition was uncertain also for the two 
late-type spirals, NGC 2139 and NGC 4618, both having an asymmetric
disk. NGC 4618 is a prototypical
one-armed SBm spiral (de Vaucouleurs \& Freeman 1972) and most probably
has no bulge at all.

\subsection{Comparison with previous studies}
In spite of the wide-spread use of the decomposition technique for
the derivation of the structural
parameters of spiral galaxies, it is difficult to find sufficient
published data to make a comparison with our results.
One reason is that it has been only recently shown that
no single shape parameter can describe the bulges of all spiral
galaxies. Also, although 2D-methods are widely used, there is no 
previous study like ours, where bulge/disk/bar decomposition using both 
the generalized S\'ersic
model for the bulge and a separate model for the bar has been applied to 
a large sample of galaxies.

In the following we compare our results with those obtained 
by M\"ollenhoff \& Heidt 
(2001=MH), who use a two-component 2D-method, and with Knapen et al. 
(2003) using a two-component 1D-method. In both
studies the observations are made in the near-IR with 
sub-arcsecond image
resolution, and also, the surface brightness profiles were modeled using 
an exponential
function for the disk and S\'ersic's function for the bulge. A typical
limiting surface brightness in the study by Knapen et al. was $20.5
\ mag \ arcsec^{-2}$ in the $K$-band, which is similar to that in our sample
in the $H$-band
($20 \ mag \ arcsec^{-2}$), whereas the limiting magnitude by
MH was not given. 

We have 13 galaxies in common with the sample by MH.  For 8 of the
galaxies the bulge/disk parameters were very similar in both
studies. Dividing the parameter value measured by us by the value
given in MH, and taking a mean of the measurements for different
galaxies, we found: $<r_e / r_e(MH)> = 0.81 \pm 0.27$, $<n /n(MH)> =
1.04 \pm 0.50$, and $<h_r / h_r(MH)> = 1.02 \pm 0.19$, where the
uncertainties are standard deviations of the mean of the measured
ratios.  However, for the remaining 5 galaxies the bulge and/or
disk parameters obtained by MH deviated significantly from our values.
For two of them, NGC 4450 (see Fig. 2) and NGC 4051, inclusion of a
bar is the likely explanation for the different results.  The bar is
very large in NGC 4051 and including it in the decomposition makes
the underlying exponential disk also large ($h_r$ =
70\rlap{.}$^{\prime\prime}$0 vs. 49\rlap{.}$^{\prime\prime}$6).  In
NGC 4450 the bar is smaller and therefore does not affect $h_r$, which
is also found to be similar in the two studies. However fitting a bar
for this galaxy affects the bulge model, thus making it more
exponential ($n$=3.6 $\rightarrow$ 1.8).  The decomposition parameters
are different in the two studies also for NGC 5248, for which we
measure a much smaller $h_r$ than MH ($h_r$ = 71''.3 v.s. 194''.1),
and for the two non-barred galaxies, NGC 4254 and NGC 2196.

With the sample by Knapen et al. we have 14 galaxies in common. In spite
of the 1D-nature of their decomposition, even in this case for half of 
the galaxies they measure similar structural parameters as obtained by us.
For these galaxies  $<r_e / r_e (Knapen)>
= 0.80 \pm 0.30$, $<n / n(Knapen)> = 1.03 \pm 0.28$, and $<h_r / h_r(Knapen)>
=1.02 \pm 0.20$. But again, for half of the galaxies the structural 
parameters are quite different. This is the case for
NGC 864, 4051, 5850 and 5921 for which the differences can be explained 
by the bar. For example
NGC 864 (shown in Fig. 2) and NGC 5921 have small bars that affect 
the bulges, whereas large bars in NGC 4051 and NGC 5850 modify
also the exponential disks. The parameters are somewhat different
also for the non-barred galaxies NGC 4689, 5247 and 2775. For example,
for NGC 2775 we measure a significantly larger $h_r$ than Knapen et al. ($h_r$
= 43''.1 v.s. 26''.2), but the value we get is similar to that obtained 
by MH ($h_r$ = 44''.3).

As expected 
the structural parameters derived from 2-component models quite 
often disagree with the parameters derived using 3-component models:
fitting a large bar in the decomposition
increases $h_r$ and modifies the bulge model, while fitting 
a small bar only makes the shape of the bulge
to appear more exponential. But bars do not explain
all the differences found between the various studies:
also the image quality may easily affect the decomposition results.

\section{Method for calculating the perturbation strengths}
Perturbation strengths induced by non-axisymmetric structures
in galactic disks are calculated using the gravitational 
torque method (GTM), which quantifies bar and spiral strengths 
using a simple force ratio (Sanders \& Tubbs 1980; Combes \& Sanders 
1981) based on gravitational potentials inferred from near-IR
light distributions. BB first applied the GTM
to a large sample of galaxies using a Cartesian potential evaluation
(Quillen, Frogel, \& Gonzalez 1994). Here we instead use a polar grid 
approach to infer the potentials, mainly because it reduces the noise,
thus better taking into account the faint
outer parts of the images. The method is explained and applied to
a large sample of 2MASS images by LS. The polar approach was initially 
applied by Salo et al. (1999) to an early-type ringed barred spiral 
IC 4214. The refined method used in this study is explained in detail 
by Salo, Buta \& Laurikainen (2004).
The analysis provides two-dimensional maps of radial ($F_R$) and
tangential ($F_T$) forces in the galaxy plane. A radial
profile of the maximum relative tangential force at each distance is 
calculated as
$$Q_T(r) = { ~~~~~|F_T(r,\phi)|_{max} \over <|F_R(r,\phi)|>},$$

\noindent 
where $<|F_R(r,\phi)|>$ denotes the azimuthally averaged axisymmetric
force at each radius. While constructing the $|F_T|_{max}$ at each radius
we use the mean of the maximum $|F_T|$ over azimuth found separately
in four different image quadrants. The maximum in the radial 
$Q_T$-profile then gives a single measure of bar strength, $Q_g$, which is 
equivalent to the maximum gravitational bar torque per unit mass
per unit square of the circular speed. 
Another useful parameter is $r_{Qg}$, which gives the radial distance 
of the maximum perturbation strength. $Q_g$ is generally associated with a
bar, but in some cases it can also be related to spiral arms. It is also 
possible that the bar and the spiral are partially overlapping in
some galaxies and a method to 
separate the two components, based on Fourier techniques,
has been developed by Buta, Block, \& Knapen (2003).

The bar-induced gravitational potential is calculated from a deprojected
$H$-band image using generally 10 even Fourier components, 
but the effect of the higher order modes on $Q_g$ was also tested. 
Because we were able to decompose the bulges from the disks and bars,
the deprojected images are not affected by the bulge "deprojection
stretch" that would affect the inner regions if one were to assume that the
bulge and disk have the same flattening. The bulge model is subtracted
from the sky-plane image, the disk is deprojected to face-on orientation, 
and the radial contribution of the force 
is added back by assuming that the bulge is spherically
symmetric. Also, this contribution is calculated from the analytical
bulge model, corresponding to the seeing-deconvolved bulge image.
The deprojections were made using the orientation parameters
derived from the $B$-band images (see Section 3). The deprojected images 
for all the sample
galaxies are shown in Fig. 4. These images are logarithmic, sky-subtracted,
and in units of mag arcsec$^{-2}$ with an arbitrary zero point. The
scales are all different and the displays are designed to show as
much information as possible over a full range of surface brightness.

The main assumptions of the GTM are that the near-IR light distribution 
traces the mass of the galaxy,
the mass-to-luminosity ratio is constant in all parts of the disks in the
$H$-band, and that the vertical density
distribution of the disk can be represented by some simple function like an
exponential. The scale height $h_z$ was estimated from the empirical
correlation between $h_r/h_z$ and the de Vaucouleurs' type index $T$ 
given in RC3 (de Grijs 1998).
The measurements for the barred galaxies are shown in Table 3 and
for the non-barred galaxies in Table 4, where the errors are standard
deviations of the measurements calculated in the four 
image quadrants.
In Table 3 we give also the distance of the maximum $Q_T$ ($r_{Qg}$), and the 
maximum $m$=2 ($A_2$) and $m$=4 ($A_4$) density amplitudes in the bar region. 
If the galaxy has a bar and some other $Q_T$ maximum due to the spiral arms 
well outside the bar, the value corresponding to the bar is given in the table.
The Fourier method enables us also to estimate the length of the bar, 
$r_{bar}$,
based on the phases of the m=2 and m=4 density amplitudes: the length of the 
bar was taken to be the radius at which the m=2 and m=4 phases were maintained 
nearly constant in the bar region. The length of the bar is given for the
galaxies for which a bar was identified by Fourier methods, as explained in Section 6. 

The last column in the table lists the visual bar classifications in the 
$H$-band (EFP) and the bar classifications given in RC3.
The radial $Q_T$-profiles for all galaxies in the sample are shown 
in Fig. 5.

The uncertainties related to the bar torque method are discussed 
in several previous papers. The method was found to be rather insensitive 
to the functional form of the vertical density distribution (LS), to the 
contribution
of the dark matter halo (BLS), and to the radial variations 
in the vertical scale height (LS). At most each of these
factors can affect $Q_g$ only by $\approx 5\%$. BLS
also studied how the position angle of the bar, 
relative to the line of nodes could modify $Q_g$:
they found that $Q_g$ is slightly
weaker for the galaxies where the bar becomes ``thicker'' in deprojection.
Also, bars can be partly superimposed with the spiral arms which
can lead to an uncertainty of about 4$\%$ on $Q_g$ (Buta, Block, 
\& Knapen 2003). However, more important sources of error are the 
vertical scale height of the disk, which can be estimated only indirectly 
from observations, and the uncertainties in the orientation parameters, 
which both can account for uncertainties in $Q_g$ as much as 
$10-15 \%$ (BLS; LS). 

In the following we take a statistical approach and evaluate how much
$Q_g$ is affected on the average, if the uncertainties in the orientation
parameters, in the vertical scale heights and in the number of Fourier modes
included to the calculation, are taken into account: 
$Q_g$ measurements 
were repeated for all the galaxies in the sample by varying one 
parameter at a time.
We also evaluate how important the bulge correction is, both statistically 
and for some individual galaxies. 

The effect of orientation parameters was tested by repeating the
measurement twice: by adding $\Delta q$=0.025 to axial ratios in one
run and by subtracting the same value in another run. Similarly, two
set of runs were made by using $\Delta \phi=\pm 5^\circ$ (the range of 
uncertainties corresponds to the mean of the absolute deviations in
comparison to Garcia-Gomez et al. (2004), found in Section 2). Denoting
$\Delta (Q_g)_q = Q_g(q+0.025) - Q_g(q-0.025)$ and $\Delta (Q_g)_\phi =
Q_g(\phi+5^\circ) - Q_g(\phi-5^\circ)$, we obtain

$$ \eqalign{
<|\Delta (Q_g)_q|> &= 0.016, \cr
<|\Delta (Q_g)_\phi|> &= 0.030. \cr
}$$

\noindent Similar uncertainties are obtained if just barred galaxies are
considered.  Note that these are very conservative upper limits for
the actual uncertainties, since the expected errors for
individual galaxies have opposite signs with equal probabilities,
effectively canceling each other in sample averages.

Another possible uncertainty in our method is the sufficiency of the
adopted number of Fourier components used ($m_{max}=10$).  To check
for this, we repeated all our measurements using all even components up
to $m_{max}=20$ (denoted by $(Q_g)_{20}$). The resulting difference
turned out to be completely insignificant

$$ \eqalign{
< (Q_g)_{20}/Q_g   > &= 1.009 \pm 0.06 \cr
<|(Q_g)_{20} -Q_g| > &= 0.004. \cr
}$$

In practice a much larger uncertainty in $Q_g$ is caused by the
assumed vertical scale height, which might also have systematic
errors. In order to assess this we again repeated our measurements
twice, using for each galaxy both the maximum and minimum $h_z$ values
implied by the variation of $h_z/h_r$ among each morphological type
found by de Grijs (1998): the maximum $h_z/h_r = 1, 1/3, 1/5$ for
morphological types $T \le 1, 2 \le T \le 4, T \ge 5$, respectively,
while the minimum is $h_z/h_r = 1/5, 1/7, 1/12$ (standard average values
used were $h_z/h_r = 1/4, 1/5, 1/9$).  In comparison to using the
standard values of $h_z$ we obtain the mean ratios (and standard
deviation of individual ratios)

$$ \eqalign{
<({Q_g})_{\rm min ~ h_z}/Q_g> &= 1.08 \pm 0.04 \cr
<({Q_g})_{\rm max ~ h_z}/Q_g> &= 0.81 \pm 0.12 \cr 
}$$

\noindent The relative uncertainty in $Q_g$ increases toward earlier Hubble
types, reflecting the larger range of uncertainty in the vertical scale height.
This trend is depicted in Fig. 6: for example, for $T \le 1$ the
systematic use of the minimal $h_z/h_r$ ratio yields $Q_g$ values about 
twice as large in comparison to what the maximal $h_z/h_r$ gives.
Nevertheless, the trend of average $Q_g$ increasing with $T$ (BLS) is
clearly not affected by this uncertainty.
The estimation of the vertical scale height is particularly 
uncertain for galaxies with prominent outer rings (NGC 210, NGC 1808 and
NGC 5701), because it is not clear whether the empirical
relation between the vertical scale height and the radial 
scale length of the disk is valid also for this type of galaxy.

The applied bulge correction is statistically not very important: 
for the barred galaxies in our sample we
obtain

$$ \eqalign{
<{Q_g}_{no bulge}/Q_g> &= 1.07 \pm 0.32, \cr
}$$

\noindent where ${Q_g}_{no bulge}$ refers to values calculated with no special
treatment of bulges, except that the $Q_T$ maxima occurring
conspicuously near the center have been suppressed, by limiting the
search of the maximum beyond 5 image pixels from the center (similar
treatment as applied in LS and in Block et al. (2004); in the case when
the bulge correction was applied no such caution was needed).  However,
the influence of artificial bulge stretch can be large for some
individual galaxies, especially if the bars/ovals reside inside a
large bulge. Fig. 7 (left panel) displays the difference in the $Q_q$
values with and without correcting for the spherical bulge (but using
the above defined 5 pixel limit), as a function of $r_{bar}/r_{\rm eff}$.
Clearly, for
$r_{bar}/r_{\rm eff}$ less than about 10, the bulge stretch starts to
affect the measured force ratio.  Note that the bulge stretch can
either increase or decrease $Q_g$ depending on whether the uncorrected
bulge projects along or perpendicular to the bar in deprojection to
face on orientation. On the other hand, if no limitation were placed
on the location of the $Q_T$ maximum (Fig. 7, right panel), a slight
systematic increase in the case of no bulge correction starts to be
evident $(<{Q_g}_{no bulge}/Q_g> = 1.16 \pm 0.45>)$, due to strong
artificial force peaks sometimes produced near the center.
Altogether, although the effect of the bulge correction is minor, for
example, when calculating the distribution of bar strengths (BLS),
it enables us to better analyze the force profiles also in the bulge
region. The correction assuming a spherical
bulge light distribution, and the case when the bulge light is not
separated from the disk, represent two extreme treatments of a bulge.
The small difference in the obtained results suggests that, at least
for statistical purposes, there is no obvious need for a more refined
bulge correction. Any uncertainty in $Q_g$, possibly originated using 
the Ferrer's function
for the bar model in the decomposition, is included in the estimated 
uncertainty due to the bulge.

\section{Bar morphologies}

As not all $Q_T$-maxima are associated with bars we have to specify
what do we mean by a bar. Our main criteria were 
that significant $A_2$ and $A_4$ amplitudes of density 
could be detected, and that their phases were maintained 
nearly constant in the bar region (the weakest detected bar in our 
sample has $A_2$=0.12). Another powerful tool to
identify bars is to inspect the polar angle maps
presented in logarithmic radial scale: bars generally have clear 
density condensations in their outer
parts and the polar angle is not 
changed much in the radial direction. Also, in the presence of
a bar two well defined maxima and two minima appear in the
$Q_T$ map in the form of a ``butterfly pattern''. The main limitation of these
diagnostics is that no difference is made between
a bar and an oval, which have to be distinguished by other means.
Ovals are more round structures and generally
also have lower surface brightnesses than bars.

In most cases the above criteria unambiguously distinguished the bar, but 
there are some galaxies which had to be inspected more carefully. 
There were 7 galaxies (NGC 3646, 3810, 3949, 4212, 4242, 5054 and
5085) that were not classified as barred, although the m=2 phases
were maintained nearly constant at a certain radius of the disk. 
In NGC 3646, 3949,
3810 and 5054, the elongated structure appeared to be an
oval, which was evident in the original images and
also while looking at the ``butterfly patterns''.
The ``butterfly patterns'' of these galaxies showed structures where 
the angle between the location of the $Q_T$-peak and the bar/oval axis, 
$\alpha$, was $55^0$, $42^0$, $59^0$ and $45^0$, respectively. These
kinds of large angles 
are typical for ovals. For NGC 4212 $\alpha$ is only $\ 30^0$, but there is
no m=4 density amplitude typical for bars in the assumed bar region.
For NGC 5085 the ``butterfly pattern'' shows a typical spiral like 
nature with four symmetrically distributed curved structures, which 
was the main reason why this galaxy was classified as non-barred.
NGC 4242 looks like a peculiar galaxy with extremely diffuse structure,
so that if there is a bar, it must also be diffuse. However, because of
the lack of any clear sign of a bar it was classified as non-barred.

In RC3 bars are assumed to be strong when they belong to de
Vaucouleurs' class B, 
and weak when they belong to AB-type systems. Similar classifications
were made in the near-IR by EFP using the OSUBGS sample. In the optical
71$\%$ of the galaxies in our sample were found to have bars (B + AB), 
whereas in the
near-IR the fraction of barred galaxies is 78 $\%$, which means that 
a small fraction of bars in the optical are obscured by dust in the near-IR. 
However, the main difference between the 
classifications in the two wavelength regions is that in the 
near-IR bars belong to the category 
B much more often than bars in the optical region (62 $\%$ v.s 37 $\%$ of all
galaxies). 
In principle this can be an artifact, because there are only three
bar bins, SA, SAB, and SB. Bars probably look stronger to some extent
in the near-IR, so that SAs can shift to SABs, and SABs can shift to
SBs, but there is no extra bin for SBs to shift into. Thus, near-IR
bars might pile up into the SB bin. The Fourier method picks up mainly
the SB-type bars, which form 90 $\%$ of all barred galaxies as identified by
Fourier methods. Also,
except for 4 galaxies, NGC 4242, 4504, 4772 and 6907 (which has a minibar), 
all galaxies classified as
SB by EFP are barred also according to our Fourier analysis.
Additionally, among the barred galaxies there are also 8 SAB-type systems 
and 1 SA galaxy as classified by EFP in the near-IR. It seems
that some well defined criteria is needed to judge whether a galaxy has a bar
or not.

Does the above then mean that the Fourier method is capable of
picking up only the strong bars? The amplitude limit 
of the m=2 Fourier component for the weakest detectable bar in 
our sample is $A_2$=0.12, which is very similar to the lower
limit of $A_2$=0.09 for the SAB-type galaxies. The weakest detected
main bar in our sample has $Q_g$=0.06, which is approximately
the lower limit of the weakest bars in the classification by BB ($Q_g$=0.05).
This means that the Fourier method is capable of detecting both strong
and weak bars. An interesting question is then why 68 $\%$ of 
the visually detected SAB-type galaxies in the near-IR
turn out to be non-barred based on the Fourier method.
In the following we discuss the morphological properties of some typical
barred and non-barred galaxies in our sample.

\vskip 0.5cm
{\it Barred galaxies}:
\vskip 0.25cm

{\bf NGC 4151:} (Fig. 8a) This 
galaxy is classified as barred by EFP, and has an intermediate  
type bar in RC3. According to all our indicators this galaxy has a bar, 
but it seems not to be well developed: only two material condensations 
at the outer edges of the bar along the bar major axis are seen,
giving it ``ansae" type characteristics. This type of bar is rare
in our sample, but is more typical for very early-type galaxies 
(see for example Buta 1995, 1996). 
\vskip 0.25cm

{\bf NGC 5701:} (Fig. 8b) is a galaxy with a bar and a well-defined
outer pseudo-ring of type R$_2^{\prime}$ (see Appendix 3, RC3, and
Buta 1995). The ring is better defined
in the B-band image. The bar itself appears
inside an oval/lens feature. The bar has clear 
$A_2$ and $A_4$ amplitude peaks and their phases 
are maintained nearly constant in the bar region. Also, the ``butterfly 
pattern'' forms four symmetric well defined regions, as expected
for a bar.
The m=2 amplitude peak in the polar angle map is very wide,
which is the reason that also the $Q_T$-profile is wide. In this case no
spiral arms are visible so that the wide amplitude and $Q_T$-profiles
seem to be characteristic for the bar itself.
\vskip 0.25cm

The galaxies NGC 5101, 3275 and 3504 are all cases having a
classical well developed bar and an inner ring located at the 
distance of the outer edge of the bar. But they also have other
characteristics that make them unique compared with the other galaxies. 
\vskip 0.25cm

{\bf NGC 5101:} (Fig. 8c; see also Fig. 2) This is another case of an "ansae"-type bar 
with two strong
material condensations at the outer edges. The galaxy
has an inner ring filled by the bar along its major axis.
This galaxy also has two amplitude maxima in the polar angle map,  
appearing along the bar major axis. The outer maximum
is located at the radius of the inner ring, showing also short spiral-like 
arm segments, which might be relics of the ring formation. 
The $A_2$-profile for this galaxy is asymmetric, 
declining rather rapidly after the maximum, which is manifested also
in the $Q_T$-profile. A natural 
explanation for this decline is that the bar ends near to the inner ring.
Like NGC 5701, NGC 5101 has an R$_2^{\prime}$ outer pseudo-ring in
blue light (RC3 Appendix 3).

\vskip 0.25cm
{\bf NGC 3275:} (Fig. 8d) is also an example of a bar/ring system,
but in this case there are also large scale spiral arms outside 
the bar. The $Q_T$-profiles in NGC 5101 and NGC 3275
have different shapes. The broad hump in the
$Q_T$-profile of NGC 3275 is caused by the strong material condensations
at the two ends of the bar coinciding with the radius of the inner ring.
It is not clear in the direct image whether these blobs are part of the 
inner ring or part of the large scale spiral arms.
\vskip 0.25cm

{\bf NGC 3504:} (Fig. 8e; see also Fig. 2) is another example of a 
bar/ring system
with prominent spiral arms, which are forming a ring. This is a good example
of galaxies showing a bar with two blobs at the ends of the bar, manifesting
also some spiral-like characteristics. The amplitude profile also 
has two peaks in the bar region. In this case the
tiny spiral arm segments are not related to the outer spiral
arm structure, being rather part of the bar/inner ring system.
The $Q_T$-profile is asymmetric, but in the opposite direction 
than in NGC 5101. 
The outer pseudo-ring of NGC 3504 is type R$_1^{\prime}$ (RC3 Appendix 3).
\vskip 0.25cm

{\bf NGC 4548:} (Fig. 8f) is a galaxy with a bar and spiral arms, but
no inner ring attached at the radius of the bar. The spiral arms
start as arc-like structures at the two ends of the bar and continue
to form global spiral patterns. It seems that the formation of the 
inner ring is not yet completed. In this case the $A_2$ amplitude profile is 
affected by the partial superposition of the global spiral arms.
This kind of bar/spiral structure is
quite typical in the OSUBGS sample, being found also in NGC 150,
289, 864, 1832, 2566, 3261, 2548, 4579, 5493, and 4781. For this 
kind of galaxy it would be valuable to apply the bar/spiral
separation method (Buta, Block, \& Knapen 2003).
\vskip 0.25cm

The two barred galaxies, NGC 7552 and NGC 7479, are examples of systems
having a bar and prominent two-armed spirals.  
\vskip 0.25cm

{\bf NGC 7552:} (Fig. 8g; see also Fig. 2) is a peculiar case of a bar and two spiral arms 
forming an outer pseudo-ring (type R$_1^{\prime}$, Buta 1995). One of 
the spiral arms, starting from the
end of the bar, is more prominent than the other arm, which seems to
have no clear continuation with the bar. The bar itself is very dusty 
and shows a lot of structure. The polar angle map shows how the spiral 
arm attached
to the bar is superimposed with the bar.
While looking at the original image this galaxy has  
similarities with NGC 3504: in both cases there are blobs at the two ends 
of the bar
showing some characteristics of spiral arms. Also, as in NGC 3504, 
the small spiral arm segments are not a continuation of the 
global spiral patterns. The main difference as compared to NGC 3504 is that 
NGC 7552 has no inner ring.

{\bf NGC 7479:} (Fig. 8h) is a late-type galaxy having a bar and two strong
spiral arms starting from the two ends of the bar. In this case two spiral arms
form a global pattern in the disk. The arms are also forming a ring
at the radius of the bar major axis. The $A_2$ density maximum is probably
a combination of both bar and spiral arms.

\vskip 0.5cm
{\it Non-barred SAB-type galaxies}:
\vskip 0.25cm

The two galaxies, NGC 1371 and NGC 5054 are chosen to illustrate 
typical SAB-type galaxies as defined by EFP in the near-IR. Some of the 
SAB-type bars in the infrared might also be spiral arms, that look
like a bar in projection. Such an example is NGC 4504, which is
classified as non-barred in RC3. 
\vskip 0.25cm

{\bf NGC 1371:} (Fig. 8i; see also Fig. 2) is classified as SAB both in RC3 and by EFP
in the near-IR. It has a peak in the $Q_T$-profile,
typical for barred galaxies, but our main bar indicators do not
show a bar: the m=2  phase is not maintained 
constant in the assumed bar region, and the ``butterfly pattern'' shows
a spiral-like nature. Also, in the direct image the structure in the region
of the $Q_T$-peak shows a spiral-like nature. However, while looking at 
the polar angle map it is interesting to 
note that this galaxy has blobs at r=20'', which coincide with
the peak in the $Q_T$ -profile. Therefore it is possible that this galaxy has 
a curved weak bar. Other similar cases in our sample
are NGC 1042 and possibly also NGC 3596.
\vskip 0.25cm

{\bf NGC 5054:} (Fig. 8j) Another more typical group of SAB-type 
galaxies are those
clearly having no bar, the non-axisymmetric forces being rather 
induced by ovals. 
Such galaxies are for example NGC 3893, 4254, 5054, 5085, 6215,
7205 and 7412, of which NGC 5054 is shown as an example.
The oval is seen at $r<20''$ having nearly constant m=2 phase
in the oval region and a bar-like ``butterfly pattern''. The
oval can be distinguished from a bar by its small major to minor axis ratio,
and also by the relatively large angle between the two amplitude maxima
in the butterfly pattern ($\alpha = 45^0$). The tangential forces 
induced by the oval 
are relatively weak for this galaxy, which is generally the case also 
for the other galaxies with ovals in our sample.

\section{Summary}

We have calculated perturbation strengths for a well defined magnitude limited
sample of 180 galaxies, based on 158 galaxies from the Ohio State University
Bright Galaxy Survey (OSUBGS) and 22 galaxies from the 2 Micron 
All Sky Survey (2MASS). We use a gravitational bar torque
method, in which the ratio of the tangential force to the azimuthally
averaged mean axisymmetric
radial force is used as a measure of the perturbation strength. The 
gravitational potential is inferred from a two-dimensional $H$-band light 
distribution using a Polar method, assuming that the light traces the
mass and that the M/L-ratio is constant throughout the disk.
The data presented here have been previously used to derive
the distribution of bar strengths for spiral galaxies (BLS)
and to compare bar strengths in active and non-active galaxies 
(LSB). 

Special attention has been given for correcting the deprojected images
for artificial stretching of the bulge light.
For this purpose the orientation parameters of the disks were measured
from the relatively deep $B$-band images by fitting ellipses to the 
outer isophotes. We found that for some of the galaxies the orientation 
parameters given in RC3 deviated significantly from our values. However,
the correlation between our measurements and those by Garcia-Gomez
et al. (2004) was quite good.
Artificial stretching of the bulges
was avoided by separating the mass of the bulge from that of the disk.
The bulge mass was then added back after deprojecting the disk to face-on 
orientation by assuming that the mass of the bulge is spherically distributed. 
This correction is important for the estimation
of bar strengths for galaxies in which the bars appear inside 
large bulges. 

We used a three-component 2D decomposition method where the disks were
modeled by an exponential function, the bulges by 
the generalized S\'ersic function (1968), and the bar by
a Ferrer's function. To our knowledge
this is the first time that a 3-component 2D-method has been applied to a 
large sample of galaxies. We found that fitting the bar model 
in the decomposition is important: if a bar is not taken into
account in the decomposition, the flux of the bulge and the B/D-ratio are 
easily overestimated. And if the bar is large, omission of the bar model might
even underestimate the scale length of the disk. 

We have shown the importance of using some well defined criteria 
while identifying bars in galaxies.  
A Fourier method is used requiring that the phases of
the density amplitudes of the main modes of bars, $m$=2 and $m$=4, are
maintained nearly constant in the bar region. This method largely selects
SB-type galaxies (as classified in the near-IR), which form 90$\%$
of the barred galaxies in our sample. Additionally, 32 $\%$ of the
SAB-type systems were found to be barred. The advantage of our method
is that it is capable of identifying also weak bars, which in the
ellipticity profiles might be overshadowed by the luminous disks.

The most remarkable result of this study is that even 68 $\%$ of the 
SAB galaxies, as classified in the near-IR, most probably are
not barred galaxies at all. They are typically systems having a 
central oval and in some cases manifesting strong two-armed or 
multiarmed spiral arms, which all are capable of inducing tangential
forces at some level. The perturbation strengths induced by ovals
are relatively weak being peaked at $Q_g$=0.15, whereas the tangential forces induced by 
spiral arms can be either very weak (multiarmed spirals) or strong
(two-armed spirals) amounting up to $Q_g$=0.6. However, compared
to bar induced perturbations, similar tangential forces induced by 
spirals are effectively weaker in the central parts of the galaxies, 
because the force maxima appear at larger distances from the 
galaxy centers. As noticed by ESB, the distributions of $Q_g$ 
and $r_{Qg}$ for SA and SAB type galaxies are remarkably similar,
except for the innermost bins (see their fig. 9) thus indicating 
a similar origin of the tangential forces. In fact, 30 \% 
of the SA type galaxies in our sample have ovals inducing 
similar perturbation strengths as found for many 
SAB-type systems. This is consistent with the picture
outlined by Kormendy (2004), where ovals in all de Vaucouleurs's 
family classes might play some role for the origin of the non-axisymmetric 
forces in galaxies. 

In addition to these typical cases, there are also galaxies
among SAB-type systems that might have weak non-classical bars.
These bars have a spiral-like nature (see, for example,
Jogee et al. 2002), and might form in
weakly centrally concentrated disks. However, more thorough investigation
is needed to verify their true nature.

\section*{Acknowledgments}

This publication makes use of data products from the Two Micron All
Sky Survey, which is a joint project of the University of
Massachusetts and the Infrared Processing and Analysis
Center/California Institute of Technology, funded by the National
Aeronautics and Space Administration and the National Science
Foundation. It also uses the NASA/IPAC 
Extragalactic Database (NED), operated by the Jet Propulsion
Laboratory in Caltech. 
EL and HS  acknowledge the Magnus Ehrnrooth Foundation and the
Academy of Finland of significant
financial support. RB and SV acknowledge the support of NSF Grant
AST-0205143 to the University of Alabama. We also acknowledge
Paul Eskridge of many useful discussions.

\vfill
\eject

\subsection{Figure captions}

\hskip 0.5cm {\bf Figure 1.} The orientation parameters measured in this study are
compared with the RC3 values and with those obtained by Garcia-Gomez
et al.  (2004). We plot the differences in the position angle $|\Delta
\phi|$ and in the minor-to-major axis ratio $|\Delta q|$, versus the
mean of $q$ for each pair of sources. The two upper panels compare our
values with those measured by Garcia-Gomez et al, and the middle
panels with the RC3 values. Comparison between RC3 values and those by
Garcia-Gomez et al. is shown in the lower panels.
\vskip 0.25cm

{\bf Figure 2.} Decomposition results using
a 2D-method, where the disks are modeled by an exponential function, the
bulges using the generalized Sersic's model, and bars by a Ferrer's
function. Shown is the observed brightness of each pixel vs. its
distance from the center, measured in the sky plane (black points), and
the corresponding model components: bulge (gray points falling on a
curve), disk (gray points in a wedge shaped region) and bar (points in
a region limited by two curves). The total model is shown by dark
gray points on top of observations.  As no flux calibration was done,
the magnitude scale in the y-axis has an arbitrary zero-point.
In a) the effect of including/excluding the bar model
in the decomposition is shown for NGC 1187 and NGC 1302, while b)
shows other selected cases discussed in the text.
\vskip 0.25cm

{\bf Figure 3.} Images related to decomposition of NGC 1302: in the
upper row a bar component is included to the decomposition whereas in
the lower row it is ignored. The frames show the residual images where
the bulge model is subtracted from the original image (right), and
when the complete galaxy model is subtracted (left).
\vskip 0.25cm

{\bf Figure 4.} The deprojected images for the whole sample. 
The images are logarithmic, sky subtracted, and in units of 
$mag / arcsec^2$ with an arbitrary zero point.
\vskip 0.25cm

{\bf Figure 5.} Radial profiles of the perturbation strength 
$Q_T(r) = {|F_T(r,\phi)|_{max} \over <|F_R(r,\phi)|>},$
where  $|F_T(r,\phi)|_{max}$ is the maximum 
tangential force, and $<|F_R(r,\phi)|>$ is the mean 
azimuthally averaged axisymmetric force at each radius.
The vertical bar denotes the length of the bar, estimated
from the phases of the m=2 amplitude of density, assuming
that it is maintained nearly constant in the bar region.
\vskip 0.25cm

{\bf Figure 6.} 
The maximal effect of the $h_z/h_r$ uncertainty as a function
of morphological type. The solid line shows the average value of $Q_g$
obtained using the assumed standard $h_z/h_r$ ratios (error bars
correspond to the standard deviation of the mean), while the dashed
(dotted) lines correspond to the mean values obtained when using
systematically the minimum (maximum) ratio of $h_z/h_r$ for each
galaxy (see text for details).
\vskip 0.25cm

{\bf Figure 7.} 
 Effect of the applied spherical bulge correction for the
barred galaxies of our sample. The difference between $Q_g$ values in
the case without and with the bulge correction is shown as a function
of $r_{bar}/r_{eff}$, where $r_{eff}$ is the effective radius of the
bulge and $r_{bar}$ is the bar length. On the left, the $Q_T$ maxima
occurring very near the center have been suppressed for the
non-corrected measurements, by limiting the search of maximum beyond 5
image pixels from the center. On the right, no such limitation was
placed, leading on average to somewhat larger excesses.
\vskip 0.25cm

{\bf Figure 8.} For some example galaxies results are shown of the
Fourier method used to calculate the perturbation strengths.  In the
{\it upper left corner} is shown the original galaxy image in the $H$-band,
and {\it in the middle upper panel} the image where the m=0 component
is subtracted. Overlayed on the image is the ``butterfly pattern'',
which shows the regions of the maximum relative tangential forces in contours
divided to intervals of 0.1 bar strength units.  The dotted lines
indicate the regions where the tangential forces chance the sign.  In
the {\it upper right corner} are shown the radial $Q_T$-profiles
calculated in the four image quadrants (thin lines) and the average
profile is indicated by a thick line. The {\it lower left panel} is
the butterfly plot in log-polar coordinates: the contours and dotted
lines are the same as in the upper middle panel. In the {right lower
corner} are shown the m=2 and m=4 Fourier amplitudes and phases. In
all figures the radial distances are in arcseconds.
\vskip 0.25cm

\vfill
\eject

\begintable*{1}
\caption{{\bf Table 1.} Orientation parameters.}
\halign{%
\rm#\hfil&\qquad\rm#\hfil&\qquad\rm\hfil#&\qquad\rm\hfil
#&\qquad\rm\hfil#&\qquad\rm\hfil#&\qquad\rm#\hfil
&\qquad\rm\hfil#&\qquad\rm#\hfil&\qquad\hfil\rm#\cr
Galaxy    & $q \pm sd$ (B)    & $\phi \pm sd$ & range   & $q$   & $\phi$ &$d \phi$  & & & \cr
          &                   &               & [arcsec]& (RC3) & (RC3)  &               & & & \cr
\noalign{\vskip 10pt}

OSUBGS: &&&&&&&&& \cr
        &&&&&&&&& \cr
 ESO  138     & 0.708    $\pm$ 0.015    &  62.4    $\pm$ 3.7    & 420-460 & 0.72   &   55   &-0.22 & & & \cr
 IC  4444$^5$ & 0.826    $\pm$ 0.018    &  74.6    $\pm$ 1.9    & 120-140 & 0.83   &   -    &+3.66   & & & \cr
 IC  5325     & 0.837    $\pm$ 0.006    &  27.5    $\pm$ 2.2    & 210-250 & 0.91   &    -   &  0.00  & & & \cr
 NGC 0150     & 0.498    $\pm$ 0.009    & 107.6    $\pm$ 2.7    & 260-320 & 0.48   &  118   & +0.13  & & & \cr
 NGC 0157     & 0.724    $\pm$ 0.008    &  36.5    $\pm$ 1.2    & 180-210 & 0.65   &   40   &+1.44   & & & \cr
 NGC 0210     & 0.654    $\pm$ 0.009    & 163.2    $\pm$ 1.5    & 275-310 & 0.66   &  160   &-0.54   & & & \cr
 NGC 0278$^4$ & 0.927    $\pm$ 0.005    & 65.8     $\pm$ 3.6    & 39-44   & 0.95   &  -     &    & & & \cr
 NGC 0289$^3$ & 0.789    $\pm$ 0.014    & 141.5    $\pm$ 2.5    & 400-420 & 0.71   &  130   &0.00  & & & \cr
 NGC 0428     & 0.750    $\pm$ 0.013    & 100.5    $\pm$ 1.8    & 230-245 &  0.76  &  120   & +1.54& & & \cr
 NGC 0488     & 0.770    $\pm$ 0.010    &   5.6    $\pm$ 1.9    & 260-300 &  0.74  &   15   &+1.95 & & & \cr
 NGC 0578     & 0.589    $\pm$ 0.014    & 101.7    $\pm$ 1.5    & 345-375 &  0.63  &  110   &0.00  & & & \cr
 NGC 0613     & 0.772    $\pm$ 0.008    & 121.6    $\pm$ 1.5    & 360-430 &  0.76  &  120   &-2.09 & & & \cr
 NGC 0685     & 0.787    $\pm$ 0.019    & 104.3    $\pm$ 3.7    & 320-345 &  0.89  &   -    &+0.60 & & & \cr
 NGC 0864     & 0.842    $\pm$ 0.008    &  28.7    $\pm$ 4.7    & 160-172 &  0.76  &   20   &-0.05 & & & \cr
 NGC 0908     & 0.456    $\pm$ 0.008    &  72.4    $\pm$ 0.3    & 440-460 &  0.44  &   75   &+0.12 & & & \cr
 NGC 1042     & 0.781    $\pm$ 0.013    &   4.5    $\pm$ 1.9    & 300-330 &  0.78  &   15   &+1.57 & & & \cr
 NGC 1058     & 0.877    $\pm$ 0.016    &  23.3:   $\pm$ 3.3    & 190-200 &  0.93  &    -   &+1.52 & & & \cr
 NGC 1073     & 0.875    $\pm$ 0.013    &   1.0    $\pm$ 4.2    & 250-280 &   0.91 &   15   &+1.59 & & & \cr
 NGC 1084     & 0.753    $\pm$ 0.021    &  57.2    $\pm$ 1.1    & 210-230 & 0.56   &  115   &-0.07 & & & \cr
 NGC 1087     & 0.609    $\pm$ 0.012    &   2.8    $\pm$ 0.1    & 190-210 & 0.60   &    5   &0.00  & & & \cr
 NGC 1187     & 0.780    $\pm$ 0.019    & 132.0    $\pm$ 5.1    & 330-400 & 0.74   &  130   &+0.20 & & & \cr
 NGC 1241     & 0.575    $\pm$ 0.011    & 147.5    $\pm$ 1.4    & 145-160 & 0.60   &  145   &-0.17 & & & \cr
 NGC 1300$^1$ & 0.760                   & 150.0                 & 355     & 0.66   &  106   &-2.18 & & & \cr
 NGC 1302     & 0.941    $\pm$ 0.012    & 179.0:   $\pm$ 4.3    & 280-310 & 0.95   &   -    & -0.08 & & & \cr
 NGC 1309     & 0.926    $\pm$ 0.023    &  65.3    $\pm$10.2    & 170-190 & 0.93   &   45   &+0.27 & & & \cr
 NGC 1317     & 0.893    $\pm$ 0.013    &  64.5    $\pm$ 3.5    & 185-215 & 0.87   &   78   & -0.07 & & & \cr
 NGC 1350     & 0.516    $\pm$ 0.004    &   4.1    $\pm$ 0.4    & 430-500 & 0.54   &    0   & -2.25 & & & \cr
 NGC 1371     & 0.770    $\pm$ 0.009    & 127.6    $\pm$ 2.7    & 350-385 & 0.69   &  135   &+0.30  & & & \cr
 NGC 1385     & 0.626    $\pm$ 0.011    & 171.9    $\pm$ 2.1    & 260-300 & 0.59   &  165   &+0.25  & & & \cr
 NGC 1493     & 0.924    $\pm$ 0.022    &  51.3:   $\pm$ 6.7    & 300-340 & 0.93   &   -    &-0.09  & & & \cr
 NGC 1559     & 0.559    $\pm$ 0.009    &  63.2    $\pm$ 0.8    & 300-320 & 0.57   &   64   &+0.20  & & & \cr
 NGC 1617     & 0.489    $\pm$ 0.004    & 111.5    $\pm$ 0.4    & 280-340 & 0.49   &  107   &-      & & & \cr
 NGC 1637     & 0.823    $\pm$ 0.006    &  34.6    $\pm$ 1.6    & 200-230 & 0.81   &   15   &+1.67  & & & \cr
 NGC 1703     & 0.869    $\pm$ 0.019    & 165.1    $\pm$ 9.5    & 150-180 & 0.89   &  -     &+3.48  & & & \cr
 NGC 1792     & 0.453    $\pm$ 0.003    & 136.9    $\pm$ 0.1    & 300-330 & 0.50   &  137   &-0.44  & & & \cr
 NGC 1808$^2$ & 0.744    $\pm$ 0.013    & 119.5    $\pm$ 1.5    & 570     & 0.60   &  133   &+1.38  & & & \cr
 NGC 1832     & 0.702    $\pm$ 0.023    &  11.0    $\pm$ 3.4    & 140-160 & 0.66   &   10   &+0.41  & & & \cr
 NGC 2090     & 0.465    $\pm$ 0.000    &  18.1    $\pm$ 1.5    & 380-470 & 0.49   &   13   &-0.36  & & & \cr
 NGC 2139     & 0.859    $\pm$ 0.024    & 121.6    $\pm$ 8.3    & 140-180 & 0.74   &  140   &+3.56  & & & \cr
 NGC 2196     & 0.788    $\pm$ 0.018    &  44.3    $\pm$ 2.2    & 170-200 & 0.78   &   35   &+3.54  & & & \cr
 NGC 2207     & 0.716    $\pm$ 0.017    & 136.9    $\pm$ 1.5    & 285-300 & 0.65   &  141   &-0.36  & & & \cr
 NGC 2442     & 0.925    $\pm$ 0.009    & 110.7    $\pm$ 3.4    & 380-410 & 0.89   &   -    &-0.37  & & & \cr
 NGC 2559     & 0.527    $\pm$ 0.005    &   2.1    $\pm$ 0.4    & 230-260 & 0.46   &    6   &+3.46  & & & \cr
 NGC 2566     & 0.734    $\pm$ 0.018    & 115.8    $\pm$ 2.9    &  345-390& 0.68   &  110   &+3.57  & & & \cr
 NGC 2775     & 0.801    $\pm$ 0.009    & 163.5    $\pm$ 1.8    & 210-280 & 0.78   &  155   &+2.95  & & & \cr
 NGC 2964     & 0.566    $\pm$ 0.004    &  96.8    $\pm$ 0.4    & 160-180 & 0.55   &   97   &+1.45  & & & \cr
 NGC 3059     & 0.920    $\pm$ 0.022    &   3.3:   $\pm$ 8.6    & 245-280 & 0.89   &   -    &+3.37  & & & \cr
 NGC 3166     & 0.586    $\pm$ 0.013    &  82.5    $\pm$ 1.2    & 380-400 & 0.49   &   87   &+1.66  & & & \cr
 NGC 3169     & 0.776    $\pm$ 0.020    &  56.4    $\pm$ 2.7    & 370-400 & 0.63   &   45   &+1.64  & & & \cr
 NGC 3223     & 0.691    $\pm$ 0.018    & 125.0    $\pm$ 1.3    & 240-260 & 0.60   &  135   &+3.53  & & & \cr
 NGC 3227     & 0.661    $\pm$ 0.013    & 153.1    $\pm$ 1.0    & 155-170 & 0.67   &  155   &+1.96  & & & \cr
 NGC 3261     & 0.732    $\pm$ 0.024    &  71.0    $\pm$ 2.9    & 160-240 & 0.76   &   85   &+3.47  & & & \cr
 NGC 3275     & 0.936    $\pm$ 0.017    & 152.6    $\pm$ 5.7    & 125-155 & 0.76   &   -    &+3.49  & & & \cr
 NGC 3319     & 0.554    $\pm$ 0.012    &  32.8    $\pm$ 1.4    & 380-400 & 0.55   &   37   &+4.36  & & & \cr
 NGC 3338     & 0.495    $\pm$ 0.008    &  93.6    $\pm$ 1.3    & 230-280 & 0.62   &  100   &+4.36  & & & \cr
 NGC 3423     & 0.769    $\pm$ 0.012    &  31.2    $\pm$ 1.1    & 260-280 & 0.85   &   10   &+3.67  & & & \cr
 NGC 3504     & 0.980    $\pm$ 0.012    &  -                    & 145-170 & 0.78   &   -    &1.32   & & & \cr
 NGC 3507     & 0.944    $\pm$ 0.056    & 91.9     $\pm$ 0.9    & 160-180 & 0.85   & 110    &+1.43  & & & \cr
 NGC 3513     & 0.768    $\pm$ 0.016    &  75.4    $\pm$ 12.0   & 240-260 & 0.79   &   75   &+3.51  & & & \cr
&&&&&&&&& \cr
}
\endtable

\vfill
\eject

\begintable*{1}
\caption{{\bf Table 1.} continued}
\halign{%
\rm#\hfil&\qquad\rm#\hfil&\qquad\rm\hfil#&\qquad\rm\hfil
#&\qquad\rm\hfil#&\qquad\rm\hfil#&\qquad\rm#\hfil
&\qquad\rm\hfil#&\qquad\rm#\hfil&\qquad\hfil\rm#\cr
Galaxy    & $q \pm sd$ (B)    & $\phi \pm sd$ & range   & $q$   & $\phi$& $d \phi$  & & & \cr
          &                   &               & [arcsec]& (RC3) & (RC3)  &               & & & \cr
\noalign{\vskip 10pt}

 NGC 3583     & 0.744    $\pm$ 0.019    & 119.2    $\pm$ 2.3    & 155-170 & 0.65   &  125   &+2.35  & & & \cr
 NGC 3593     & 0.486    $\pm$ 0.021    &  86.2    $\pm$ 0.8    & 300-340 & 0.37   &   92   &+4.27  & & & \cr
 NGC 3596$^6$ & 0.829    $\pm$ 0.006    & 92.5     $\pm$ 5.3    & 235-260 & 0.95   &   -    &+1.45  & & & \cr
 NGC 3646     & 0.599    $\pm$ 0.010    & 50.4     $\pm$ 1.3    & 320-350 & 0.58   &  50    &+0.53  & & & \cr
 NGC 3675     & 0.494    $\pm$ 0.003    & 178.1    $\pm$ 0.3    & 240-270 & 0.52   &  178   &+2.54  & & & \cr
 NGC 3681     & 0.901    $\pm$ 0.017    &  34.9    $\pm$ 5.7    & 140-150 & 0.79   &   -    &+2.53  & & & \cr
 NGC 3684     & 0.704    $\pm$ 0.008    & 119.1    $\pm$ 1.0    & 185-200 & 0.69   &  130   &+1.47  & & & \cr
 NGC 3686     & 0.753    $\pm$ 0.008    &18.2      $\pm$ 1.0    & 150-190 & 0.78   &   15   &+1.31  & & & \cr
 NGC 3726     & 0.624    $\pm$ 0.011    &  15.6    $\pm$ 0.0    & 340-350 & 0.69   &   10   &+4.39  & & & \cr
 NGC 3810     & 0.680    $\pm$ 0.007    & 21.4     $\pm$ 1.1    & 260-300 & 0.71   &  15    &+1.71  & & & \cr
 NGC 3887     & 0.710    $\pm$ 0.012    &   4.6    $\pm$ 1.1    & 240-260 & 0.76   &   20   &+3.50  & & & \cr
 NGC 3893$^3$ & 0.595    $\pm$ 0.010    & 170.1    $\pm$ 0.3    & 247-260 & 0.62   &  165   &+1.77  & & & \cr
 NGC 3938     & 0.914    $\pm$ 0.020    &  37.2    $\pm$ 0.8    & 300-320 & 0.91   &    -   &+1.90  & & & \cr
 NGC 3949     & 0.904    $\pm$ 0.020    & 103.4    $\pm$ 9.6    & 184-200 & 0.57   &  120   &+4.35  & & & \cr
 NGC 4027     & 0.753    $\pm$ 0.009    & 163.1    $\pm$ 1.2    & 200-240 & 0.76   &  167   &+3.43  & & & \cr
 NGC 4030     & 0.729    $\pm$ 0.009    &  26.7    $\pm$ 3.1    & 260-300 & 0.72   &   27   &+3.47  & & & \cr
 NGC 4051     & 0.846    $\pm$ 0.154    & 128.0    $\pm$ 1.3    & 300-340 & 0.74   & 135    &+0.69  & & & \cr
 NGC 4123$^6$ & 0.677    $\pm$ 0.017    & 125.7    $\pm$ 1.4    & 270-320 & 0.74   & 135    &0.00   & & & \cr
 NGC 4136$^6$ &0.958     $\pm$ 0.015    & -                     & 290-320 & 0.93   &  -     &+0.71  & & & \cr
 NGC 4138$^4$ &0.598     $\pm$ 0.006    &148.2     $\pm$ 0.7    & 50-62   & 0.66   & 150    &      & & & \cr
 NGC 4145     & 0.572    $\pm$ 0.007    & 101.5    $\pm$ 0.5    & 380-410 & 0.72   & 100    &+0.59 & & & \cr
 NGC 4151$^1$ & 0.92                    & -                     & 400     & 0.71   & 50     &+0.73 & & & \cr
 NGC 4212$^6$ & 0.663    $\pm$ 0.017    &  75.7    $\pm$ 0.8    & 220-260 & 0.62   &   75   &+0.60 & & & \cr
 NGC 4242     & 0.672    $\pm$ 0.012    &  28.3    $\pm$ 1.8    & 250-310 & 0.75   &   25   &+4.38 & & & \cr
 NGC 4254     & 0.868    $\pm$ 0.012    &  57.4    $\pm$ 5.6    & 380-400 & 0.87   &   -    &+2.03 & & & \cr
 NGC 4293     & 0.463    $\pm$ 0.007    &  65.1    $\pm$ 0.4    & 310-390 & 0.46   &   72   &+0.73 & & & \cr
 NGC 4303     & 0.861    $\pm$ 0.011    & 146.9    $\pm$ 1.8    & 390-440 & 0.89   &   -    &+0.73 & & & \cr
 NGC 4314     & 0.959    $\pm$ 0.019    &  61.8    $\pm$ 15.1   & 240-280 & 0.89   &   -    &+1.90 & & & \cr
 NGC 4394     & 0.902    $\pm$ 0.009    & 103.0    $\pm$ 3.7    & 185-210 & 0.89   &   -    &+1.47 & & & \cr
 NGC 4414$^6$ & 0.644    $\pm$ 0.011    & 160.0    $\pm$ 0.9    & 290-320 & 0.56   &  155   &+0.33 & & & \cr
 NGC 4450     & 0.720    $\pm$ 0.009    &   2.2    $\pm$ 3.0    & 360-420 & 0.74   &  175   &+0.64 & & & \cr
 NGC 4457$^6$ & 0.883    $\pm$ 0.017    &  80.8    $\pm$ 2.1    & 230-250 & 0.85   &   -    &+0.53 & & & \cr
 NGC 4487     & 0.659    $\pm$ 0.008    &  73.8    $\pm$ 0.8    & 290-340 & 0.68   &  100   &+3.60 & & & \cr
 NGC 4490     & 0.441    $\pm$ 0.004    & 123.5    $\pm$ 0.5    & 320-380 & 0.49   &  125   & +0.66& & & \cr
 NGC 4496A$^6$& 0.788    $\pm$ 0.011    &  71.7    $\pm$ 1.6    & 280-300 & 0.47   &   -    &+0.52 & & & \cr
 NGC 4504     & 0.631    $\pm$ 0.025    & 142.5    $\pm$ 3.4    & 200-330 & 0.62   &   30   &+3.50 & & & \cr
 NGC 4527     & 0.456    $\pm$ 0.007    &  67.1    $\pm$ 0.6    & 420-440 & 0.34   &   67   &+0.68 & & & \cr
 NGC 4548     & 0.744    $\pm$ 0.009    & 153.2    $\pm$ 1.7    & 315-350 & 0.79   &  150   &+3.87 & & & \cr
 NGC 4571$^6$ & 0.821    $\pm$ 0.012    &  34.9    $\pm$ 4.6    & 245-300 & 0.89   &   55   &+0.50 & & & \cr
 NGC 4579     & 0.783    $\pm$ 0.008    &  94.8    $\pm$ 1.3    & 340-380 & 0.79   &   95   &+0.54 & & & \cr
 NGC 4580$^6$ & 0.712    $\pm$ 0.012    & 163.1    $\pm$ 1.5    & 115-140 & 0.78   &  165   &-3.09 & & & \cr
 NGC 4593     & 0.742    $\pm$ 0.012    &  98.1    $\pm$ 4.6    & 260-330 & 0.74   &   55   &+3.50 & & & \cr
 NGC 4618$^6$ & 0.807    $\pm$ 0.008    &  36.6    $\pm$ 0.0    & 250-280 & 0.81   &   25   &-1.68 & & & \cr
 NGC 4643     & 0.818    $\pm$ 0.012    &  56.0    $\pm$ 3.1    & 225-250 & 0.74   &  130   &+2.40 & & & \cr
 NGC 4647$^6$ & 0.663    $\pm$ 0.019    & 119.0    $\pm$ 1.2    & 220-240 & 0.79   &  125   &-0.60 & & & \cr
 NGC 4651$^6$ & 0.612    $\pm$ 0.127    &  73.1    $\pm$ 1.0    & 280-380 & 0.66   &   80   &+0.53 & & & \cr
 NGC 4654     & 0.563    $\pm$ 0.017    & 123.1    $\pm$ 3.2    & 200-400 & 0.58   &  128   &+0.65 & & & \cr
 NGC 4665     & 0.891    $\pm$ 0.013    & 102.6    $\pm$ 1.4    & 175-210 & 0.83   &   -    &+2.65 & & & \cr
 NGC 4689     & 0.734    $\pm$ 0.027    & 167.4    $\pm$ 1.4    & 230-350 & 0.81   &   -    &+0.66 & & & \cr
 NGC 4691$^6$ & 0.842    $\pm$ 0.023    &  41.2    $\pm$ 3.8    & 210-290 & 0.81   &   15   &+0.57 & & & \cr
 NGC 4698     & 0.566    $\pm$ 0.016    & 174.6    $\pm$ 1.6    & 280-420 & 0.62   &  170   &+0.67 & & & \cr
 NGC 4699     & 0.720    $\pm$ 0.016    &  41.2    $\pm$ 3.4    & 190-300 & 0.69   &   45   &+3.69 & & & \cr
 NGC 4772     & 0.503    $\pm$ 0.011    & 144.8    $\pm$ 1.5    & 120-200 & 0.50   &  147   &+2.49 & & & \cr
 NGC 4775     & 0.912    $\pm$ 0.026    &  66.9    $\pm$ 7.9    & 130-190 & 0.93   &   -    &+1.87 & & & \cr
 NGC 4781     & 0.452    $\pm$ 0.018    & 116.4    $\pm$ 2.3    & 150-280 & 0.45   &  120   &+3.62 & & & \cr
 NGC 4900$^6$ & 0.925    $\pm$ 0.015    &  96.1    $\pm$ 2.0    & 155-200 & 0.93   &   -    &+0.55 & & & \cr
 NGC 4902     & 0.915    $\pm$ 0.033    &  81.4    $\pm$ 16.1   & 160-200 & 0.89   &   70   &+3.54 & & & \cr
 NGC 4930     & 0.798    $\pm$ 0.005    &  55.0    $\pm$ 4.9    & 250-380 & 0.83   &   40   &+3.51 & & & \cr
 NGC 4939     & 0.534    $\pm$ 0.007    &   4.5    $\pm$ 0.4    & 420-440 & 0.51   &    5   &0.00  & & & \cr
 NGC 4941$^6$ & 0.470    $\pm$ 0.007    &  15.9    $\pm$ 0.7    & 250-280 & 0.54   &   15   &-1.09 & & & \cr
 NGC 4995$^6$ & 0.616    $\pm$ 0.005    &  96.4    $\pm$ 1.0    & 140-160 & 0.66   &   92   &-1.68 & & & \cr
 NGC 5005     & 0.444    $\pm$ 0.023    &  63.5    $\pm$ 0.6    & 180-300 & 0.48   &   65   &+4.41 & & & \cr
&&&&&&&&& \cr
}
\endtable

\vfill
\eject

\begintable*{1}
\caption{{\bf Table 1.} continued}
\halign{%
\rm#\hfil&\qquad\rm#\hfil&\qquad\rm\hfil#&\qquad\rm\hfil
#&\qquad\rm\hfil#&\qquad\rm\hfil#&\qquad\rm#\hfil
&\qquad\rm\hfil#&\qquad\rm#\hfil&\qquad\hfil\rm#\cr
Galaxy    & $q \pm sd$ (B)    & $\phi \pm sd$ & range   & $q$   & $\phi$ & $d \phi$  & & & \cr
          &                   &               & [arcsec]& (RC3) & (RC3)  &               & & & \cr
\noalign{\vskip 10pt}

 NGC 5054     & 0.612    $\pm$ 0.008    & 158.2    $\pm$ 0.5    & 320-380 & 0.58   &  155   &-0.34 & & & \cr
 NGC 5085     & 0.909    $\pm$ 0.013    &  49.5    $\pm$ 5.7    & 300-320 & 0.87   &   38   &-0.28 & & & \cr
 NGC 5101     & 0.934    $\pm$ 0.017    &  78.2    $\pm$ 4.1    & 310-330 & 0.85   &   -    &-0.47 & & & \cr
 NGC 5121     & 0.819    $\pm$ 0.017    &  26.8    $\pm$ 1.7    & 93-112  & 0.78   &   36   &+3.49 & & & \cr
 NGC 5247     & 0.831    $\pm$ 0.012    &  40.0    $\pm$ 3.1    & 420-470 & 0.87   &   20   &+1.34 & & & \cr
 NGC 5248$^4$ & 0.909    $\pm$ 0.036    &  104.0   $\pm$ 8.5    & 105-115 & 0.72   & 110    &      & & & \cr
 NGC 5334     & 0.760    $\pm$ 0.012    &  10.7    $\pm$ 3.2    & 250-340 & 0.72   &   15   &+1.86 & & & \cr
 NGC 5427     & 0.928    $\pm$ 0.013    & 154.0    $\pm$ 5.9    & 230-240 & 0.85   &  170   &+0.15 & & & \cr
 NGC 5483 $^7$& 0.886    $\pm$ 0.027    &  23.0    $\pm$ 7.7    & 210-250 & 0.91   &   25   &+3.43 & & & \cr
 NGC 5643     & 0.898    $\pm$ 0.027    & 131.2    $\pm$ 13.5   & 240-380 & 0.87   &   -    &0.00  & & & \cr
 NGC 5676     & 0.442    $\pm$ 0.005    &  45.6    $\pm$ 0.9    & 140-230 & 0.48   &   47   &+2.57 & & & \cr
 NGC 5701     & 0.913    $\pm$ 0.018    &  52.0    $\pm$ 4.2    & 330-400 & 0.95   &   -    &+3.65 & & & \cr
 NGC 5713     & 0.863    $\pm$ 0.029    &   3.9    $\pm$ 0.0    & 190-250 & 0.89   &   10   &+1.67 & & & \cr
 NGC 5850     & 0.866    $\pm$ 0.024    & 181.6    $\pm$ 6.8    & 320-400 & 0.87   &  140   &+2.99 & & & \cr
 NGC 5921     & 0.705    $\pm$ 0.013    & 130.9    $\pm$ 3.4    & 240-330 & 0.81   &  130   &+4.36 & & & \cr
 NGC 5962$^6$ & 0.660    $\pm$ 0.029    & 111.5    $\pm$ 3.5    & 160-250 & 0.71   &  110   &+0.44 & & & \cr
 NGC 6215     & 0.954    $\pm$ 0.026    & 118.8    $\pm$ 10.8   & 130-170 & 0.83   &   78   &+1.51 & & & \cr
 NGC 6221     & 0.654    $\pm$ 0.006    &  10.6    $\pm$ 0.2    & 285-320 & 0.69   &    5   &-0.03 & & & \cr
 NGC 6300     & 0.693    $\pm$ 0.000    & 104.8    $\pm$ 0.6    & 370-480 & 0.66   &  118   &-0.21 & & & \cr
 NGC 6384     & 0.646    $\pm$ 0.023    &  28.5    $\pm$ 1.9    & 300-500 & 0.66   &   30   &+0.74 & & & \cr
 NGC 6753     & 0.852    $\pm$ 0.020    &  29.3    $\pm$ 2.0    & 160-210 & 0.87   &   30   &+0.19 & & & \cr
 NGC 6782     & 0.898    $\pm$ 0.017    &  34.9    $\pm$ 4.4    & 160-210 & 0.66   &   45   &+0.26 & & & \cr
 NGC 6902     & 0.836    $\pm$ 0.029    & 149.5    $\pm$ 16.0   & 250-280 & 0.69   & 153    &-0.60 & & & \cr
 NGC 6907     & 0.876    $\pm$ 0.012    &  82.6    $\pm$ 4.0    & 175-200 & 0.81   &   46   &-2.12 & & & \cr
 NGC 7083     & 0.553    $\pm$ 0.008    &   6.2    $\pm$ 1.9    & 240-300 & 0.60   &    5   &-0.17 & & & \cr
 NGC 7205     & 0.515    $\pm$ 0.007    &  67.1    $\pm$ 1.4    & 210-350 & 0.50   &   73   &-0.16 & & & \cr
 NGC 7213     & 0.944    $\pm$ 0.015    &   1.9    $\pm$ 6.5    & 280-370 & 0.89   &   -    &-2.34 & & & \cr
 NGC 7217     & 0.861    $\pm$ 0.021    &  93.0    $\pm$ 2.8    & 190-250 & 0.83   &   95   &0.00  & & & \cr
 NGC 7412     & 0.621    $\pm$ 0.023    &  52.4    $\pm$ 2.3    & 320-380 & 0.74   &   65   &-0.07 & & & \cr
 NGC 7418     & 0.765    $\pm$ 0.008    & 132.5    $\pm$ 1.8    & 220-320 & 0.74   &  139   &+0.23 & & & \cr
 NGC 7479     & 0.741    $\pm$ 0.018    &  35.7    $\pm$ 2.5    & 195-240 & 0.76   &   25   &-0.04 & & & \cr
 NGC 7552     & 0.873    $\pm$ 0.027    & 169.9    $\pm$ 1.4    & 230-280 & 0.79   &    1   &+0.21 & & & \cr
 NGC 7582     & 0.471    $\pm$ 0.007    & 152.9    $\pm$ 1.0    & 300-450 & 0.42   &  157   &-2.31 & & & \cr
 NGC 7713     & 0.427    $\pm$ 0.007    & 171.7    $\pm$ 1.3    & 200-240 & 0.41   &  168   &-0.06 & & & \cr
 NGC 7723     & 0.693    $\pm$ 0.010    &  38.5    $\pm$ 1.4    & 160-210 & 0.68   &   35   &+1.65 & & & \cr
 NGC 7727     & 0.892    $\pm$ 0.014    & 158.1    $\pm$ 11.4   & 165-190 & 0.76   &   35   &+1.65 & & & \cr
 NGC 7741     & 0.690    $\pm$ 0.020    & 161.9    $\pm$ 4.1    & 160-210 & 0.68   &  170   &+1.47 & & & \cr
}
\tabletext{\noindent {$^1$} Manual determination.

\noindent {$^2$} From the outer ring.

\noindent {$^3$} The image field is too small for a good measurement.

\noindent {$^4$} The measurement was done using the H-band image.

\noindent {$^5$} The pixel size of the B-image is 0.40 arcsec instead of 0.27 aresec given in the image header.  

\noindent {$^6$} The pixel size of the B-image is 0.72 arcsec instead of 0.36 aresec given in the image header.

\noindent {$^7$} The pixel size of the B-image is 0.50 arcsec instead of 0.39 aresec given in the image header.
}

\endtable

\vfill
\eject

\begintable*{2}
\caption{{\bf Table 2.} Structural parameters from 2D-decompositions}
\halign{%
\rm#\hfil&\qquad\rm#\hfil&\qquad\rm\hfil#&\qquad\rm\hfil
#&\qquad\rm\hfil#&\qquad\rm\hfil#&\qquad\rm#\hfil
&\qquad\rm\hfil#&\qquad\rm#\hfil&\qquad\hfil\rm#\cr
Galaxy    & $r_{eff}$    & $\beta$ & $h_r$   & $B/D$   & bar/oval & & & & \cr
          &  [arcsec]    &         & [arcsec]&         &          & & & & \cr
\noalign{\vskip 10pt}

OSUBGS: & & & & & & & & & \cr 
 & & & & & & & & & \cr 

ESO  138         & 8.822    &0.495     &  43.8     & 0.098      &          & & & & \cr      
IC 4444          & 1.439    &1.466     &  11.0     & 0.020      & bar/oval  & & & & \cr
IC 5325          & 1.887    &1.223     &  21.4     & 0.017      & bar/oval  & & & & \cr
NGC 150          & 1.420    &0.904     &  23.7     & 0.065      & bar/oval  & & & & \cr
NGC 157          & 1.847    &1.418     &  25.0     & 0.022      &          & & & & \cr
NGC 210$^7$      & 3.928    &0.770     & 132.1     & 0.309      & bar/oval  & & & & \cr
NGC 278          & 3.117    &1.125     &  14.4     & 0.054      & bar/oval  & & & & \cr
NGC 289          & 2.346    &1.062     &  19.1:    & 0.049      & bar/oval  & & & & \cr
NGC 428          & 0.982    &0.483     &  26.7     & 0.002      & bar/oval  & & & & \cr
NGC 488          & 9.305    &0.395     &  38.7     & 0.266      &        & & & & \cr
NGC 578          & 1.860    &1.559     &  39.5     & 0.011      & bar/oval  & & & & \cr 
NGC 613          & 4.166    &1.051     &  48.5     & 0.122      & bar/oval  & & & & \cr 
NGC 685          & 2.104    &1.408     &  40.4     & 0.007      & bar/oval   & & & & \cr
NGC 864          & 1.826    &0.943     &  28.2     & 0.024      & bar/oval   & & & & \cr
NGC 908          & 3.021    &0.583     &  40.9     & 0.052      &            & & & & \cr
NGC 1042         & 2.184    &0.775     &  42.8     & 0.019      & bar/oval  & & & & \cr 
NGC 1058$^5$     & 1.458:   &1.857:    &  19.6:    & 0.017:     &           & & & & \cr 
NGC 1073         & 3.997    &1.484     &  47.5     & 0.030      & bar/oval  & & & & \cr 
NGC 1084         & 9.104    &0.689     &  28.5     & 0.336      & bar/oval  & & & & \cr 
NGC 1087         & 1.788    &1.690     &  26.8     & 0.013      & bar/oval  & & & & \cr 
NGC 1187         & 1.359    &0.701     &  32.2     & 0.046      & bar/oval  & & & & \cr  
NGC 1241         & 2.055    &1.601     &  19.8     & 0.126      & bar/oval  & & & & \cr 
NGC 1300$^2$     & 3.407:   &0.770:    &  74.3:    & 0.116:     & bar/oval   & & & & \cr
NGC 1302         &10.664    &0.275     &  45.8     & 0.580      & bar/oval  & & & & \cr 
NGC 1309         & 2.279    &0.970     &  10.9     & 0.065      &            & & & & \cr
NGC 1317         & 4.346    &0.670     &  26.0     & 0.420      & bar/oval   & & & & \cr
NGC 1350         & 4.899    &0.732     &  79.3     & 0.215      & bar/oval   & & & & \cr
NGC 1371         & 3.319    &0.834     &  27.2     & 0.111      & bar/oval   & & & & \cr
NGC 1385         &16.195    &0.904     &  37.0     & 0.563      & bar/oval   & & & & \cr
NGC 1493         & 1.249    &0.513     &  29.9     & 0.006      & bar/oval   & & & & \cr
NGC 1559         &  -       &  -       &  22.4     &  -         & bar/oval   & & & & \cr
NGC 1617         & 4.075    &0.697     &  33.9     & 0.169      & bar/oval   & & & & \cr
NGC 1637         & 2.119    &0.723     &  29.9     & 0.058      & bar/oval   & & & & \cr
NGC 1703         & 1.927    &1.165     &  19.7     & 0.033      & bar/oval   & & & & \cr
NGC 1792         & 2.106    &1.100     &  33.7     & 0.024      &            & & & & \cr
NGC 1808$^{2,7}$ & 6.646:   &0.385:    & 109.3:    & 0.595:     & bar/oval   & & & & \cr
NGC 1832         & 1.831    &0.920     &  15.6     & 0.097      & bar/oval   & & & & \cr
NGC 2090$^2$     & 2.053:   &0.852:    &  18.9:    & 0.016(B)   &            & & & & \cr
NGC 2139         &20.225    &0.448     &  18.5     & 0.546      & bar/oval   & & & & \cr
NGC 2196         & 6.291    &0.508     &  20.7     & 0.415      &            & & & & \cr
NGC 2207$^5$     & 2.668:   &1.153:    &  41.4:    & 0.171:     & bar/oval   & & & & \cr
NGC 2442$^6$     & 5.281:   &0.447:    & 153.5(B)  & 0.166:     & bar/oval   & & & & \cr
NGC 2559         & 2.589    &1.275     &  26.2     & 0.056      & bar/oval   & & & & \cr
NGC 2566         & 2.470    &0.434     &  90.2     & 0.141:     & bar/oval   & & & & \cr
NGC 2775         &20.274    &0.358     &  43.1     & 0.931      &            & & & & \cr
NGC 2964         & 1.405    &1.043     &  19.9     & 0.064      & bar/oval   & & & & \cr
NGC 3059         & 1.891    &1.903     &  28.8     & 0.015      & bar/oval   & & & & \cr
NGC 3166         & 4.547    &0.594     &  39.6     & 0.554      & bar/oval   & & & & \cr
NGC 3169         & 6.863    &0.569     &  32.5     & 0.779      &            & & & & \cr
NGC 3223         & 5.528    &0.473     &  30.8     & 0.147      &            & & & & \cr
NGC 3227$^2$     & 1.811    &0.450     &  26.1:    & 0.177      & bar/oval   & & & & \cr
NGC 3261         & 2.743    &0.881     &  22.0     & 0.243      & bar/oval   & & & & \cr
NGC 3275         & 2.893    &0.750     &  23.2     & 0.211      & bar/oval   & & & & \cr
NGC 3319         & 2.036    &1.427     &  60.8     & 0.006      & bar/oval   & & & & \cr
NGC 3338         &20.647    &0.222     &  33.4     & 0.290      & bar/oval   & & & & \cr
NGC 3423         & 4.555    &0.873     &  33.0     & 0.033      & bar/oval   & & & & \cr
NGC 3504         & 2.651    &0.997     &  27.8     & 0.356      & bar/oval   & & & & \cr
NGC 3507         & 2.363    &0.633     &  26.3     & 0.072      & bar/oval   & & & & \cr
NGC 3513         & 1.678    &2.077     &  27.2     & 0.007      & bar/oval   & & & & \cr
&&&&&&&&& \cr
}
\endtable

\vfill
\eject

\begintable*{2}
\caption{{\bf Table 2.} continued}
\halign{%
\rm#\hfil&\qquad\rm#\hfil&\qquad\rm\hfil#&\qquad\rm\hfil
#&\qquad\rm\hfil#&\qquad\rm\hfil#&\qquad\rm#\hfil
&\qquad\rm\hfil#&\qquad\rm#\hfil&\qquad\hfil\rm#\cr
Galaxy    & $r_{eff}$    & $\beta$ & $h_r$   & $B/D$   & bar/oval & & & & \cr
          &  [arcsec]    &         & [arcsec]&         &          & & & & \cr
\noalign{\vskip 10pt}

NGC 3583         & 2.078    &0.853     &  17.8     & 0.122      & bar/oval   & & & & \cr
NGC 3593         & 4.097    &0.712     &  30.1     & 0.161      & bar/oval   & & & & \cr
NGC 3596         & 1.777    &0.740     &  20.3     & 0.047      & bar/oval   & & & & \cr
NGC 3646         & 2.849    &0.525     &  24.5     & 0.168      &            & & & & \cr
NGC 3675         &13.362    &0.231     &  41.0     & 0.230      & bar/oval   & & & & \cr
NGC 3681         & 1.457    &1.250     &  12.2     & 0.066      & bar/oval   & & & & \cr
NGC 3684         & 1.405    &0.591     &  13.3     & 0.022      & bar/oval   & & & & \cr
NGC 3686         & 1.673    &1.126     &  25.4     & 0.020      & bar/oval   & & & & \cr
NGC 3726         & 1.205    &0.747     &  46.4     & 0.009      & bar/oval   & & & & \cr
NGC 3810         &10.697    &0.761     &  29.6     & 0.453      &            & & & & \cr
NGC 3887         & 2.375    &1.111     &  34.7     & 0.019      & bar/oval   & & & & \cr
NGC 3893         &19.939    &0.463     &  34.2     &  -         &            & & & & \cr
NGC 3938         & 5.636    &0.610     &  32.3:    & 0.053      & bar/oval   & & & & \cr
NGC 3949$^2$     & 4.942:   &0.737:    &  14.1(B)  & 0.133      &            & & & & \cr
NGC 4027         & 1.412    &3.626     &  26.0     & -          & bar/oval   & & & & \cr
NGC 4030         & 9.761    &0.497     &  22.7     & 0.300      & bar/oval   & & & & \cr
NGC 4051$^2$     & 3.220:   &0.314:    &  70.0:    & 0.141:     & bar/oval   & & & & \cr
NGC 4123         & 2.308    &0.760     &  31.4     & 0.090      & bar/oval   & & & & \cr
NGC 4136         & 2.122    &0.642     &  25.7     & 0.019      & bar/oval   & & & & \cr
NGC 4138         & 4.013    &0.371     &  17.4     & 0.298      &            & & & & \cr
NGC 4145         &   -      & -        &  57.9     &  -         & bar/oval   & & & & \cr
NGC 4151$^2$     & 4.293:   &0.516:    &  27.1(B)  & 0.758:     & bar/oval   & & & & \cr
NGC 4212         & 2.358    &1.175     &  21.9     & 0.044      & bar/oval   & & & & \cr
NGC 4242         & 3.535    &0.569     &  53.4     & 0.003      &            & & & & \cr
NGC 4254         & 9.814    &0.700     &  33.5     & 0.151      &            & & & & \cr
NGC 4293         & 4.646    &0.698     &  52.3     & 0.077      & bar/oval   & & & & \cr
NGC 4303         & 2.956    &1.093     &  42.4     & 0.085      & bar/oval   & & & & \cr
NGC 4314         & 5.461    &0.791     &  51.1     & 0.190      & bar/oval   & & & & \cr
NGC 4394         & 3.821    &0.740     &  33.4     & 0.215      & bar/oval   & & & & \cr
NGC 4414         &10.872    &0.290     &  24.1     & 0.524      &            & & & & \cr
NGC 4450         & 4.836    &0.563     &  43.6     & 0.145      & bar/oval   & & & & \cr
NGC 4457         & 4.283    &0.556     &  28.5     & 0.733      & bar/oval   & & & & \cr
NGC 4487         &29.140:   &0.206:    &  31.6:    & 0.099:     & bar/oval   & & & & \cr
NGC 4490         &  -       &  -       &  37.6     & -          & bar/oval   & & & & \cr
NGC 4496A        &  -       &  -       &  30.8     & -          & bar/oval   & & & & \cr
NGC 4504$^2$     & 4.155:   &0.441:    &  22.9(B)  & 0.011:     & bar/oval   & & & & \cr
NGC 4527         & 3.776    &0.788     &  38.9     & 0.243      & bar/oval   & & & & \cr
NGC 4548         & 6.118    &0.589     &  59.5     & 0.176      & bar/oval   & & & & \cr
NGC 4571         & 2.947    &1.253     &  35.4     & 0.011      & bar/oval   & & & & \cr
NGC 4579         & 4.538    &0.691     &  41.8     & 0.155      & bar/oval   & & & & \cr
NGC 4580         & 1.602    &2.928     &  16.9     & 0.010      &            & & & & \cr
NGC 4593         & 4.317    &0.673     &  56.6     & 0.271      & bar/oval   & & & & \cr
NGC 4618         &26.681:   &0.455:    &  32.2:    & 0.109:     & bar/oval   & & & & \cr
NGC 4643         & 6.412    &0.727     &  46.1     & 0.431      & bar/oval   & & & & \cr
NGC 4647         &20.695    &0.477     &  27.7     & 0.395      &            & & & & \cr
NGC 4651         &17.215    &0.377     &  26.4     & 0.553      & bar/oval   & & & & \cr
NGC 4654         & 2.102    &0.757     &  30.9     & 0.012      & bar/oval   & & & & \cr
NGC 4665         & 5.578    &0.932     &  39.4     & 0.198      & bar/oval   & & & & \cr
NGC 4689         & 4.770    &1.032     &  36.4     & 0.042      &            & & & & \cr
NGC 4691         & 3.455    &1.365     &  30.9     & 0.065      & bar/oval   & & & & \cr
NGC 4698$^2$     &12.918:   &0.345:    &  43.5(B)  & 0.691:     & bar/oval   & & & & \cr
NGC 4699         & 6.122    &0.458     &  21.1     & 0.509      & bar/oval   & & & & \cr
NGC 4772         & 7.196    &0.581     &  42.7     & 0.513      &            & & & & \cr
NGC 4775         & 4.090    &1.461     &  16.0     & 0.038      &            & & & & \cr
NGC 4781         &12.379    &0.486     &  29.4     & 0.069      & bar/oval   & & & & \cr
NGC 4900$^3$     & 2.702:   &1.227:    &  14.5:    & 0.019:     & bar/oval   & & & & \cr
NGC 4902         & 2.862    &1.010     &  22.1     & 0.086      & bar/oval   & & & & \cr
NGC 4930$^2$     & 3.641:   &0.788:    &  36.8(B)  & 0.237:     & bar/oval   & & & & \cr
NGC 4939         &14.099    &0.403     &  40.6     & 0.379      & bar/oval   & & & & \cr
NGC 4941         & 2.474    &0.705     &  30.3     & 0.142      &            & & & & \cr
NGC 4995         & 1.415    &0.997     &  18.3     & 0.051      & bar/oval   & & & & \cr
NGC 5005         & 3.029    &0.686     &  34.1     & 0.144      & bar/oval   & & & & \cr
&&&&&&&&& \cr
}
\endtable

\vfill
\eject

\begintable*{2}
\caption{{\bf Table 2.} continued}
\halign{%
\rm#\hfil&\qquad\rm#\hfil&\qquad\rm\hfil#&\qquad\rm\hfil
#&\qquad\rm\hfil#&\qquad\rm\hfil#&\qquad\rm#\hfil
&\qquad\rm\hfil#&\qquad\rm#\hfil&\qquad\hfil\rm#\cr
Galaxy    & $r_{eff}$    & $\beta$ & $h_r$   & $B/D$   & bar/oval & & & & \cr
          &  [arcsec]    &         & [arcsec]&         &          & & & & \cr
\noalign{\vskip 10pt}

NGC 5054         & 4.839    &0.404     &  42.5     & 0.173      &            & & & & \cr
NGC 5085         & 3.867    &1.095     &  26.9     & 0.050      &            & & & & \cr
NGC 5101         &11.469    &0.350     & 100.0     & 0.495      & bar/oval   & & & & \cr
NGC 5121         & 2.568    &0.622     &  13.4     & 0.372      &            & & & & \cr
NGC 5247         & 8.313    &0.804     &  62.7     & 0.096      &            & & & & \cr
NGC 5248         & 7.010    &0.696     &  71.3     & 0.287      & bar/oval   & & & & \cr
NGC 5334         & 0.602    &0.785     &  32.5     & 0.001      & bar/oval   & & & & \cr
NGC 5427         & 3.418    &1.080     &  19.1     & 0.081      &            & & & & \cr
NGC 5483         & 6.438    &1.434     &  24.0     & 0.082      & bar/oval   & & & & \cr
NGC 5643         & 2.588    &0.552     &  39.7     & 0.069      & bar/oval   & & & & \cr
NGC 5676         & 1.973    &1.078     &  21.1     & 0.045      &            & & & & \cr
NGC 5701$^{2,7}$ & 7.078:   &0.439:    &  32.6(B)  & 0.352:     & bar/oval   & & & & \cr
NGC 5713         & 2.112    &1.504     &  18.5     & 0.051      & bar/oval   & & & & \cr
NGC 5850         & 4.912    &0.769     &  58.4     & 0.208      & bar/oval   & & & & \cr
NGC 5921         & 2.331    &0.863     &  35.7     & 0.108      & bar/oval   & & & & \cr
NGC 5962         & 4.549    &0.383     &  17.6:    & 0.213      & bar/oval   & & & & \cr
NGC 6215         & 2.651    &0.828     &  15.5     & 0.083      &            & & & & \cr
NGC 6221         & 2.715    &0.702     &  32.6     & 0.086      & bar/oval   & & & & \cr
NGC 6300         & 3.467    &0.656     &  39.6     & 0.050      & bar/oval   & & & & \cr
NGC 6384         & 7.149    &0.385     &  40.3     & 0.148      & bar/oval   & & & & \cr
NGC 6753         & 6.171    &0.593     &  22.7     & 0.397      & bar/oval   & & & & \cr
NGC 6782         & 3.193    &1.019     &  26.5     & 0.319      & bar/oval  & & & & \cr
NGC 6902         & 3.066    &0.794     &  33.1     & 0.131      & bar/oval  & & & & \cr
NGC 6907         & 5.230    &0.694     &  23.9     & 0.184      & bar/oval  & & & & \cr
NGC 7083         & 4.989    &0.516     &  20.0     & 0.150      &            & & & & \cr
NGC 7205         & 1.262    &0.889     &  27.0     & 0.021      & bar/oval   & & & & \cr
NGC 7213         &15.330    &0.363     &  54.3     & 1.472      &            & & & & \cr
NGC 7217         &18.423    &0.401     &  32.7     & 0.852      &            & & & & \cr
NGC 7412$^4$     & 4.272    &0.691     &  32.2     & 0.051      &            & & & & \cr
NGC 7418         & 1.279    &0.576     &  43.5     & 0.017      & bar/oval   & & & & \cr
NGC 7479         & 3.397    &0.967     &  39.6     & 0.069      & bar/oval   & & & & \cr
NGC 7552         & 3.971    &0.685     &  59.1     & 0.435      & bar/oval   & & & & \cr
NGC 7582         & 1.760    &0.454     &  45.1     & 0.178      & bar/oval   & & & & \cr
NGC 7713         &34.353    &0.355     &  37.8     & 0.056      & bar/oval   & & & & \cr
NGC 7723         & 1.770    &1.220     &  20.9     & 0.046      & bar/oval   & & & & \cr
NGC 7727         & 8.059    &0.330     &  22.5     & 0.669      & bar/oval   & & & & \cr
NGC 7741$^1$     &  -       &  -       &  37.5     &  -         & bar/oval   & & & & \cr
&&&&&&&&& \cr
}
\endtable

\vfill
\eject

\begintable*{2}
\caption{{\bf Table 2.} continued}
\halign{%
\rm#\hfil&\qquad\rm#\hfil&\qquad\rm\hfil#&\qquad\rm\hfil
#&\qquad\rm\hfil#&\qquad\rm\hfil#&\qquad\rm#\hfil
&\qquad\rm\hfil#&\qquad\rm#\hfil&\qquad\hfil\rm#\cr
Galaxy    & $r_{eff}$    & $\beta$ & $h_r$   & $B/D$   & bar/oval & & & & \cr
          &  [arcsec]    &         & [arcsec]&         &          & & & & \cr
\noalign{\vskip 10pt}

2MASS:& & & & & & & & &  \cr
& & & & & & & & &  \cr

NGC  772         &24.311    &0.320   &  49.6     & 0.713      &            & & & & \cr
NGC 1068         & 1.923    &0.817   &  20.7     & 0.128      &  bar/oval  & & & & \cr
NGC 1097         & 7.297    &0.740   &  25.0     & 0.241      &  bar/oval  & & & & \cr
NGC 1232         & 5.034    &0.484   &  35.8     & 0.029      &  bar/oval  & & & & \cr
NGC 1398         & 8.071    &0.524   &  38.0     & 0.371      &  bar/oval  & & & & \cr
NGC 2655         & 8.947    &0.519   &  28.0     & 0.811      &            & & & & \cr
NGC 2841         & 6.323    &0.566   &  50.8     & 0.195      &             & & & & \cr
NGC 2985         &10.596:   &0.413:  &  26.1(MH) & 0.685:     &  bar/oval   & & & & \cr
NGC 3031         &10.915    &0.498   &  57.1     & 0.337      &             & & & & \cr
NGC 3077         & 8.898    &0.681   &  33.1     & 0.110      &             & & & & \cr
NGC 3486         & 2.771    &0.977   &  18.0     & 0.120      &             & & & & \cr
NGC 3521         & 2.462    &0.506   &  37.2     & 0.087      &  bar/oval   & & & & \cr
NGC 3718         & 7.219    &0.376   &  27.7     & 0.487      &            & & & & \cr 
NGC 3898         & 3.420    &0.524   &  26.0     & 0.326      &  bar/oval  & & & & \cr 
NGC 4321         & 8.623:   &0.560:  &  49.5(MH) & 0.096:     &  bar/oval  & & & & \cr
NGC 4501         & 3.388:   &0.512:  &  42.3(MH) & 0.049:     &  bar/oval  & & & & \cr
NGC 4569         & 3.933    &0.445   &  61.5     & 0.143      &  bar/oval  & & & & \cr
NGC 4736         &14.586    &0.573   &  58.3     & 1.007      &  bar/oval  & & & & \cr
NGC 4753         &14.914    &0.330   &  38.4     & 0.934      &            & & & & \cr
NGC 5457         &29.366    &0.385   & 112.8     & 0.081      &            & & & & \cr
NGC 6643         & 1.910    &2.243   &  23.9     & 0.019      &            & & & & \cr
NGC 7513         & 3.776    &1.289   &  58.0     & 0.032      &  bar/oval  & & & & \cr
}

\tabletext{\noindent {$:$} The measurement is uncertain.

\noindent {$^1$} Manual decomposition fit.

\noindent {$^2$} H-band image not deep enough to determine $h_r$ well, but the bulge-model is reasonable.

\noindent {$^3$} Bulge-model is taken from the fit to the inner regions, 
and the disk parameters from the fit to the outer regions of the galaxy. 

\noindent {$^4$} No bar is fitted, although the galaxy is barred in RC3.

\noindent {$^5$} The number of iterations is limited in the decomposition.

\noindent {$^6$} The H-band image field is too small. 

\noindent {$^7$} Prominment outer ring dominates the disk.
}

\endtable

\vfill
\eject


\begintable*{3}
\caption{{\bf Table 3.} Galaxies with Fourier bars.}
\halign{%
\rm#\hfil&\qquad\rm#\hfil&\qquad\rm\hfil#&\qquad\rm\hfil
#&\qquad\rm\hfil#&\qquad\rm\hfil#&\qquad\rm#\hfil
&\qquad\rm\hfil#&\qquad\rm#\hfil&\qquad\hfil\rm#\cr
Galaxy    & $Q_g$    & $r_{Q_g}$  & $bar length$  & $A_2$  & $A_4$ & EFP/RC3 & & & \cr
          &          &  [arcsec]  & [arcsec]      &        &       &         & & & \cr
\noalign{\vskip 10pt}

OSUBGS: & & & & & & & & & \cr
& & & & & & & & & \cr
 IC 4444    &0.254    $\pm$ 0.033 &   5.5 &  22.2             & 0.300  &  0.180   &(B) (X) & & & \cr
 IC 5325    &0.219    $\pm$ 0.020 &  12.7 &  23.2             & 0.280  &  0.140   &(X) (X) & & & \cr
 NGC 150    &0.459    $\pm$ 0.085 &  26.7 &  29.0             & 0.616  &  0.249   &(B) (B) & & & \cr
 NGC 210    &0.061    $\pm$ 0.001 &  36.0 &  46.0              & 0.413  &  0.106   &(B) (X) & & & \cr
 NGC 289    &0.212    $\pm$ 0.003 &  12.8 &  19.7             & 0.388  &  0.106   &(B) (B) & & & \cr
 NGC 428    &0.251    $\pm$ 0.020 &  28.5 &  45.0             & 0.437  &  0.076   &(B) (X) & & & \cr
 NGC 578    &0.182    $\pm$ 0.014 &  10.4 &  19.7             & 0.277  &  0.053   &(B) (X) & & & \cr
 NGC 613    &0.401    $\pm$ 0.045 &  68.4 & 104.4              & 0.754  &  0.479   &(B) (B) & & & \cr
 NGC 685    &0.424    $\pm$ 0.012 &  10.4 &  20.9             & 0.400  &  0.153   &(B) (X) & & & \cr
 NGC 864    &0.360    $\pm$ 0.037 &  19.5 &  25.5             & 0.438  &  0.155   &(B) (X) & & & \cr 
 NGC 1073   &0.607    $\pm$ 0.013 &  25.5 &  37.5             & 0.703  &  0.313   &(B) (B) & & & \cr
 NGC 1087   &0.442    $\pm$ 0.020 &   7.5 &  18.0             & 0.440  &  0.172   &(B) (X) & & & \cr
 NGC 1187   &0.207    $\pm$ 0.043 &  36.0 &  29.0             & 0.345  &  0.123   &(B) (B) & & & \cr
 NGC 1241   &0.251    $\pm$ 0.028 &  22.5 &  30.0             & 0.410  &  0.136   &(B) (B) & & & \cr 
 NGC 1300   &0.537    $\pm$ 0.011 &  68.4 &  87.0             & 0.743  &  0.348   &(B) (B) & & & \cr
 NGC 1302   &0.075    $\pm$ 0.006 &  24.4 &  25.5             & 0.303  &  0.066   &(B) (B) & & & \cr
 NGC 1317   &0.085    $\pm$ 0.007 &  40.6 &  58.0             & 0.334  &  0.102   &(B) (X) &  + minibar  & & \cr
 NGC 1350   &0.243    $\pm$ 0.039 &  68.4 &  81.2             & 0.713  &  0.207   &(B) (B) & + minibar  & & \cr 
 NGC 1385   &0.319    $\pm$ 0.030 &   3.5 &   9.3             & 0.266  &  0.196   &(B) (B) & & & \cr
 NGC 1493   &0.363    $\pm$ 0.010 &  10.4 &  23.2             & 0.304  &  0.137   &(B) (B) & & & \cr
 NGC 1559   &0.334    $\pm$ 0.001 &   5.8 &  17.4             & 0.246  &  0.102   &(B) (B) & & & \cr
 NGC 1617   &0.079    $\pm$ 0.027 &   7.8 &  22.2             & 0.600  &  0.200   &(X) (B) & & & \cr
 NGC 1637   &0.202    $\pm$ 0.014 &  16.5 &  22.5             & 0.350  &  0.137   &(B) (X) & & & \cr
 NGC 1703   &0.100    $\pm$ 0.005 &   8.9 &  11.1             & 0.184  &  0.070   &(X) (B) & & & \cr
 NGC 1808   &0.274    $\pm$ 0.001 &  77.7 &  87.0             & 1.102  &  0.586   &(B) (X) & + minibar & & \cr
 NGC 1832   &0.195    $\pm$ 0.024 &  12.2 &  16.6             & 0.405  &  0.166   &(B) (B) & & & \cr
 NGC 2139   &0.398    $\pm$ 0.032 &   3.3 &  16.6             & 0.380  &  0.139   &(B) (X) & & & \cr
 NGC 2207   &0.317    $\pm$ 0.032 &  29.0 &  46.4             & 0.700  &  0.350   &(B) (X) & & & \cr
 NGC 2442   &0.669    $\pm$ 0.428 &  77.7 &  92.8             & 0.979  &  0.421   &(B) (X) & & & \cr
 NGC 2559   &0.316    $\pm$ 0.039 &  27.8 &  33.3             & 0.512  &  0.151   &(B) (B) & & & \cr
 NGC 2566   &0.316    $\pm$ 0.069 &  54.4 &  72.1             & 0.865  &  0.391   &(B) (B) & & & \cr
 NGC 2964   &0.310    $\pm$ 0.003 &  22.5 &  30.0             & 0.419  &  0.223   &(X) (X) & & & \cr
 NGC 3059   &0.544    $\pm$ 0.048 &   7.8 &  20.0             & 0.727  &  0.359   &(B) (B) & & & \cr
 NGC 3166   &0.107    $\pm$ 0.019 &  31.5 &  45.0             & 0.524  &  0.221   &(B) (X) & + oval/minibar & & \cr
 NGC 3227   &0.158    $\pm$ 0.021 &  55.5 &  75.0             & 0.444  &  0.272   &(B) (X) & & & \cr
 NGC 3261   &0.196    $\pm$ 0.009 &  18.9 &  27.7             & 0.538  &  0.233   &(B) (B) & & & \cr
 NGC 3275   &0.187    $\pm$ 0.013 &  23.3 &  41.1             & 0.533  &  0.214   &(B) (B) & & & \cr
 NGC 3319   &0.542    $\pm$ 0.019 &  13.5 &  37.5             & 0.630  &  0.250   &(B) (B) & & & \cr
 NGC 3338   &0.083    $\pm$ 0.005 &  13.5 &  22.5             & 0.118  &  0.054   &(X) (A) & & & \cr
 NGC 3504   &0.288    $\pm$ 0.030 &  28.5 &  60.0             & 0.991  &  0.476   &(B) (X) & & & \cr
 NGC 3507   &0.176    $\pm$ 0.005 &  19.5 &  22.5             & 0.379  &  0.107   &(B) (B) & & & \cr
 NGC 3513   &0.541    $\pm$ 0.069 &  14.4 &  27.7             & 0.429  &  0.226   &(B) (B) & & & \cr
 NGC 3583   &0.246    $\pm$ 0.006 &  16.5 &  22.5             & 0.649  &  0.203   &(B) (B) & & & \cr
 NGC 3593   &0.152    $\pm$ 0.002 &  10.5 &  15.0             & 0.415  &  0.059   &(A) (A) & & & \cr
 NGC 3675   &0.085    $\pm$ 0.012 &  16.5 &  30.0             & 0.300  &  0.08    &(B) (A) & & & \cr
 NGC 3681   &0.199    $\pm$ 0.011 &   7.5 &  15.0             & 0.401  &  0.144   &(B) (X) & & & \cr
 NGC 3686   &0.253    $\pm$ 0.018 &  10.5 &  18.0             & 0.316  &  0.091   &(B) (B) & & & \cr
 NGC 3726   &0.213    $\pm$ 0.024 &  25.5 &  30.0             & 0.219  &  0.079   &(B) (X) & & & \cr
 NGC 3887   &0.207    $\pm$ 0.017 &  31.5 &  40.5             & 0.263  &  0.097   &(B) (B) & & & \cr
 NGC 4027   &0.623    $\pm$ 0.008 &   3.3 &  20.0             & 0.495  &  0.183   &(B) (B) & & & \cr
 NGC 4051   &0.280    $\pm$ 0.008 &  55.5 &  45.0             & 0.655  &  0.173   &(B) (X) & & & \cr
 NGC 4123   &0.428    $\pm$ 0.070 &  37.5 &  52.5             & 0.573  &  0.251   &(B) (B) & & & \cr
 NGC 4136   &0.131    $\pm$ 0.003 &  10.5 &  15.0             & 0.249  &  0.037   &(B) (X) & & & \cr
 NGC 4145   &0.356    $\pm$ 0.002 &   4.5 &  19.5             & 0.267  &  0.087   &(B) (X) & & & \cr
 NGC 4151   &0.119    $\pm$ 0.012 &  67.5 &  97.5             & 0.746  &  0.329   &(B) (X) & & & \cr
 NGC 4293   &0.355    $\pm$ 0.003 &  49.5 &  67.5             & 0.668  &  0.271   &(B) (B) & & & \cr
 NGC 4303   &0.259    $\pm$ 0.044 &  40.5 &  30.0             & 0.443  &  0.145   &(B) (X) & & & \cr
 NGC 4314   &0.442    $\pm$ 0.024 &  52.5 &  75.0             & 0.896  &  0.571   &(B) (B) & & & \cr
 NGC 4394   &0.272    $\pm$ 0.006 &  31.5 &  45.0             & 0.577  &  0.311   &(B) (B) & & & \cr
 NGC 4450   &0.131    $\pm$ 0.011 &  37.5 &  30.0             & 0.322  &  0.125   &(B) (X) & & & \cr

&&&&&&&&& \cr
}
\endtable

\vfill
\eject

\begintable*{3}
\caption{{\bf Table 3.} continued}
\halign{%
\rm#\hfil&\qquad\rm#\hfil&\qquad\rm\hfil#&\qquad\rm\hfil
#&\qquad\rm\hfil#&\qquad\rm\hfil#&\qquad\rm#\hfil
&\qquad\rm\hfil#&\qquad\rm#\hfil&\qquad\hfil\rm#\cr
Galaxy    & $Q_g$    & $r_{Q_g}$  & $bar length$  & $A_2$  & $A_4$ & EFP/RC3 & & & \cr
          &          &  [arcsec]  & [arcsec]      &        &       & & & & \cr
\noalign{\vskip 10pt}

 NGC 4457   &0.089    $\pm$ 0.004 &  31.5 & 45.0             & 0.435  &  0.095   &(B) (X)  & & & \cr
 NGC 4487   &0.177    $\pm$ 0.035 &  10.0 & 22.2             & 0.176  &  0.071   &(B) (X)  & & & \cr
 NGC 4490   &0.334    $\pm$ 0.032 &   7.5 & 18.0             & 0.188  &  0.041   &(B) (B)  & & & \cr
 NGC 4496   &0.365    $\pm$ 0.004 &   7.5 & 25.5             & 0.283  &  0.055   &(B) (B)  & & & \cr
 NGC 4527   &0.198    $\pm$ 0.026 &  46.5 & 82.5             & 0.600  &  0.200   &(X) (X)  & & & \cr
 NGC 4548   &0.344    $\pm$ 0.017 &  55.5 & 67.5             & 0.723  &  0.338   &(B) (B)  & & & \cr
 NGC 4579   &0.197    $\pm$ 0.020 &  34.5 & 45.0             & 0.494  &  0.246   &(B) (X)  & & & \cr
 NGC 4593   &0.309    $\pm$ 0.020 &  45.5 & 61.0             & 0.765  &  0.367   &(B) (B)  & & & \cr
 NGC 4618   &0.392    $\pm$ 0.046 &  10.5 & 33.0             & 0.327  &  0.105   &(B) (B)  & & & \cr
 NGC 4643   &0.251    $\pm$ 0.004 &  43.5 & 67.5             & 0.828  &  0.516   &(B) (B)  & & & \cr
 NGC 4647   &0.117    $\pm$ 0.010 &  10.5 & 11.5             & 0.310  &  0.115   &(B) (X)  & & & \cr
 NGC 4651   &0.120    $\pm$ 0.046 &  16.5 & 22.5             & 0.207  &  0.044   &(X) (A)  & & & \cr
 NGC 4654   &0.171    $\pm$ 0.005 &   7.5 & 19.5             & 0.175  &  0.063   &(B) (X)  & & & \cr
 NGC 4665   &0.257    $\pm$ 0.023 &  37.5 & 60.0             & 0.615  &  0.306   &(B) (B)  & & & \cr
 NGC 4691   &0.504    $\pm$ 0.027 &  13.5 & 45.0             & 0.803  &  0.419   &(B) (B)  & & & \cr
 NGC 4699   &0.144    $\pm$ 0.027 &  10.0 & 14.4             & 0.382  &  0.086   &(B) (X)  & & & \cr
 NGC 4781   &0.352    $\pm$ 0.061 &  16.7 & 38.8             & 0.323  &  0.131   &(B) (B)  & & & \cr
 NGC 4900   &0.384    $\pm$ 0.041 &   7.5 & 18.0             & 0.369  &  0.103   &(B) (B)  & & & \cr
 NGC 4902   &0.277    $\pm$ 0.026 &  16.7 & 22.2             & 0.526  &  0.277   &(B) (B)  & & & \cr
 NGC 4930   &0.207    $\pm$ 0.022 &  34.4 & 44.4             & 0.607  &  0.313   &(B) (B)  & & & \cr
 NGC 4939   &0.128    $\pm$ 0.052 &  15.1 & 17.4             & 0.284  &  0.109   &(X) (A)  & & & \cr
 NGC 4995   &0.278    $\pm$ 0.047 &  22.5 & 22.5             & 0.377  &  0.124   &(B) (X)  & & & \cr
 NGC 5005   &0.152    $\pm$ 0.000 &  28.5 & 45.0             & 0.362  &  0.122   &(B) (X)  & & & \cr
 NGC 5101   &0.187    $\pm$ 0.015 &  45.2 & 69.6             & 0.708  &  0.386   &(B) (B)  & & & \cr
 NGC 5334   &0.364    $\pm$ 0.010 &  10.5 & 18.0             & 0.285  &  0.108   &(B) (B)  & & & \cr
 NGC 5483   &0.174    $\pm$ 0.003 &   7.8 & 13.3             & 0.210  &  0.090   &(B) (A)  & & & \cr
 NGC 5643   &0.415    $\pm$ 0.013 &  33.6 & 46.4             & 0.433  &  0.257   &(B) (X)  & & & \cr
 NGC 5701   &0.143    $\pm$ 0.000 &  30.0 & 49.9             & 0.466  &  0.196   &(B) (B)  & & & \cr
 NGC 5713   &0.357    $\pm$ 0.033 &  10.5 & 30.0             & 0.600  &  0.210   &(B) (X)  & & & \cr
 NGC 5850   &0.318    $\pm$ 0.010 &  61.5 & 90.0             & 0.693  &  0.366   &(B) (B)  & + minibar & & \cr
 NGC 5921   &0.416    $\pm$ 0.023 &  46.5 & 52.5             & 0.704  &  0.361   &(B) (B)  &  & & \cr
 NGC 5962   &0.148    $\pm$ 0.052 &  13.5 & 15.0             & 0.249  &  0.096   &(B) (A)  &  & & \cr
 NGC 6221   &0.436    $\pm$ 0.112 &  27.8 & 40.6             & 0.617  &  0.277   &(B) (B)  &   & & \cr
 NGC 6300   &0.187    $\pm$ 0.002 &  33.6 & 46.4             & 0.414  &  0.163   &(B) (B)  &   & & \cr
 NGC 6384   &0.136    $\pm$ 0.020 &  16.5 & 33.0             & 0.352  &  0.069   &(B) (X)  &  & & \cr 
 NGC 6782   &0.165    $\pm$ 0.008 &  24.4 & 46.4             & 0.631  &  0.278   &(B) (X)  &   & & \cr
 NGC 6902   &0.075    $\pm$ 0.004 &  13.9 & 17.4             & 0.140  &  0.04    &(B) (A)  &   & & \cr
 NGC 7418   &0.192    $\pm$ 0.029 &  15.1 & 15.1             & 0.273  &  0.076   &(B) (X)  &   & & \cr
 NGC 7479   &0.696    $\pm$ 0.060 &  43.5 & 60.0             & 0.867  &  0.555   &(B) (B)  &   & & \cr
 NGC 7552   &0.395    $\pm$ 0.044 &  45.2 & 69.6             & 1.148  &  0.724   &(B) (B)  &   & & \cr
 NGC 7582   &0.436    $\pm$ 0.069 &  56.8 & 92.8             & 0.932  &  0.551   &(B) (B)  &   & & \cr
 NGC 7723   &0.349    $\pm$ 0.030 &  16.5 & 24.0             & 0.386  &  0.184   &(B) (B)  &   & & \cr
 NGC 7727   &0.096    $\pm$ 0.024 &  10.5 & 27.0             & 18.00  &  7.70    &(Spec) (X)&  & & \cr
 NGC 7741   &0.687    $\pm$ 0.006 &  10.5 & 52.5             & 0.602  &  0.283   &(B) (B)  &   & & \cr
& & & & & & & & & \cr
2MASS:& & & & & & & & & \cr
& & & & & & & & & \cr
 NGC 1068  &0.165     $\pm$ 0.010 &  11.0 & 15.0             & 0.391  &  0.149   &(A)     &   & & \cr
 NGC 1097  &0.279     $\pm$ 0.048 &  75.0 &100.0             & 0.821  &  0.421   &(B)     & + minbar   & & \cr
 NGC 1232  &0.210     $\pm$ 0.002 &   7.0 & 12.0             & 0.295  &  0.066   &(X)     &   & & \cr
 NGC 1398  &0.202     $\pm$ 0.011 &  39.0 & 55.0             & 0.406  &  0.275   &(B)     &   & & \cr
 NGC 3521  &0.096     $\pm$ 0.019 &  25.0 & 28.0             & 0.368  &  0.269   &(X)     &   & & \cr
 NGC 4321  &0.183     $\pm$ 0.027 &  61.0 & 90.0             & 0.335  &  0.230   &(X)     & + minbar  & & \cr
 NGC 4569(R)&0.175    $\pm$ 0.064 &  10.0 & 30.0             & 0.434  &  0.188   &(X)     &   & & \cr
 NGC 4736  &0.048     $\pm$ 0.004 &  11.0 & 15.0             & 0.156  &  0.070   &(A)     &   & & \cr
 NGC 7513  &0.483     $\pm$ 0.031 &  27.0 & 55.0             & 0.649  &  0.354   &(B)     &   & & \cr
&&&&&&&&& \cr
}
\endtable

\vfill
\eject

\begintable*{4}
\caption{{\bf Table 4.} Galaxies without Fourier bars.}
\halign{%
\rm#\hfil&\qquad\rm#\hfil&\qquad\rm\hfil#&\qquad\rm\hfil
#&\qquad\rm\hfil#&\qquad\rm\hfil#&\qquad\rm#\hfil
&\qquad\rm\hfil#&\qquad\rm#\hfil&\qquad\hfil\rm#\cr
Galaxy    & $Q_g$    & $r_{Q_g}$  & EFP/RC3 & & & & & & \cr
          &          &  [arcsec]  &         & & & & & & \cr
\noalign{\vskip 10pt}

OSUBGS: & & & & & & & & & \cr
& & & & & & & & & \cr
ESO 138  & 0.148    $\pm$ 0.001 &  80.0 &(A) (A)  & & & & & & \cr 
NGC 157  & 0.326    $\pm$ 0.174 &  31.5 &(A) (X)  & & & & & & \cr  
NGC 278  & 0.063    $\pm$ 0.021 &  28.5 &(A) (X)  &  & & & & & \cr
NGC 488  & 0.032    $\pm$ 0.003 &  19.5 &(A) (A)   & & & & & & \cr
NGC 908  & 0.183    $\pm$ 0.005 &  75.4 &(A) (A)   & & & & & & \cr
NGC 1042 & 0.533    $\pm$ 0.209 &  31.5 &(X) (X)   & & & & & & \cr
NGC 1058 & 0.138    $\pm$ 0.001 &  22.5 &(A) (A)   & & & & & & \cr
NGC 1084 & 0.212    $\pm$ 0.025 &  35.5 &(A) (A)   & & & & & & \cr
NGC 1309 & 0.148    $\pm$ 0.060 &  15.1 &(X) (A)   & & & & & & \cr
NGC 1371 & 0.113    $\pm$ 0.003 &  19.7 &(X) (X)    &  & & & & & \cr
NGC 1792 & 0.151    $\pm$ 0.092 &  36.0 &(A) (A)    &  & & & & & \cr
NGC 2090 & 0.114    $\pm$ 0.005 &  10.4 &(A) (A)    &  & & & & & \cr
NGC 2196 & 0.070    $\pm$ 0.005 &   7.8 &(A) (A)    &  & & & & & \cr
NGC 2775 & 0.050    $\pm$ 0.010 &  46.5 &(A) (A)   &  & & & & & \cr 
NGC 3169 & 0.090    $\pm$ 0.005 &  16.5 &(A) (A)   &  & & & & & \cr
NGC 3223 & 0.038    $\pm$ 0.003 &  69.9 &(A) (A)    & & & & & & \cr
NGC 3423 & 0.099    $\pm$ 0.075 &  67.7 &(A) (A)   &  & & & & & \cr
NGC 3596 & 0.157    $\pm$ 0.044 &  46.5 &(X) (X)    & & & & & & \cr
NGC 3646 & 0.241    $\pm$ 0.025 &  64.5 &(X)(RING) &  & & & & & \cr
NGC 3684 & 0.085    $\pm$ 0.020 &  10.5 &(X) (A)    & & & & & & \cr
NGC 3810 & 0.128    $\pm$ 0.022 &  16.5 &(X) (A)   &  & & & & & \cr
NGC 3893 & 0.148    $\pm$ 0.001 &  19.5 &(X) (X)   &  & & & & & \cr
NGC 3938 & 0.070    $\pm$ 0.005 &  58.5 &(A) (A)   &  & & & & & \cr
NGC 3949 & 0.276    $\pm$ 0.084 &  25.5 &(X) (A)    & & & & & & \cr
NGC 4030 & 0.060    $\pm$ 0.013 &  16.7 &(A) (A)    & & & & & & \cr
NGC 4138 & 0.046    $\pm$ 0.007 &   7.5 &(A) ( )   &  & & & & & \cr
NGC 4212 & 0.234    $\pm$ 0.051 &  28.5 &(X) (A)   &  & & & & & \cr
NGC 4242 & 0.237    $\pm$ 0.062 &  40.5 &(B) (X)   &  & & & & & \cr
NGC 4254 & 0.122    $\pm$ 0.029 &  22.5 &(X) (A)   &  & & & & & \cr
NGC 4414 & 0.149    $\pm$ 0.003 &  28.5 &(A) (A)   &  & & & & & \cr
NGC 4504 & 0.136    $\pm$ 0.019 &  25.5 &(B) (A)   &  & & & & & \cr
NGC 4571 & 0.067    $\pm$ 0.027 &  40.5 &(A) (A)   &  & & & & & \cr
NGC 4580 & 0.109    $\pm$ 0.012 &  13.5 &(A) (X)   &  & & & & & \cr
NGC 4689 & 0.068    $\pm$ 0.001 &  58.5 &(A) (A)   &  & & & & & \cr
NGC 4698 & 0.084    $\pm$ 0.040 &  64.5 &(A) (A)    & & & & & & \cr
NGC 4772 & 0.042    $\pm$ 0.013 &  70.5 &(B) (A)    & & & & & & \cr
NGC 4775 & 0.134    $\pm$ 0.013 &   7.5 &(A) (A)    & & & & & & \cr
NGC 4941 & 0.056    $\pm$ 0.008 &  49.5 &(X) (X)   &  & & & & & \cr
NGC 5054 & 0.090    $\pm$ 0.023 &  80.0 &(X) (A)   &  & & & & & \cr
NGC 5085 & 0.152    $\pm$ 0.021 &  22.0 &(X) (A)   &  & & & & & \cr
NGC 5121 & 0.024    $\pm$ 0.007 &  27.8 &(A) (A)   &  & & & & & \cr
NGC 5247 & 0.329    $\pm$ 0.150 &  75.4 &(A) (A)   &  & & & & & \cr
NGC 5248 & 0.269    $\pm$ 0.064 &  76.5 &(A) (X)   &  & & & & & \cr
NGC 5427 & 0.231    $\pm$ 0.074 &  38.3 &(A) (A)   &  & & & & & \cr
NGC 5676 & 0.102    $\pm$ 0.014 &  16.5 &(X) (A)   &  & & & & & \cr
NGC 6215 & 0.239    $\pm$ 0.130 &  26.7 &(X) (A)   &  & & & & & \cr
NGC 6753 & 0.039    $\pm$ 0.009 &  12.8 &(A) (A)  &  & & & & & \cr
NGC 6907 & 0.329    $\pm$ 0.154 &  29.0 &(B) (B)   &   & & \cr
NGC 7083 & 0.073    $\pm$ 0.001 &  26.7 &(A) (A)   & & & & & & \cr
NGC 7205 & 0.060    $\pm$ 0.017 &  63.8 &(X) (A)  &  & & & & & \cr
NGC 7213 & 0.023    $\pm$ 0.002 & 100.9 &(A) (A)  &  & & & & & \cr
NGC 7217 & 0.036    $\pm$ 0.001 &  13.5 &(A) (B)   & & & & & & \cr
NGC 7412 & 0.415    $\pm$ 0.183 &  52.2 &(X) (B)  &  & & & & & \cr
NGC 7713 & 0.099    $\pm$ 0.015 &  26.7 &(A) (B)  &  & & & & & \cr
& & & & & & & & & \cr
2MASS:  & & & & & & & & & \cr
& & & & & & & & & \cr
NGC 772    & 0.066 $\pm$ 0.020 & 25.0  &(A)  &   & & & & & \cr
NGC 2655   & 0.128 $\pm$ 0.004 & 19.0  &(X)  &  & & & & & \cr
NGC 2841   & 0.071 $\pm$ 0.008 & 37.0  &(A)  &  & & & & & \cr

&&&&&&&&& \cr
}
\endtable

\vfill
\eject

\begintable*{4}
\caption{{\bf Table 4.} continued}
\halign{%
\rm#\hfil&\qquad\rm#\hfil&\qquad\rm\hfil#&\qquad\rm\hfil
#&\qquad\rm\hfil#&\qquad\rm\hfil#&\qquad\rm#\hfil
&\qquad\rm\hfil#&\qquad\rm#\hfil&\qquad\hfil\rm#\cr
Galaxy    & $Q_g$    & $r_{Q_g}$  & EFP/RC3 & & & & & & \cr
          &          &  [arcsec]  &         & & & & & & \cr
\noalign{\vskip 10pt}

NGC 2985   & 0.056 $\pm$ 0.001 & 11.0  &(A)  &  & & & & & \cr
NGC 3031   & 0.091 $\pm$ 0.031 & 29.0  &(A)  &  & & & & & \cr
NGC 3077   & 0.119 $\pm$ 0.016 &  9.0  &(IO)  & & & & & & \cr
NGC 3486(R)& 0.108 $\pm$ 0.002 & 15.0  &(X)  &  & & & & & \cr
NGC 3718   & 0.106 $\pm$ 0.007 & 73.0  &(B)  &  & & & & & \cr
NGC 3898   & 0.047 $\pm$ 0.000 & 13.0  &(A)  &  & & & & & \cr
NGC 4501   & 0.072 $\pm$ 0.026 & 89.0  &(A)  &  & & & & & \cr
NGC 4753   & 0.106 $\pm$ 0.019 & 59.0  &(IO) &  & & & & & \cr
NGC 5457   & 0.225 $\pm$ 0.001 & 73.0  &(X)  &  & & & & & \cr
NGC 6643   & 0.118 $\pm$ 0.006 & 17.0  &(A)  &  & & & & & \cr

&&&&&&&&& \cr
}
\endtable

\vfill
\eject


\begintable*{5}
\caption{{\bf Table 5.} Number of SA, SAB and SB galaxies in different classifications.}
\halign{%
\rm#\hfil&\qquad\rm#\hfil&\qquad\rm\hfil#&\qquad\rm\hfil
#&\qquad\rm\hfil#&\qquad\rm\hfil#&\qquad\rm#\hfil
&\qquad\rm\hfil#&\qquad\rm#\hfil&\qquad\hfil\rm#\cr
          & SA    & SAB & SB & & & & & & \cr
\noalign{\vskip 10pt}

RC3          &46 &50 & 59 & & & & & & \cr
H (visual)   &33 &26 & 98  & & & & & & \cr
H (Fourier)  &53 &   & 105 & & & & & & \cr
&&&&&&&&& \cr
}
\endtable

\vfill
\eject

\psfig{file=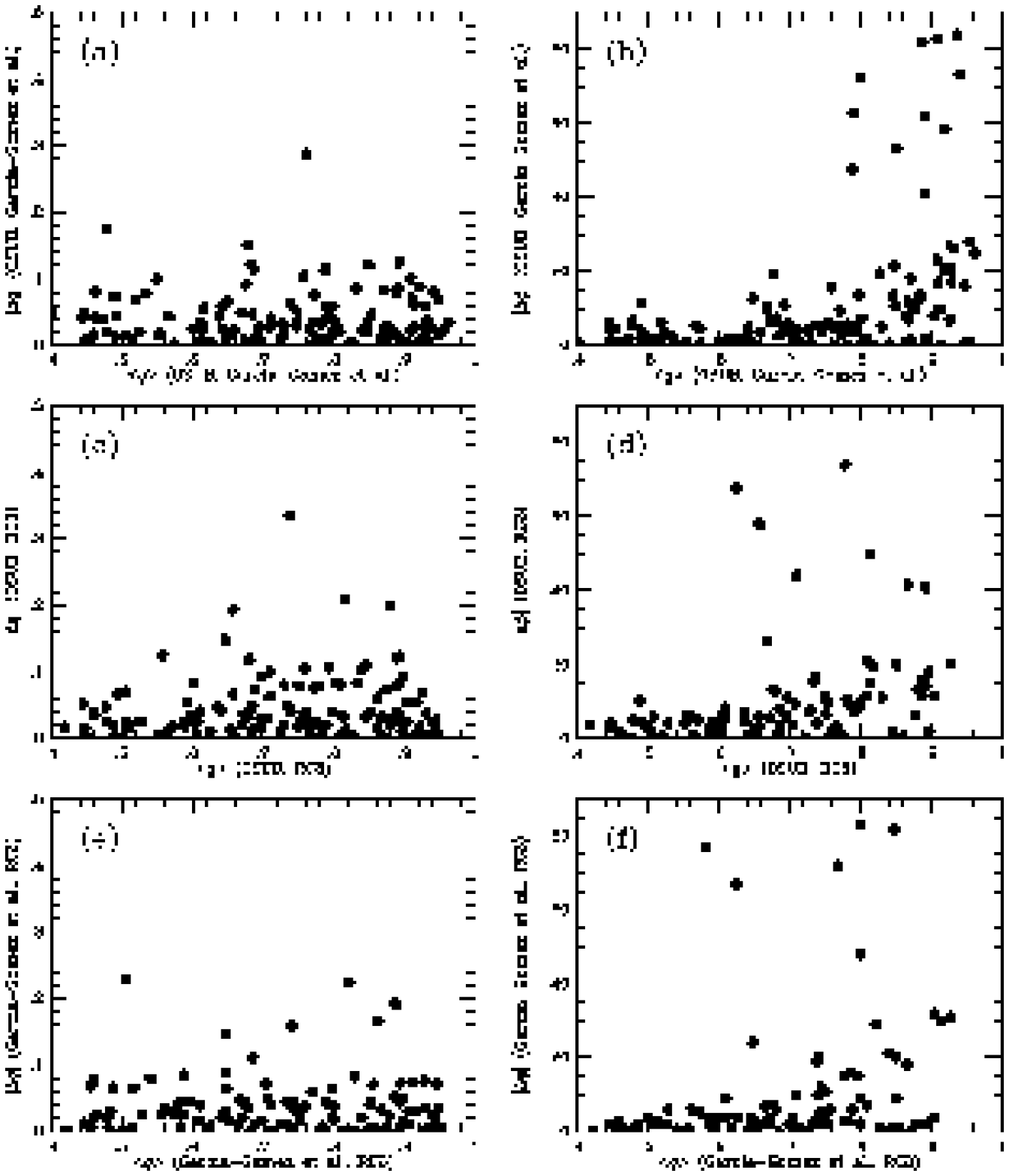,width=14cm}
Fig. 1
\vfill
\eject

\psfig{file=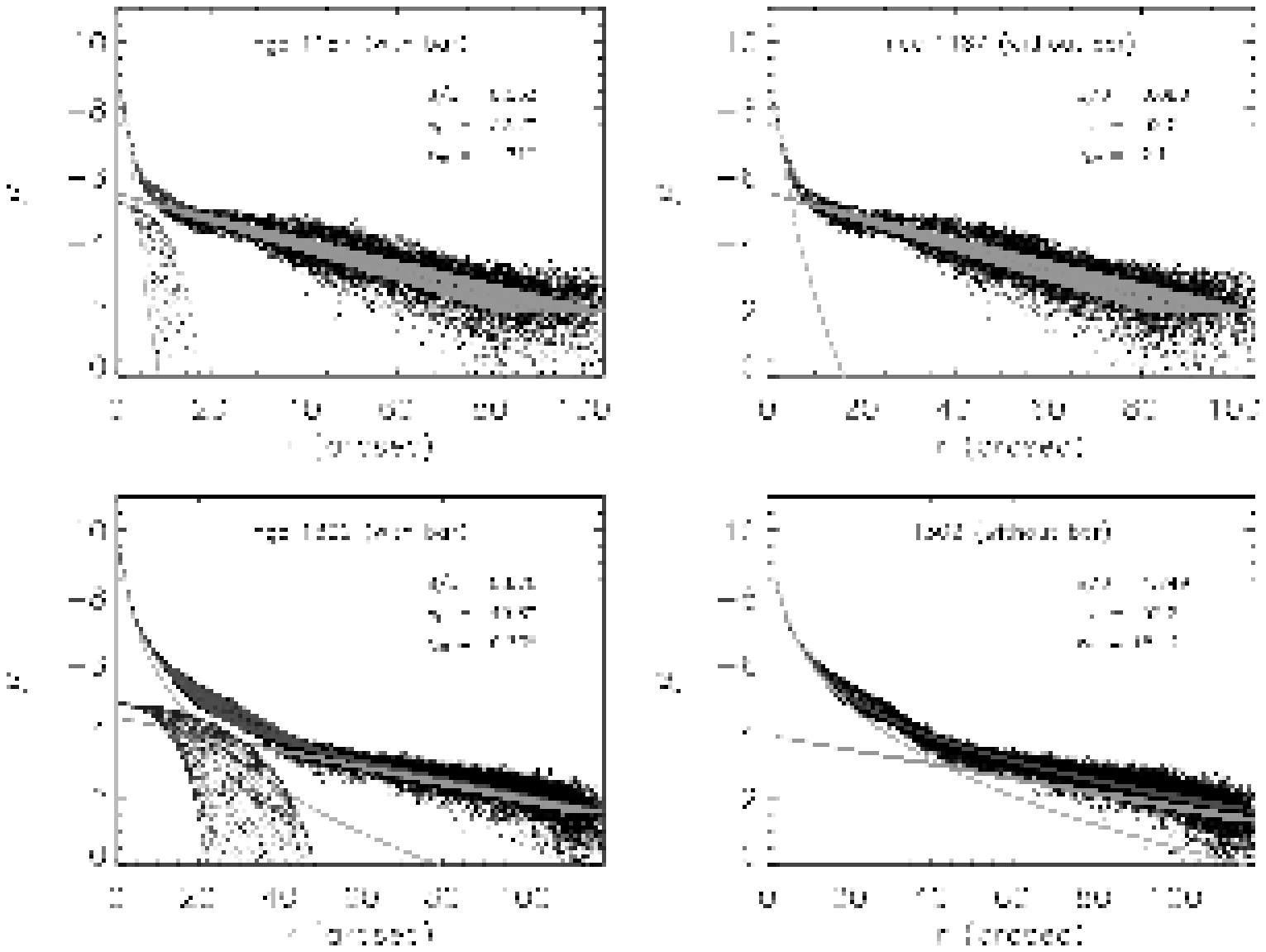,width=14cm}
Fig. 2a
\vfill
\eject

\psfig{file=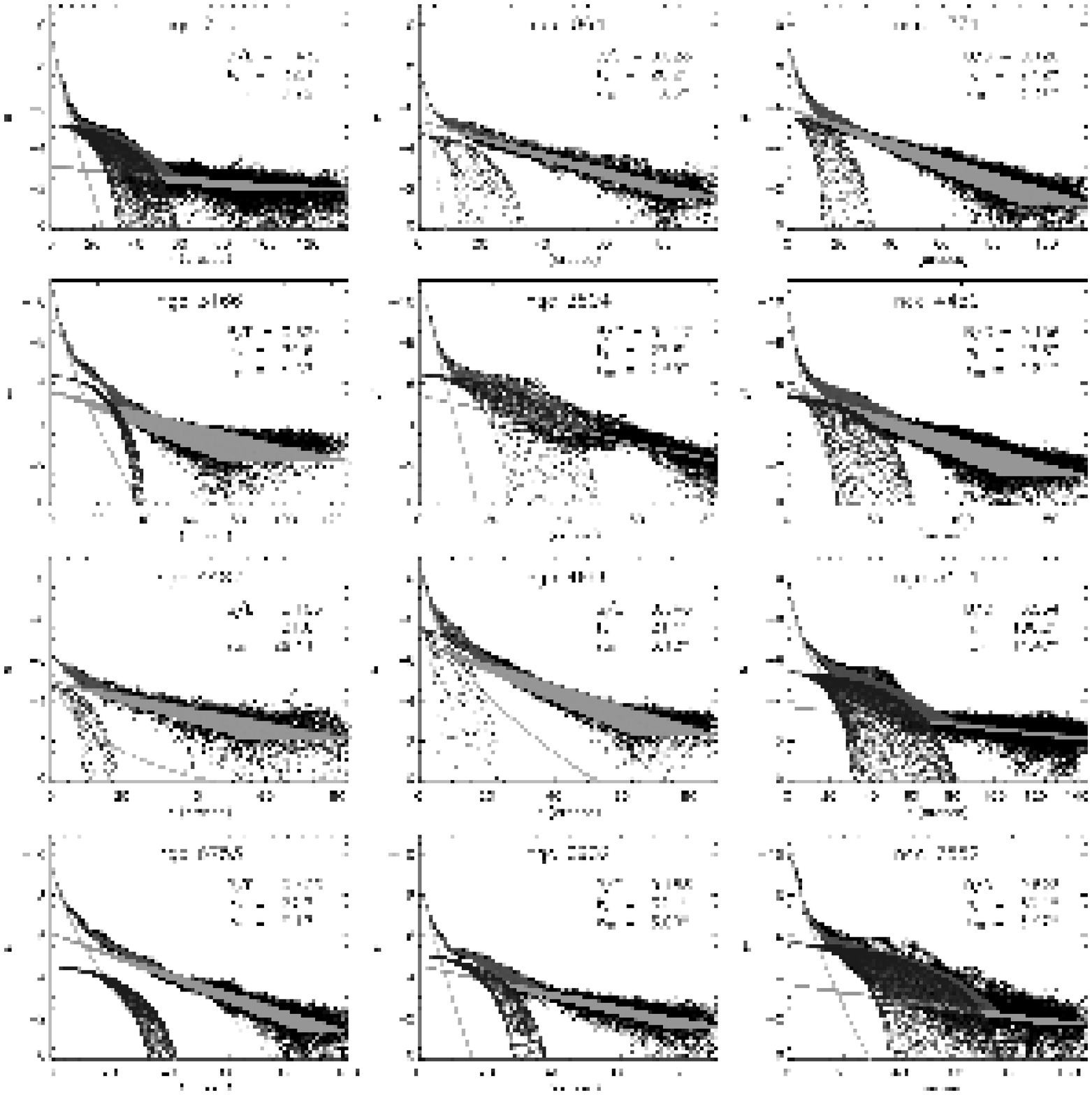,width=14cm}
Fig. 2b
\vfill
\eject

\psfig{file=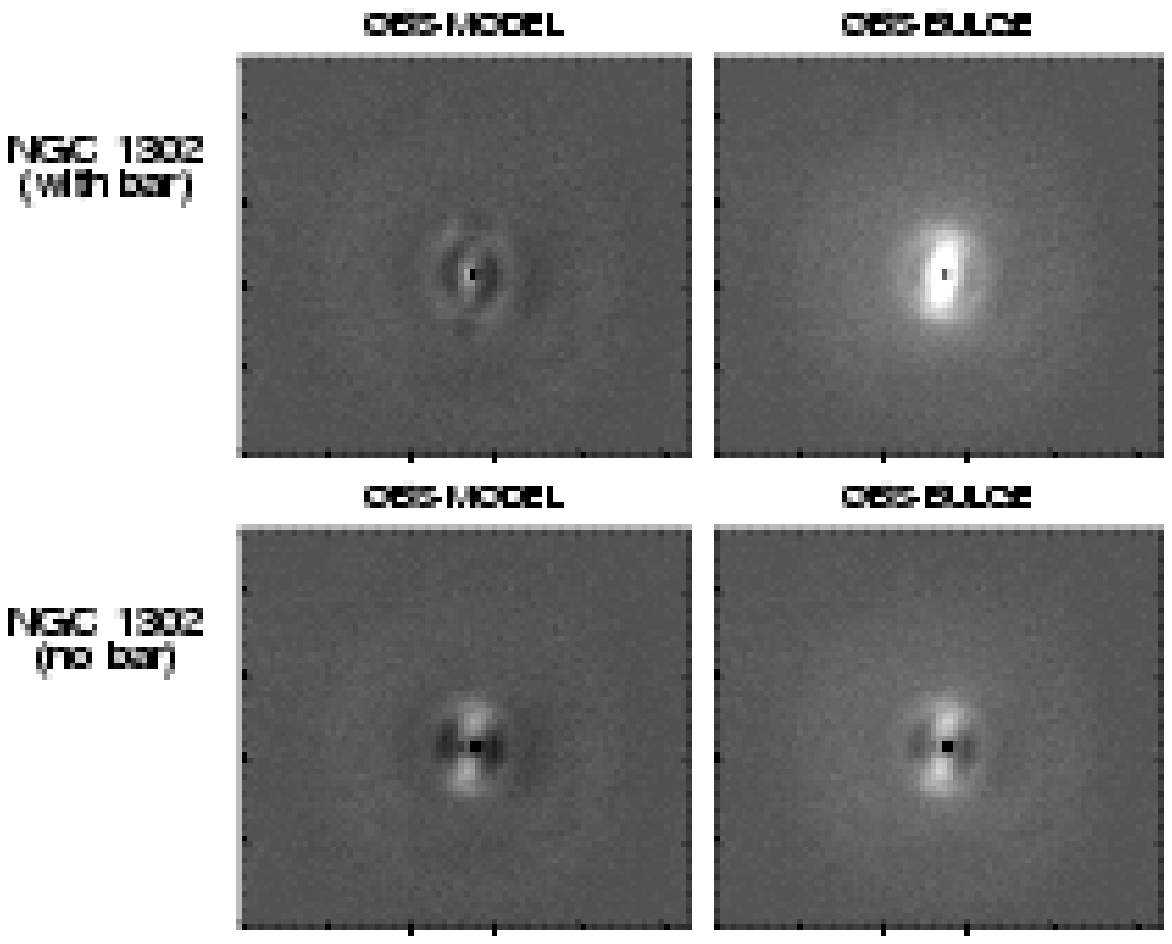,width=16cm}
Fig. 3
\vfill
\eject

\psfig{file=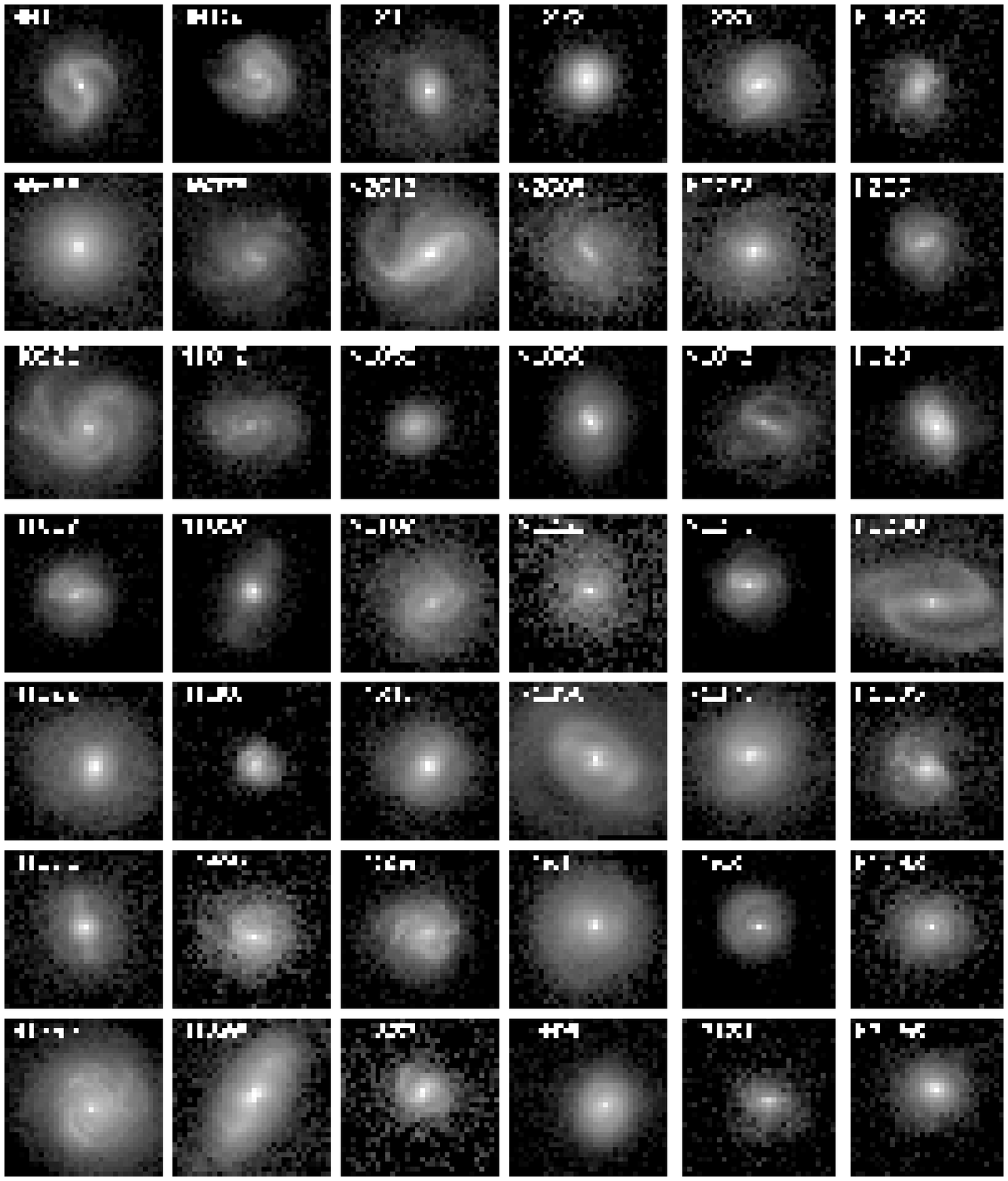,width=16cm}
Fig. 4a
\vfill
\eject

\psfig{file=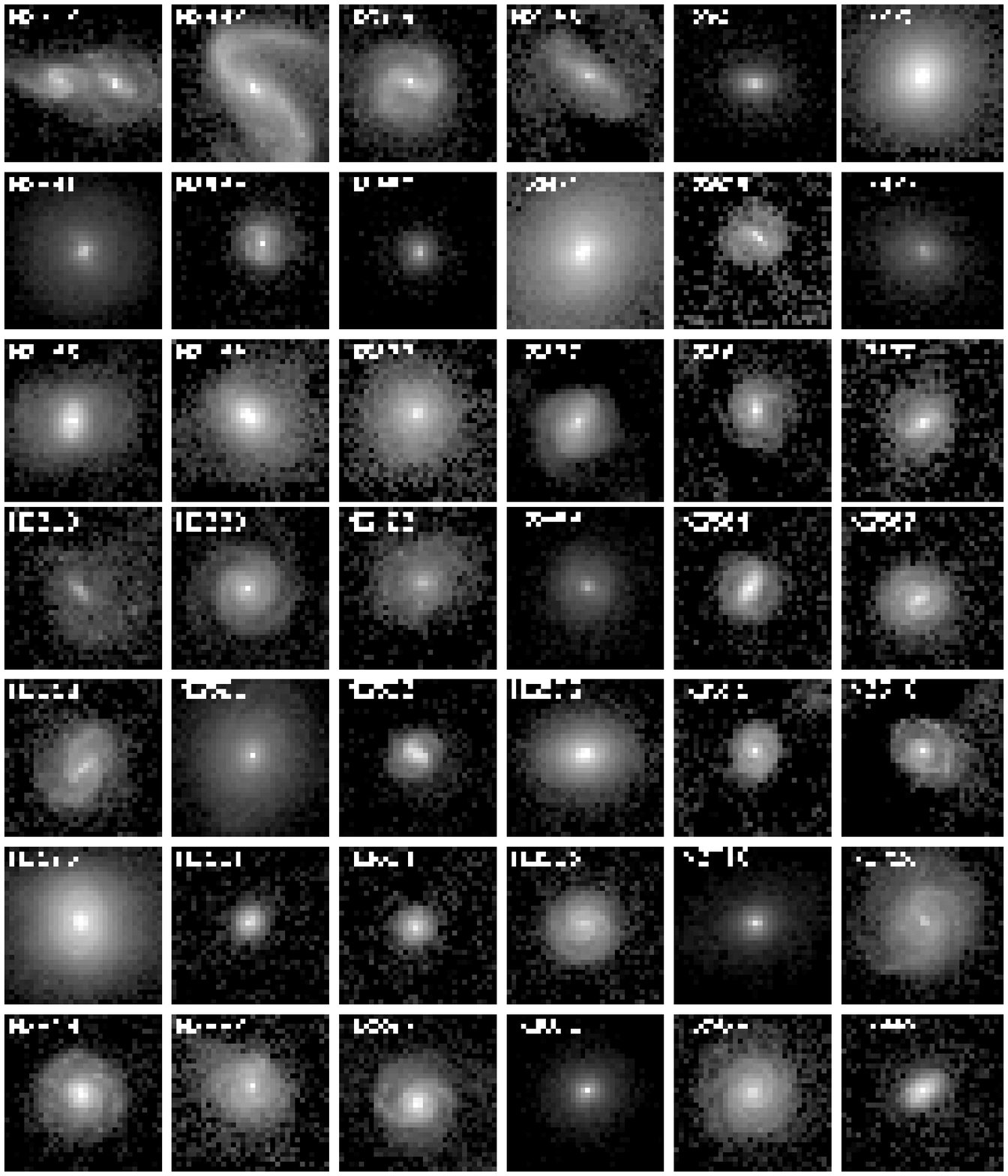,width=16cm}
Fig. 4b
\vfill
\eject

\psfig{file=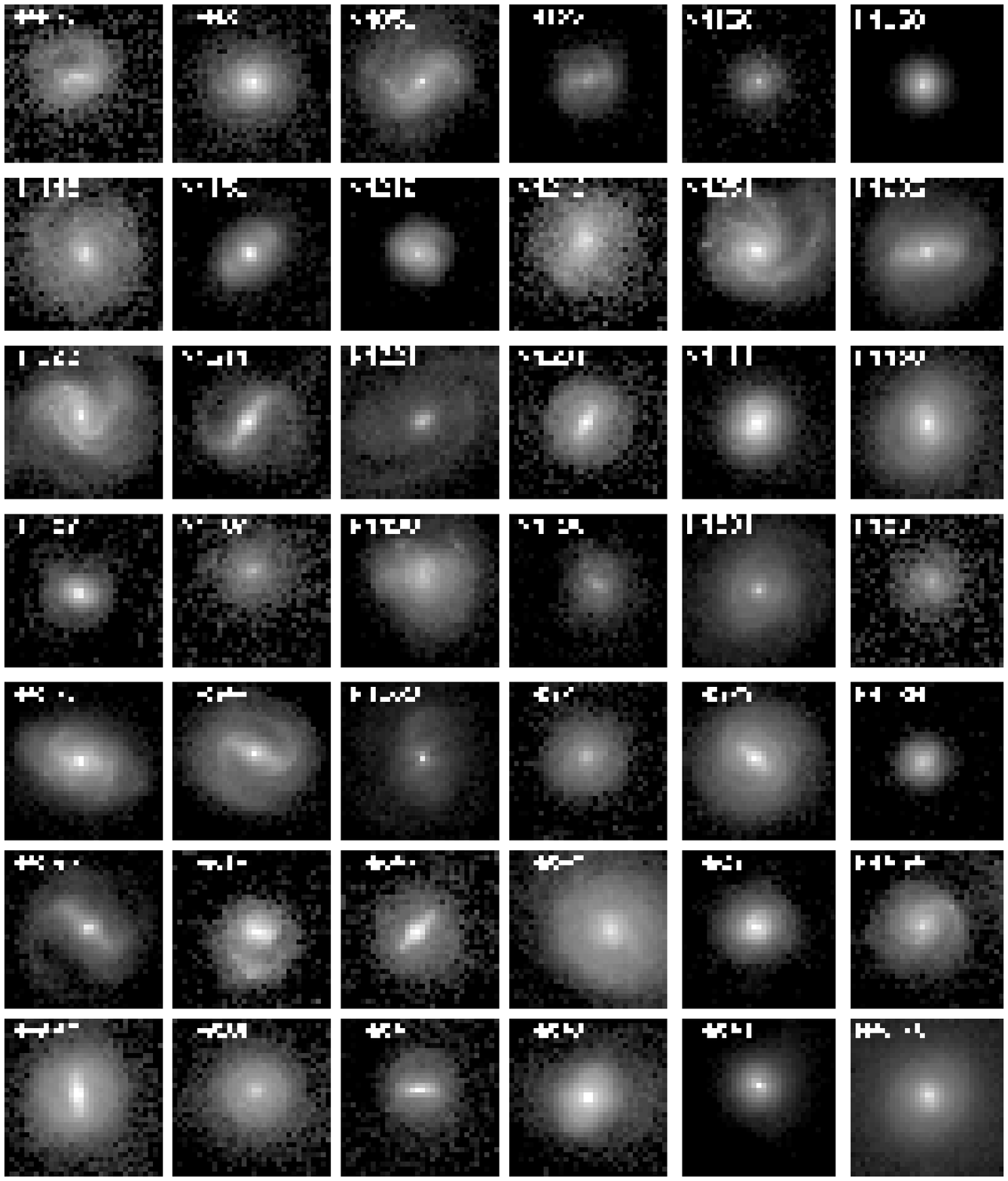,width=16cm}
Fig. 4c
\vfill
\eject

\psfig{file=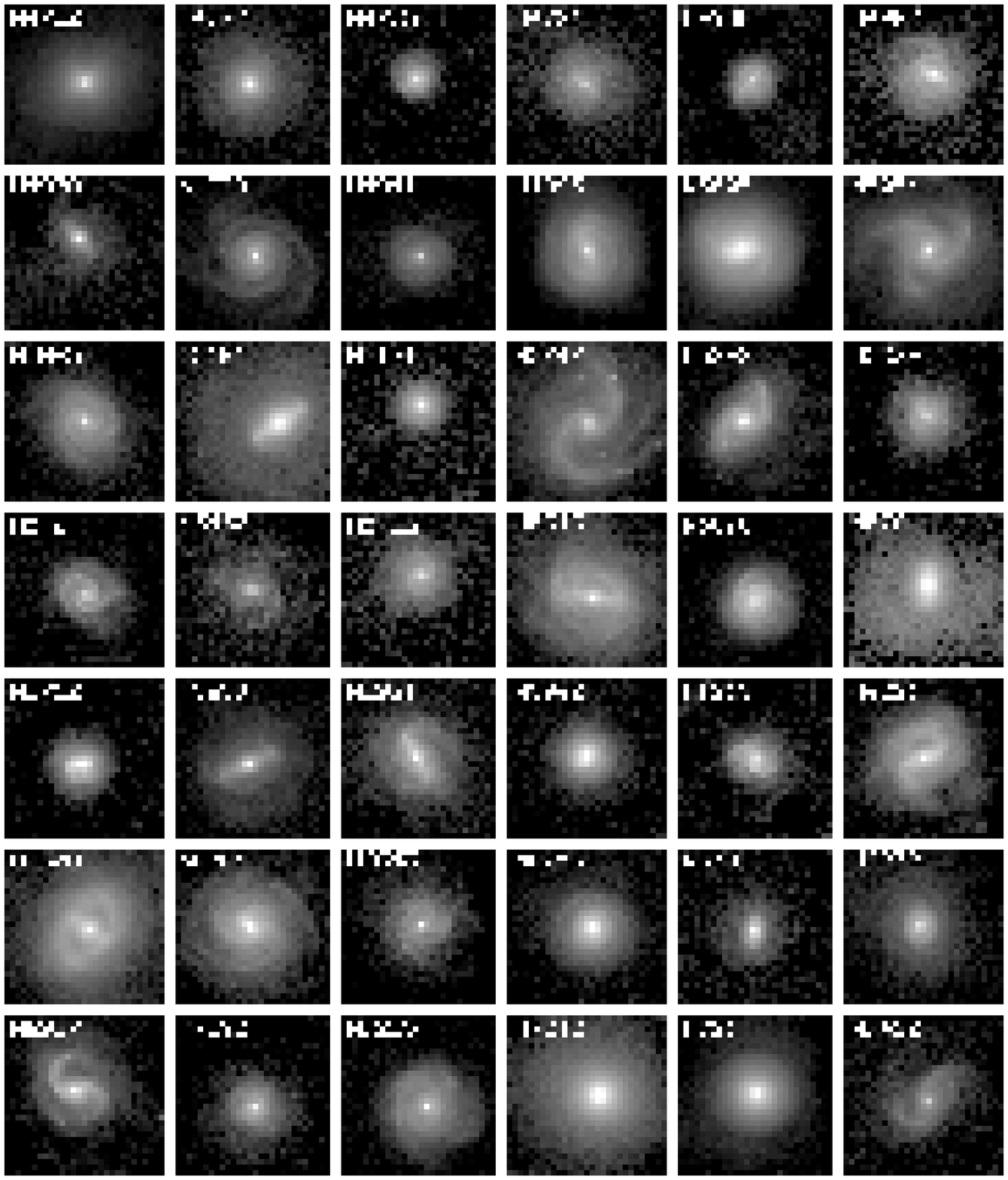,width=16cm}
Fig. 4d
\vfill
\eject

\psfig{file=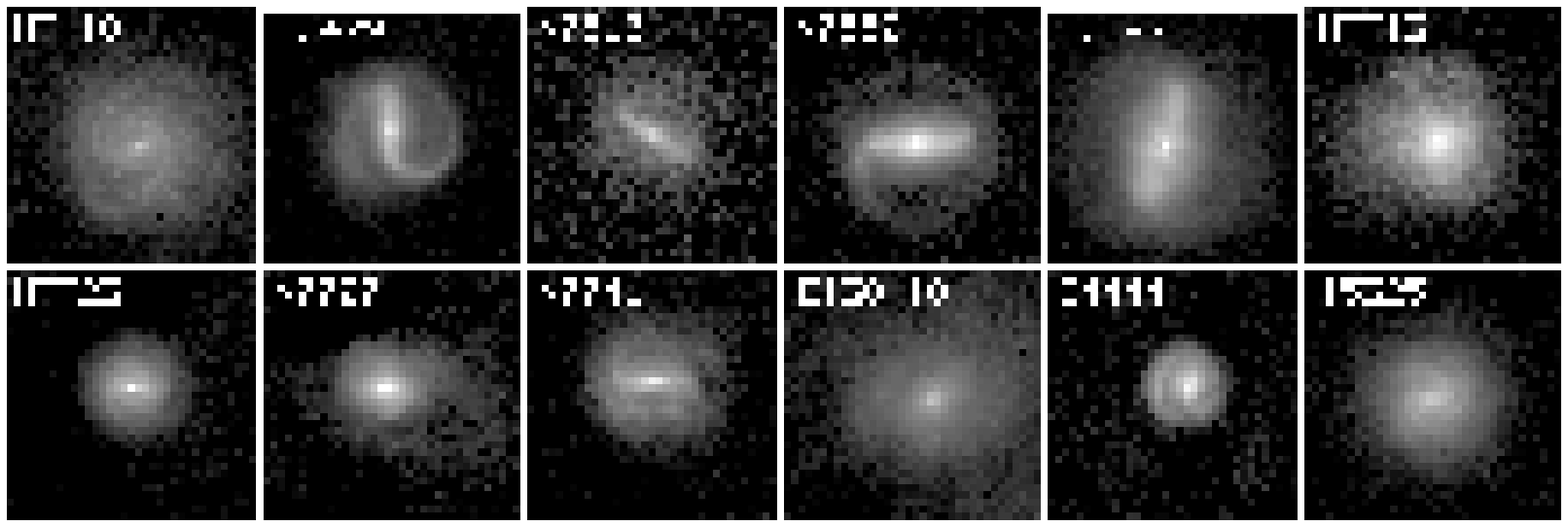,width=16cm}
Fig. 4e
\vfill
\eject

\psfig{file=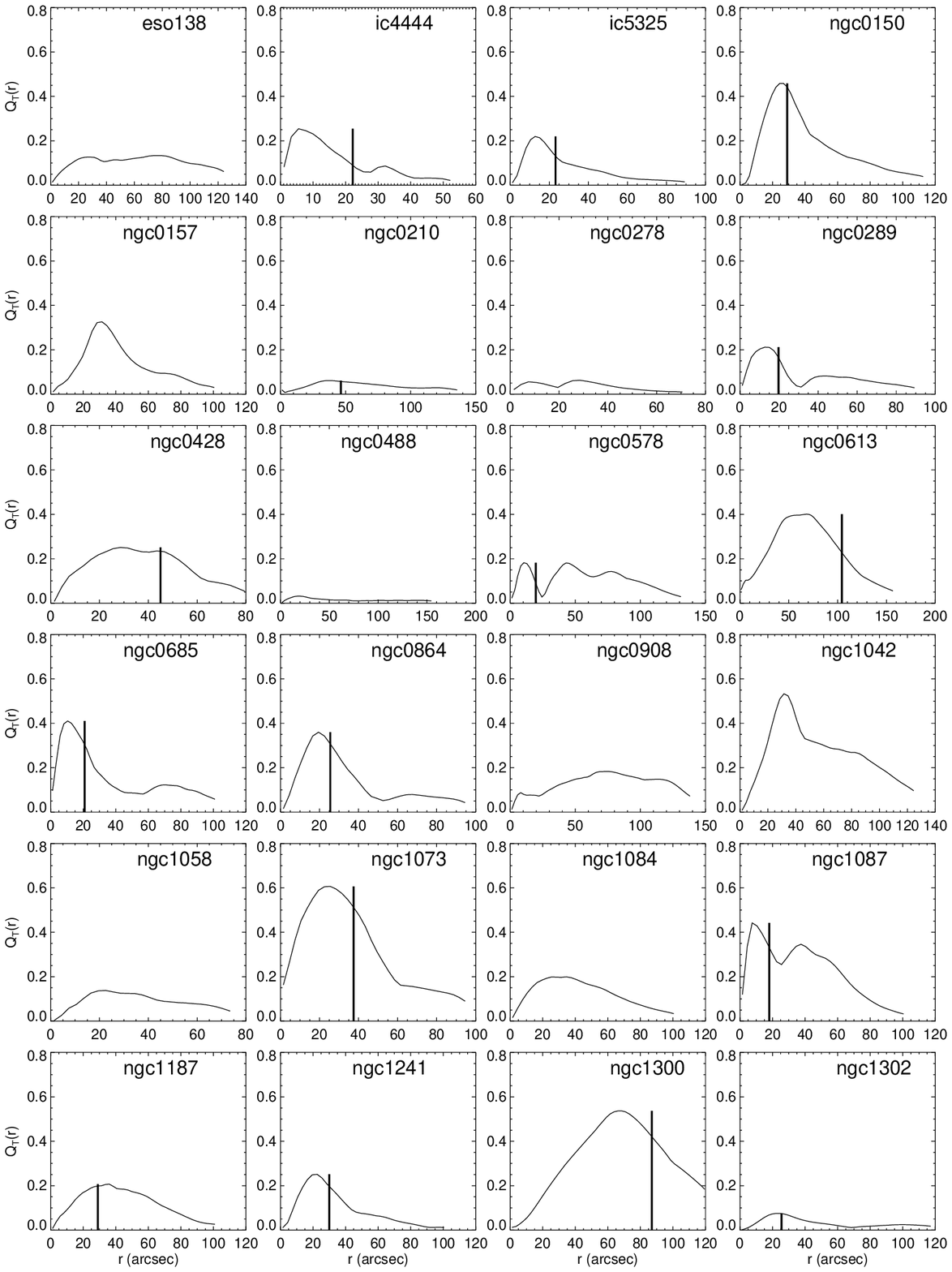,width=16cm}
Fig. 5a
\vfill
\eject

\psfig{file=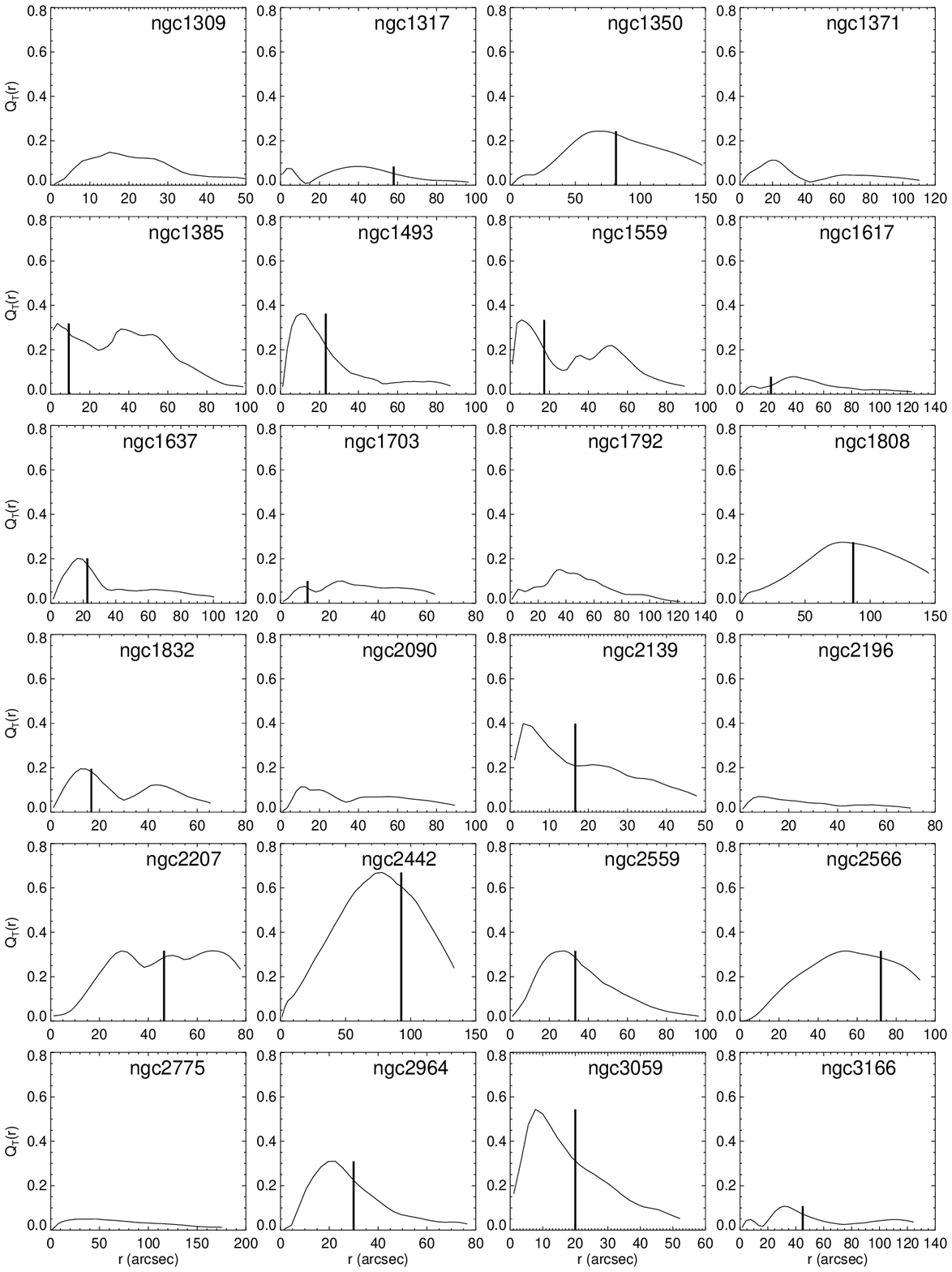,width=16cm}
Fig. 5b
\vfill
\eject

\psfig{file=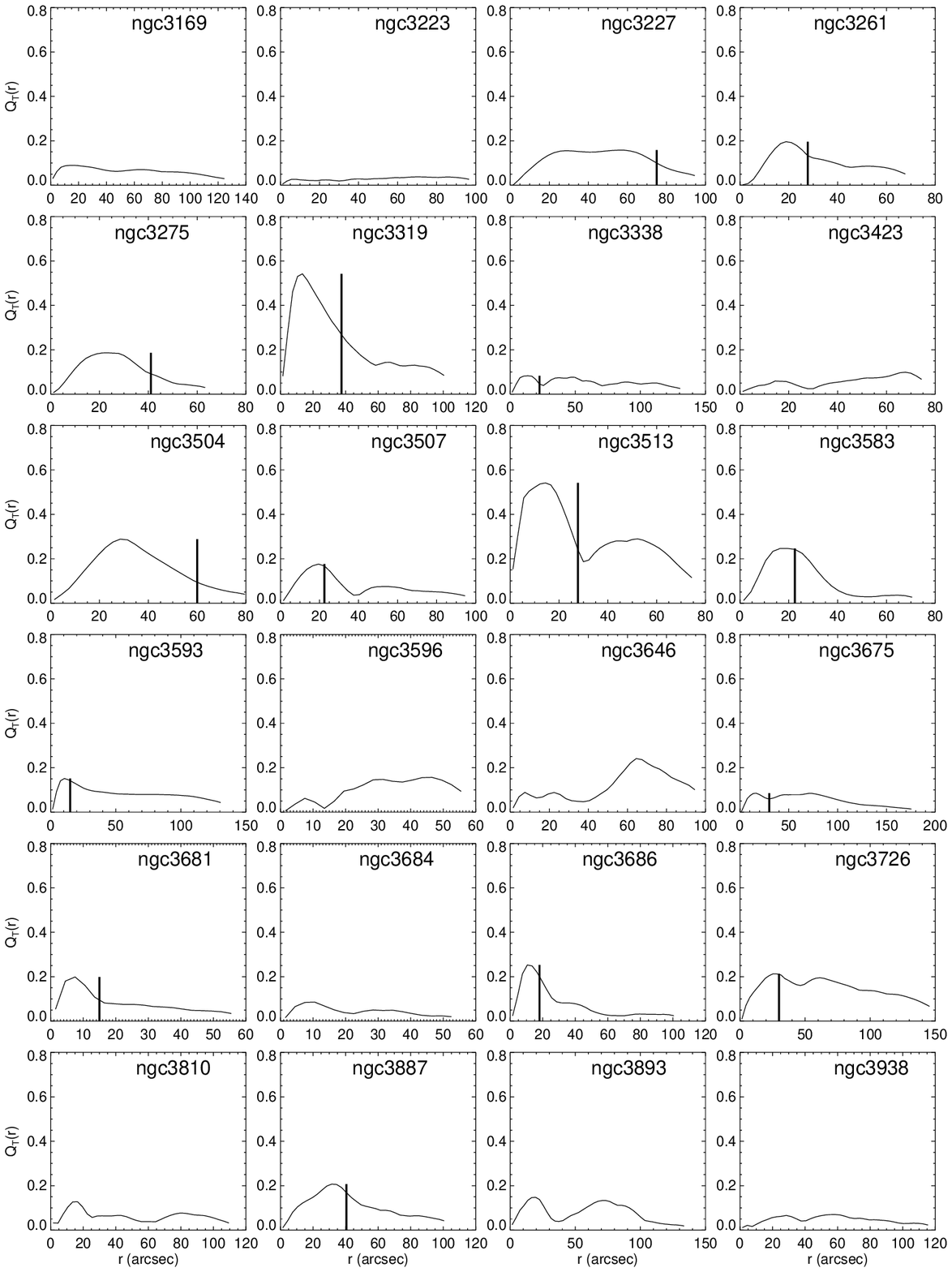,width=16cm}
Fig. 5c
\vfill
\eject

\psfig{file=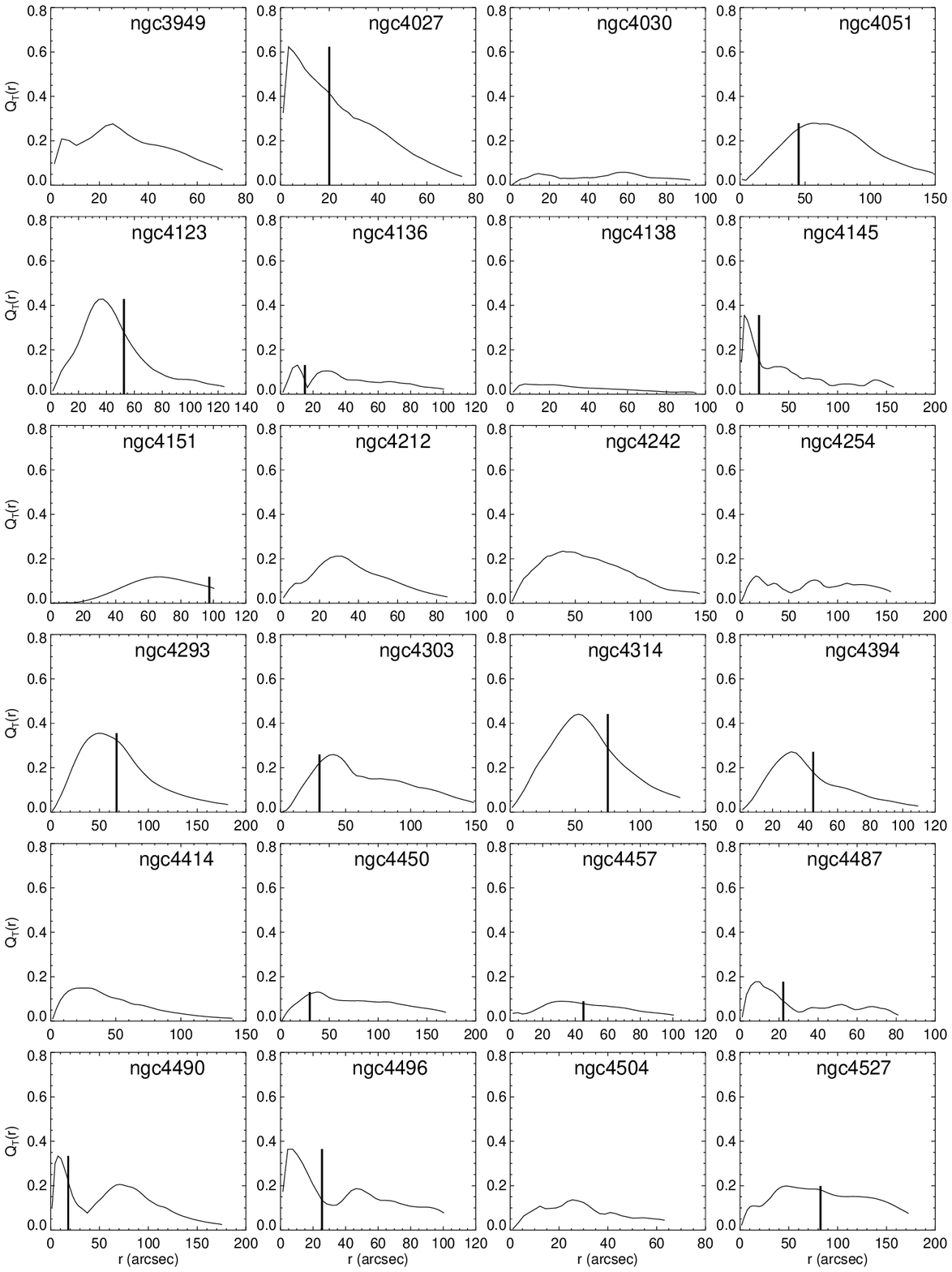,width=16cm}
Fig. 5d
\vfill
\eject

\psfig{file=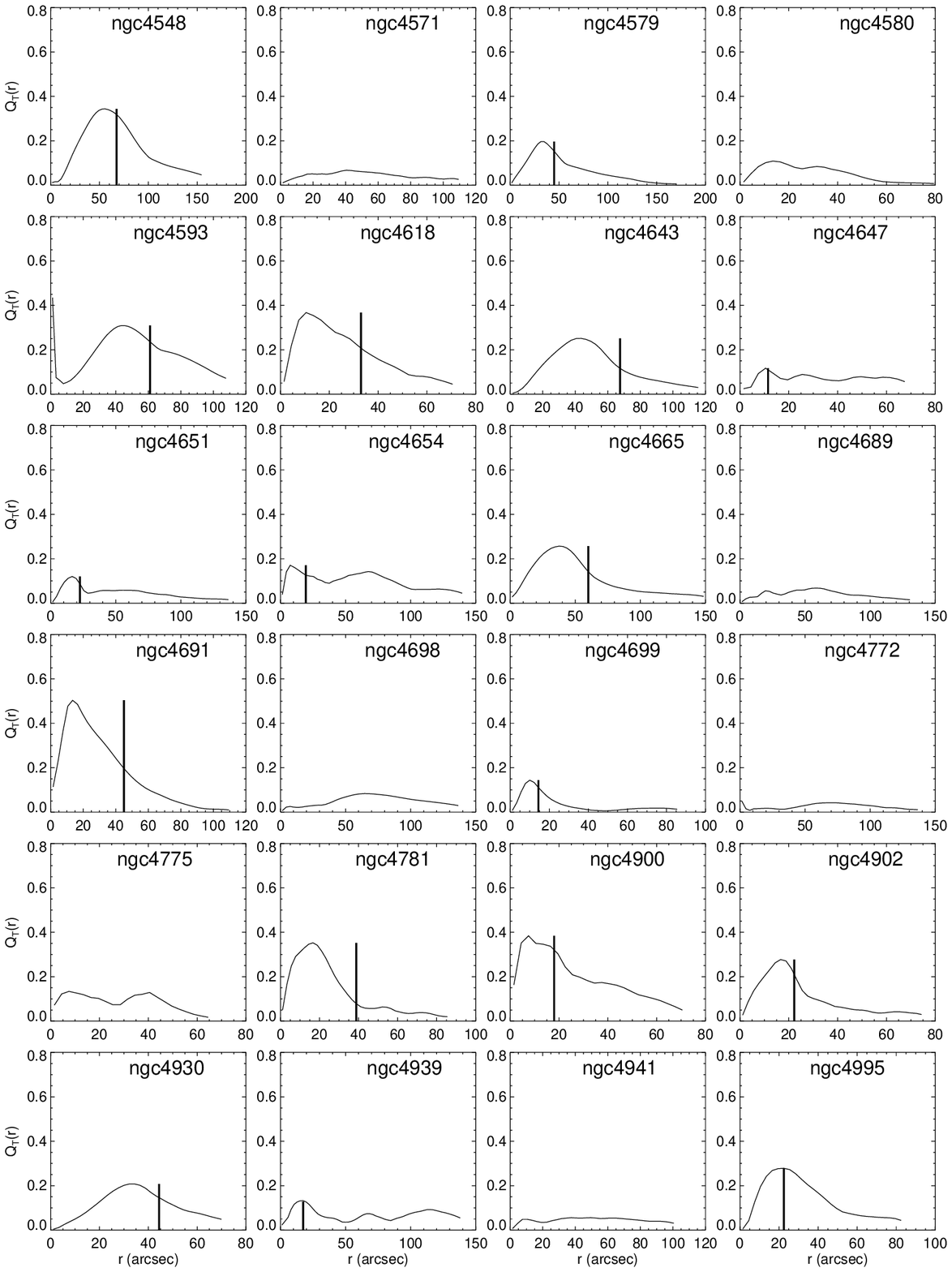,width=16cm}
Fig. 5e
\vfill
\eject

\psfig{file=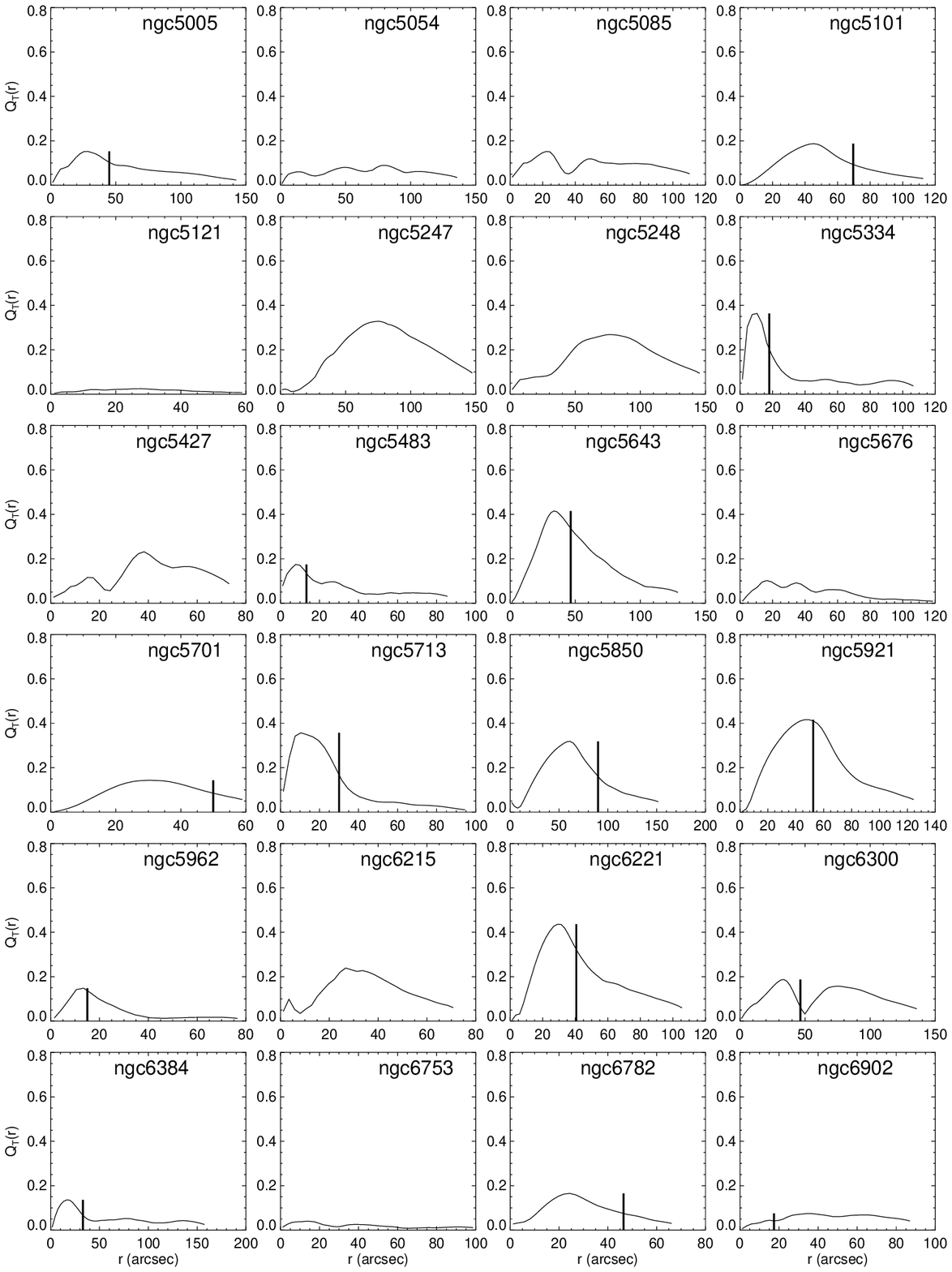,width=16cm}
Fig. 5f
\vfill
\eject

\psfig{file=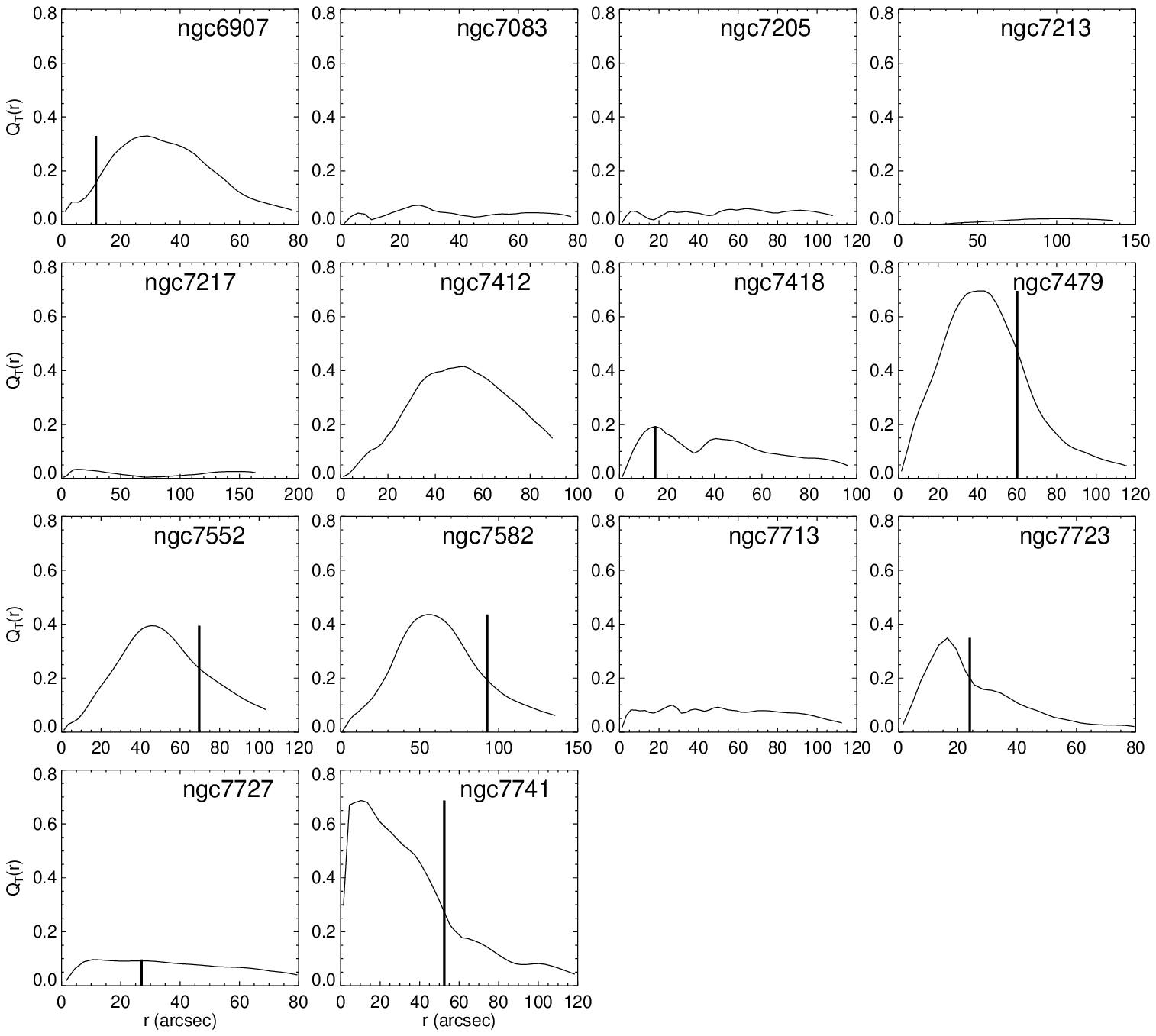,width=16cm}
Fig. 5g
\vfill
\eject

\psfig{file=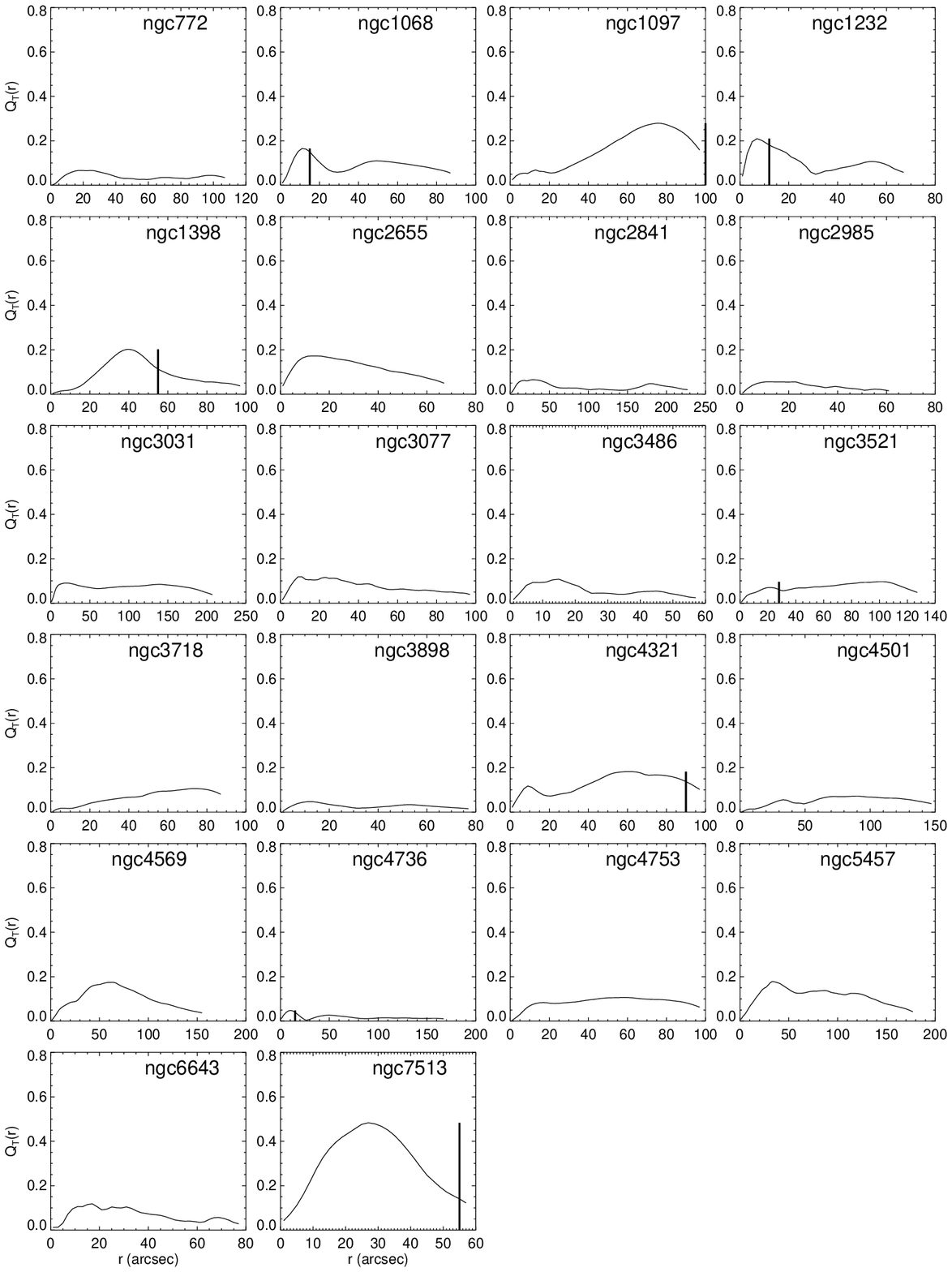,width=16cm}
Fig. 5h
\vfill
\eject

\psfig{file=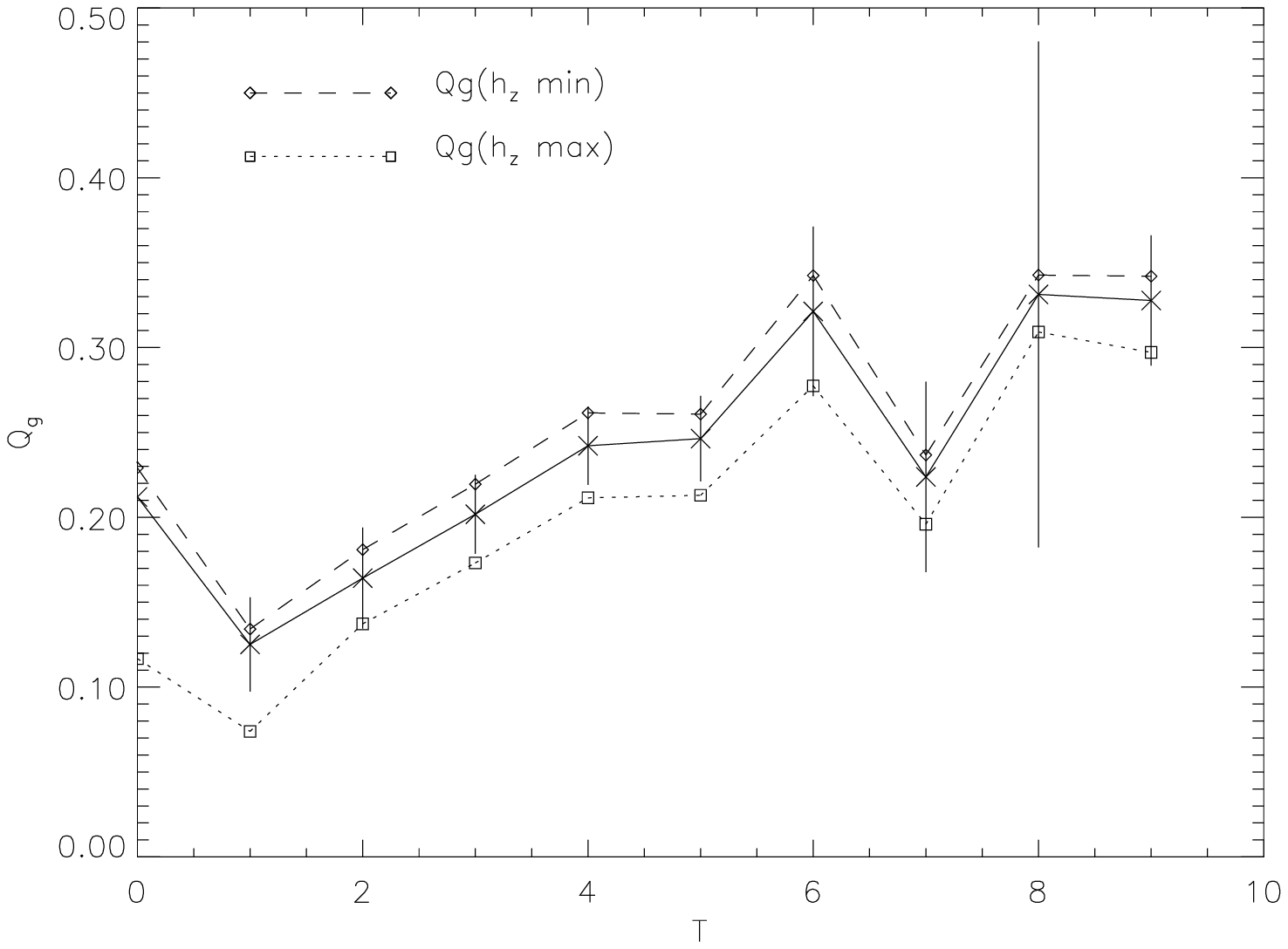,width=16cm}
Fig. 6
\vfill
\eject

\psfig{file=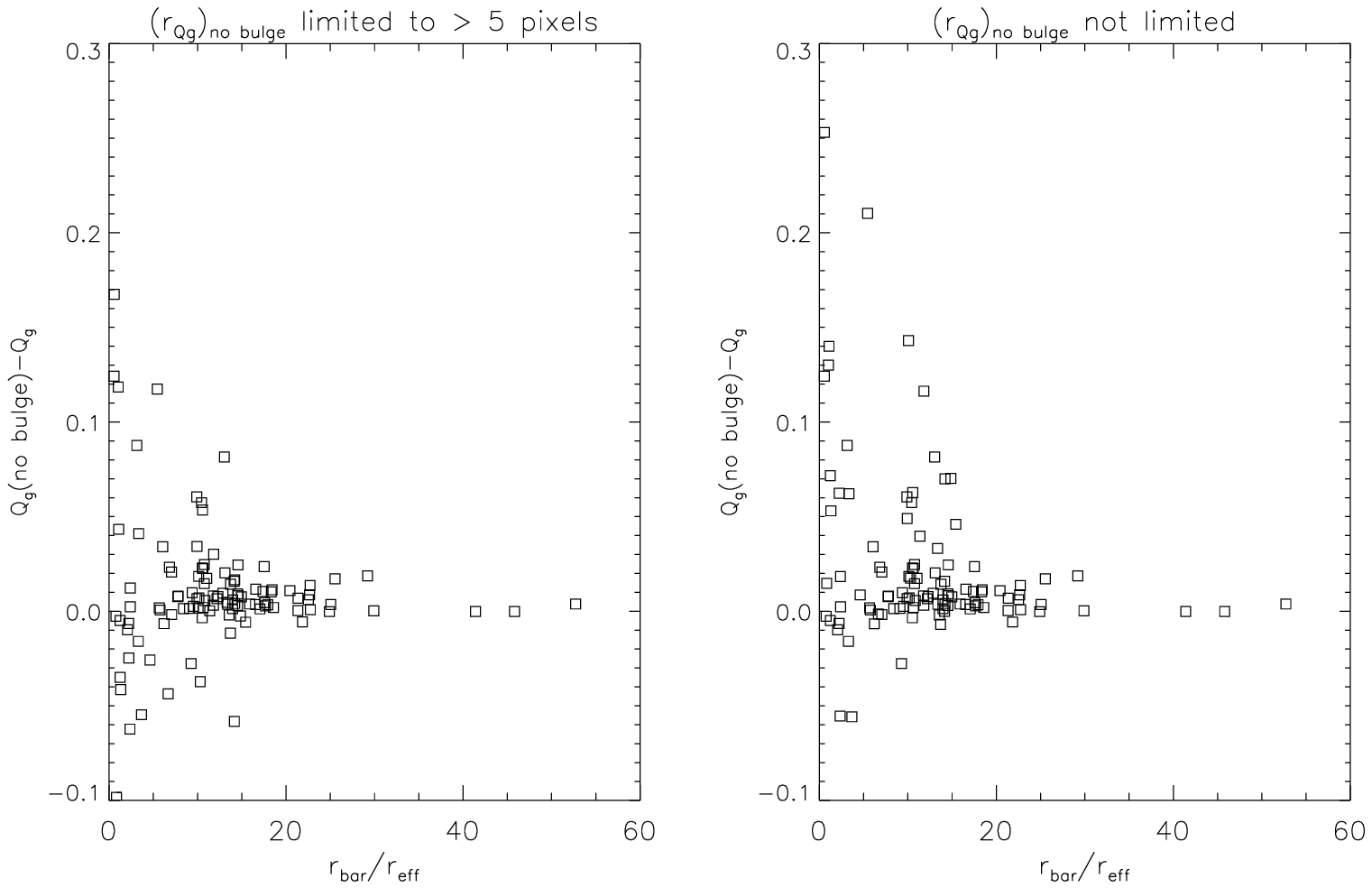,width=16cm}
Fig. 7
\vfill
\eject

\psfig{file=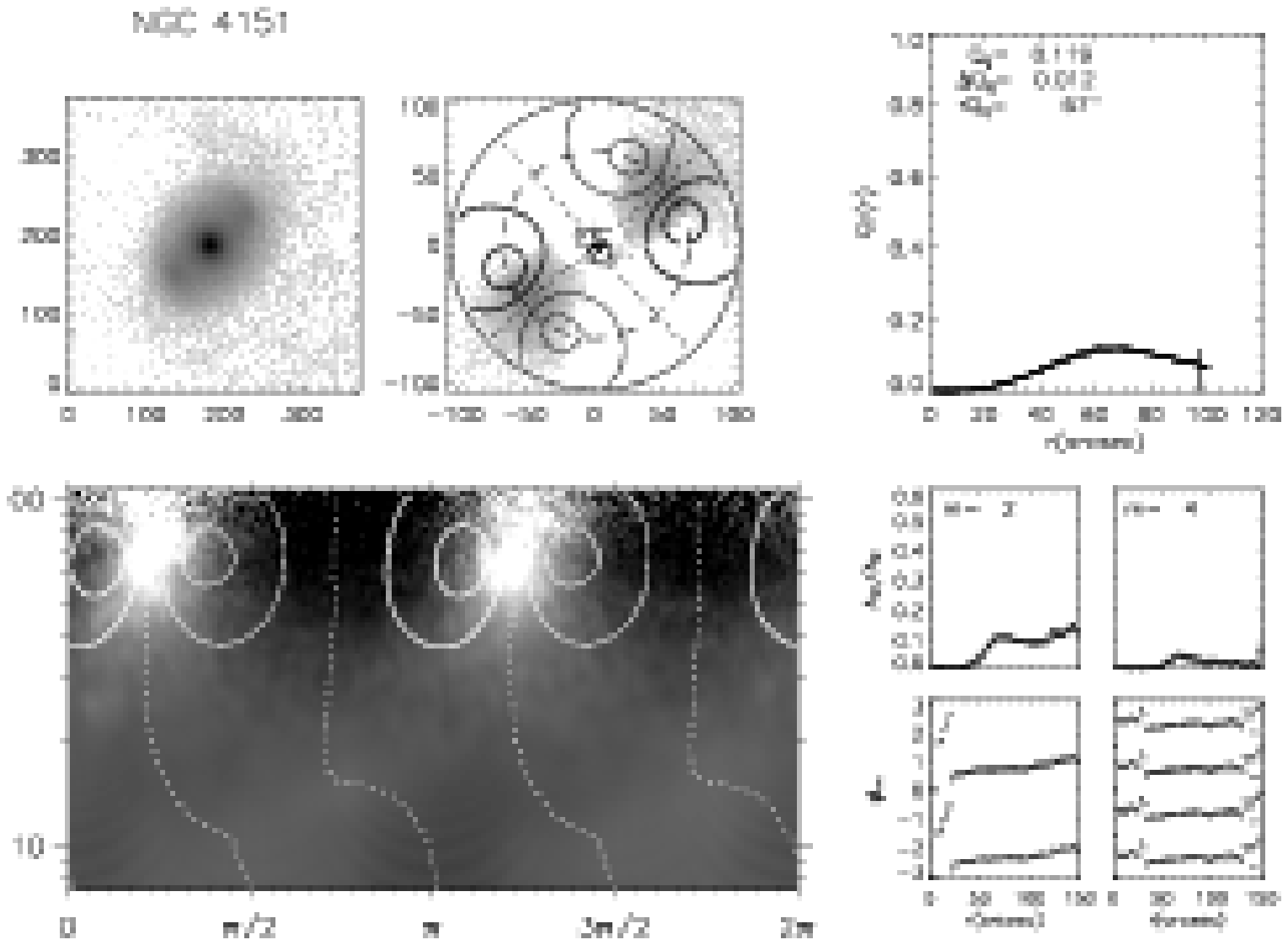,width=16cm}
Fig. 8a
\vfill
\eject

\psfig{file=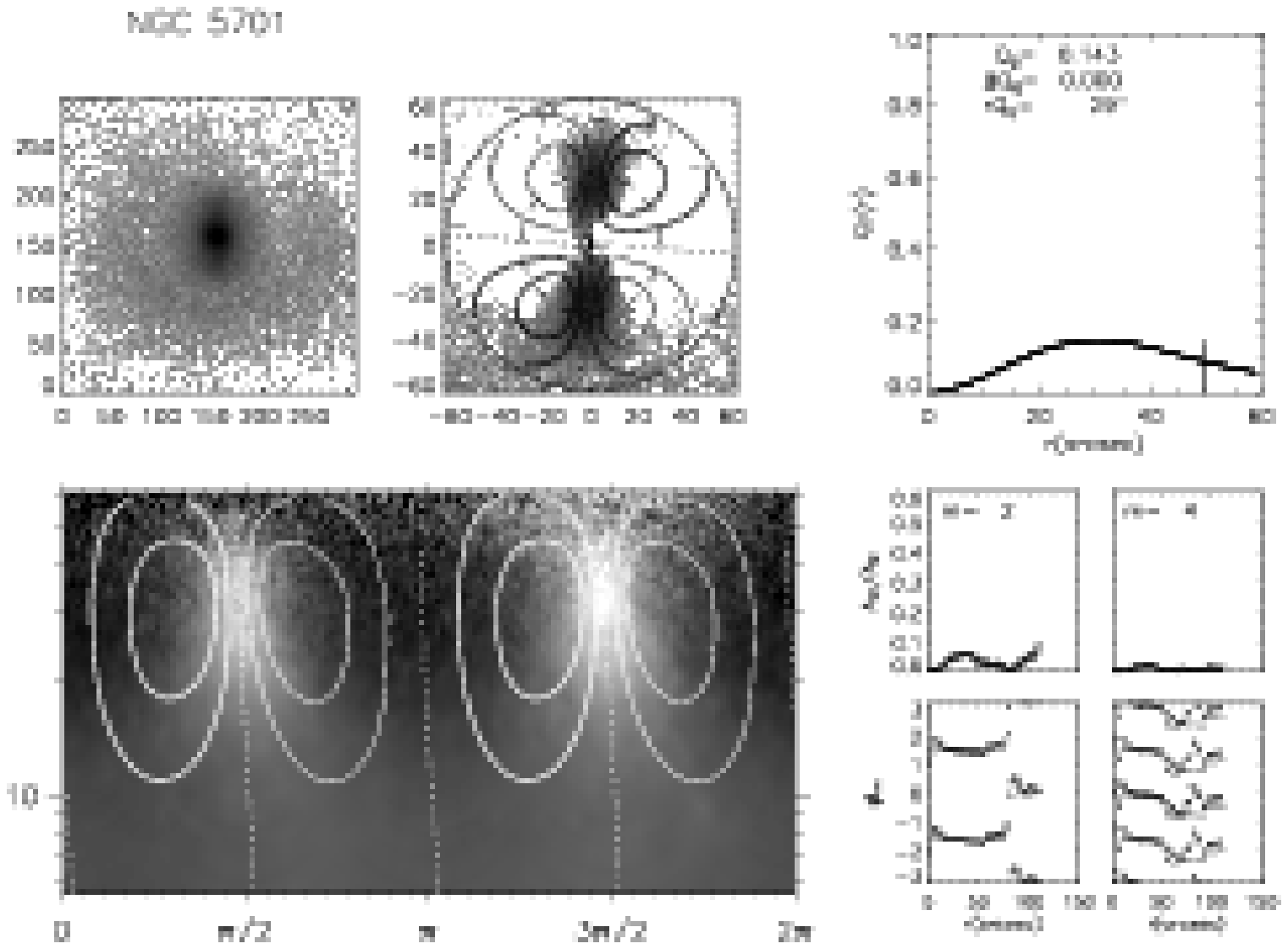,width=16cm}
Fig. 8b
\vfill
\eject

\psfig{file=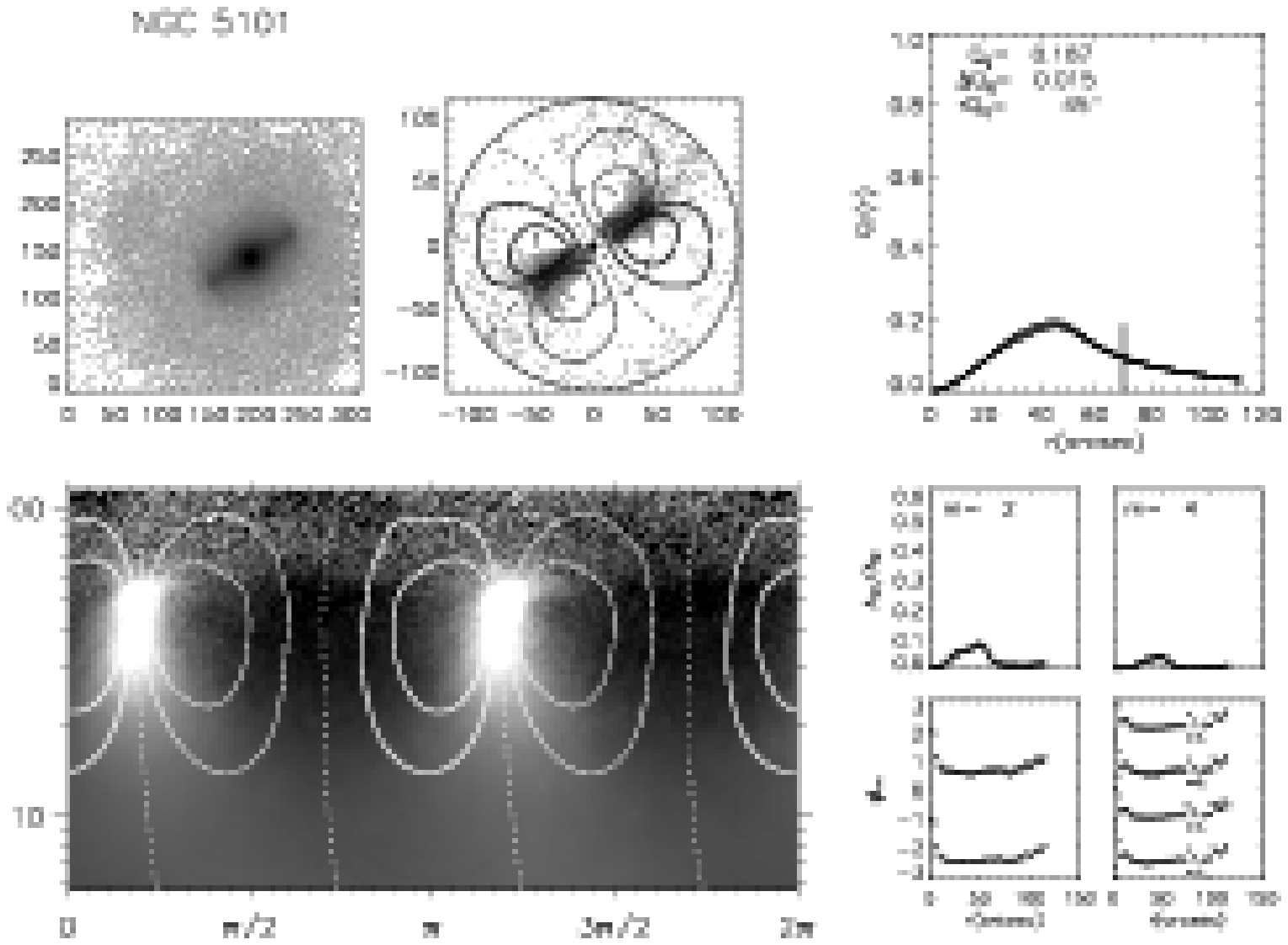,width=16cm}
Fig. 8c
\vfill
\eject

\psfig{file=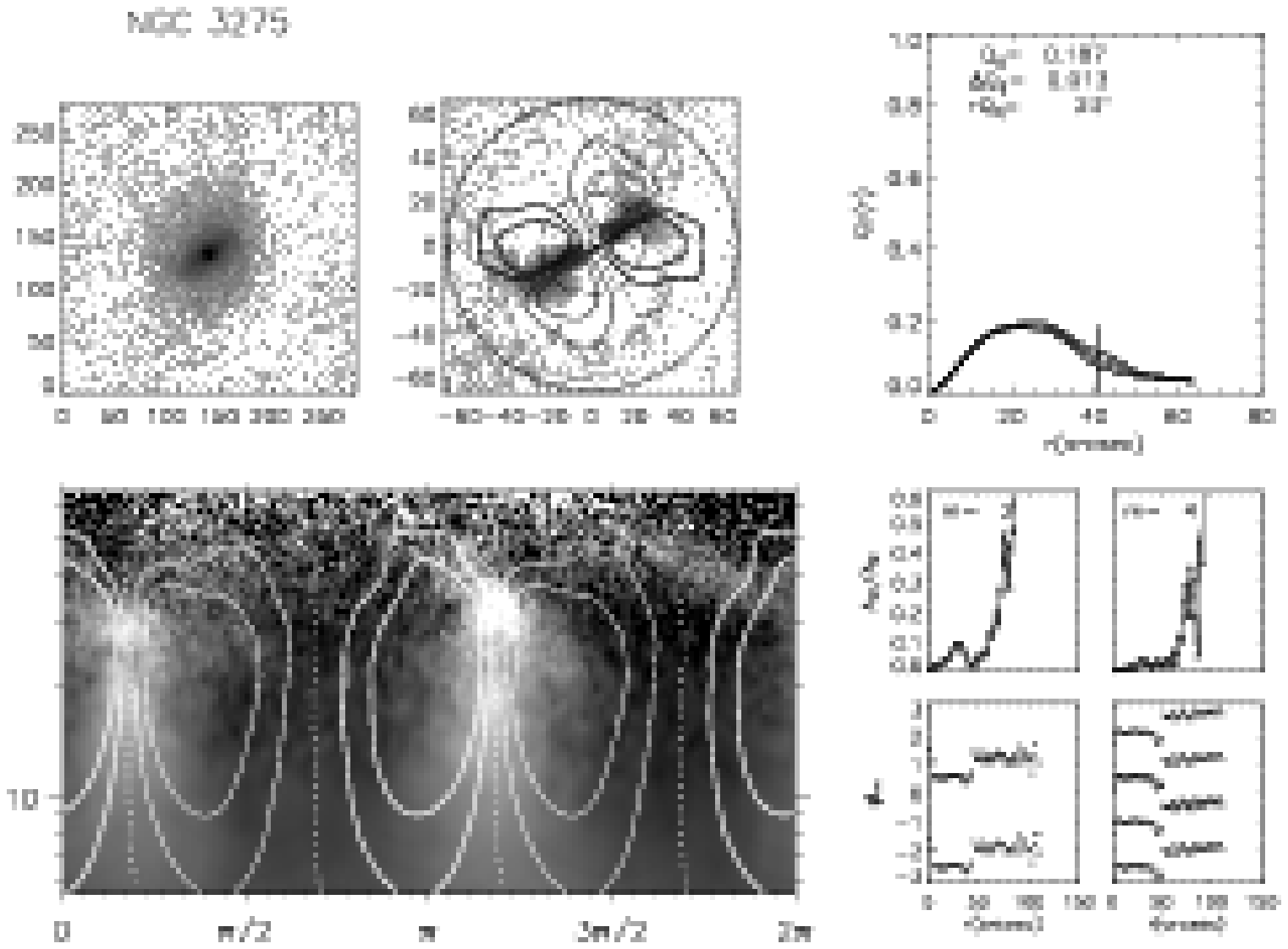,width=16cm}
Fig. 8d
\vfill
\eject

\psfig{file=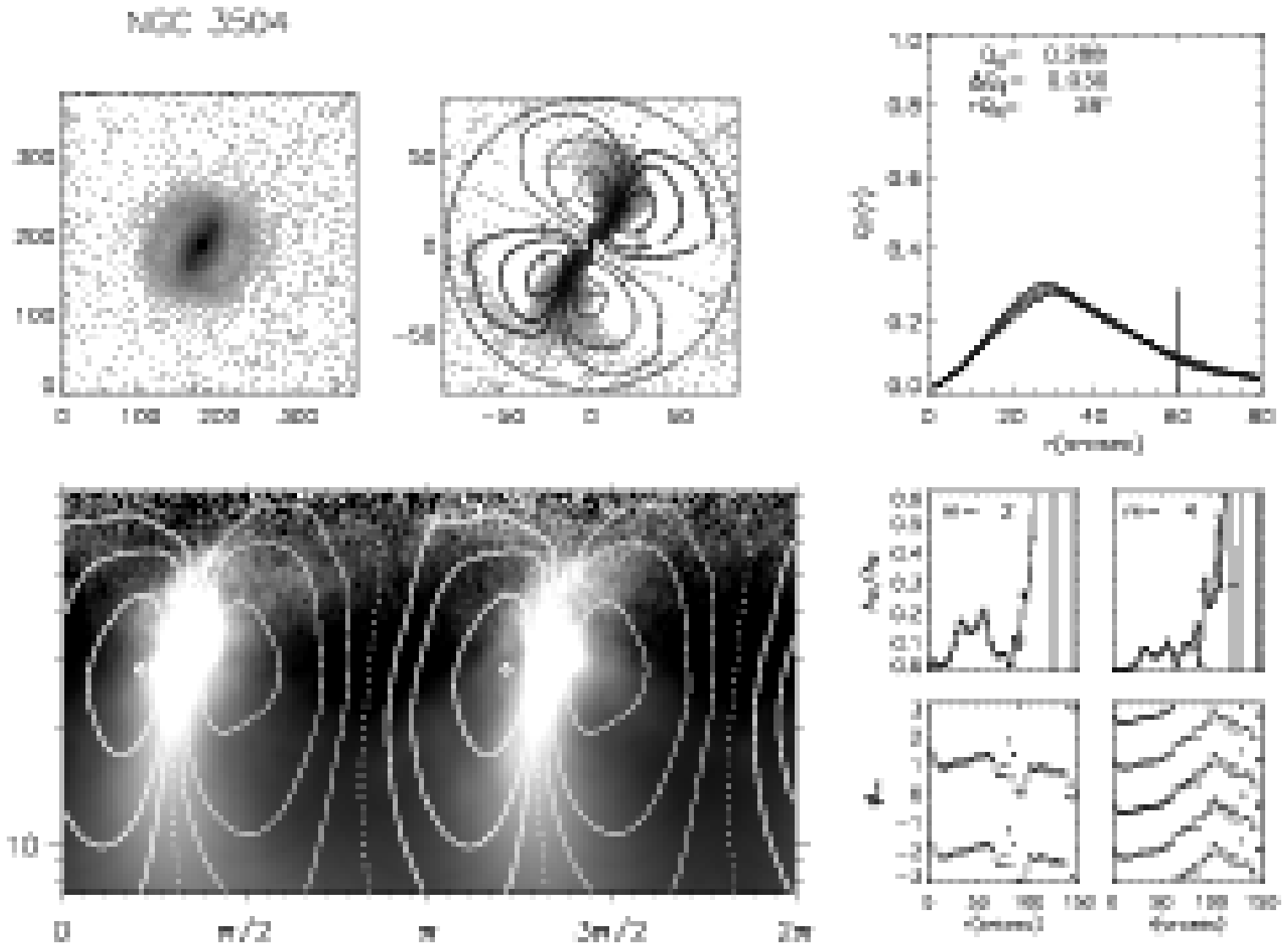,width=16cm}
Fig. 8e
\vfill
\eject

\psfig{file=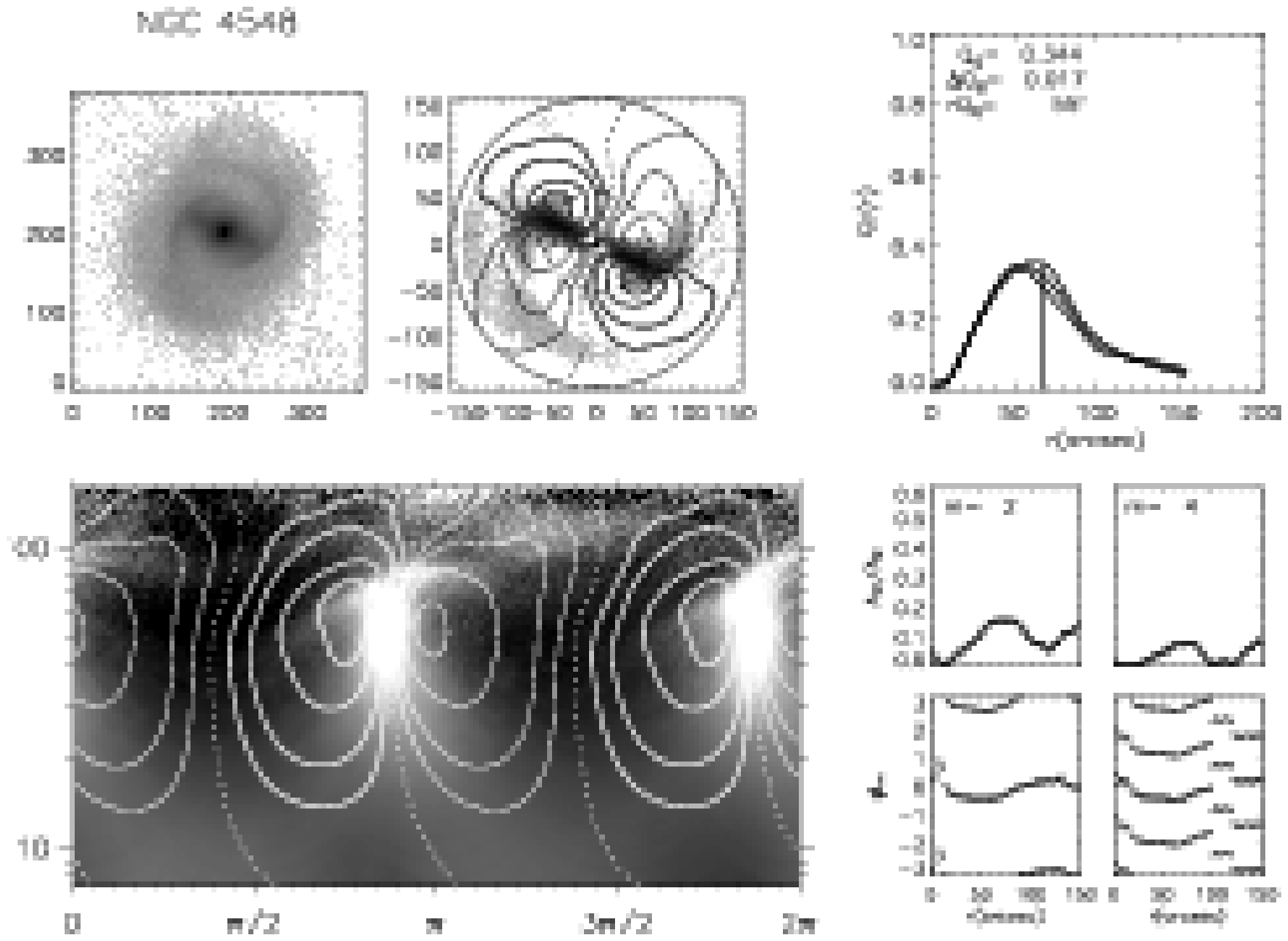,width=16cm}
Fig. 8f
\vfill
\eject

\psfig{file=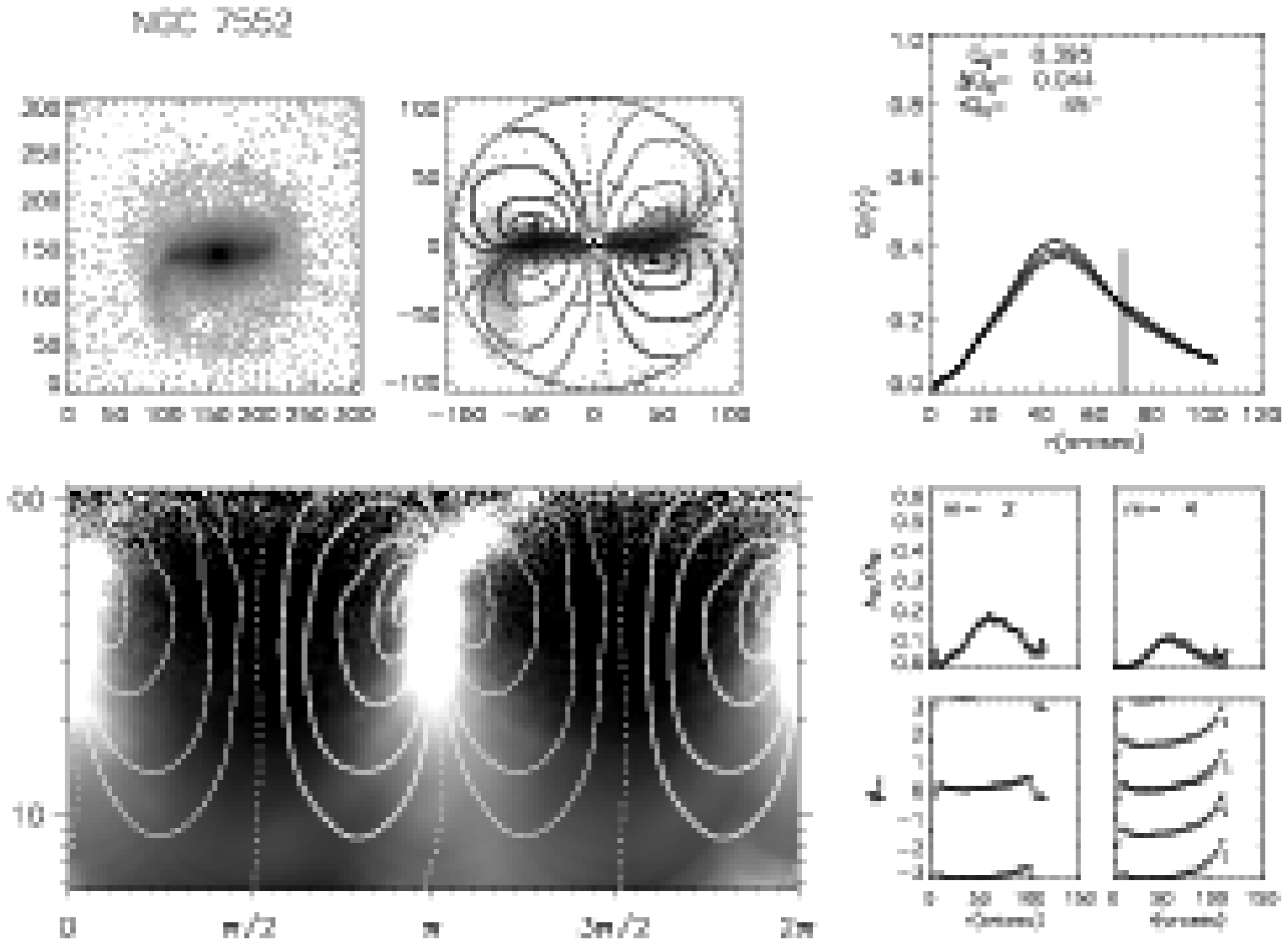,width=16cm}
Fig. 8g
\vfill
\eject

\psfig{file=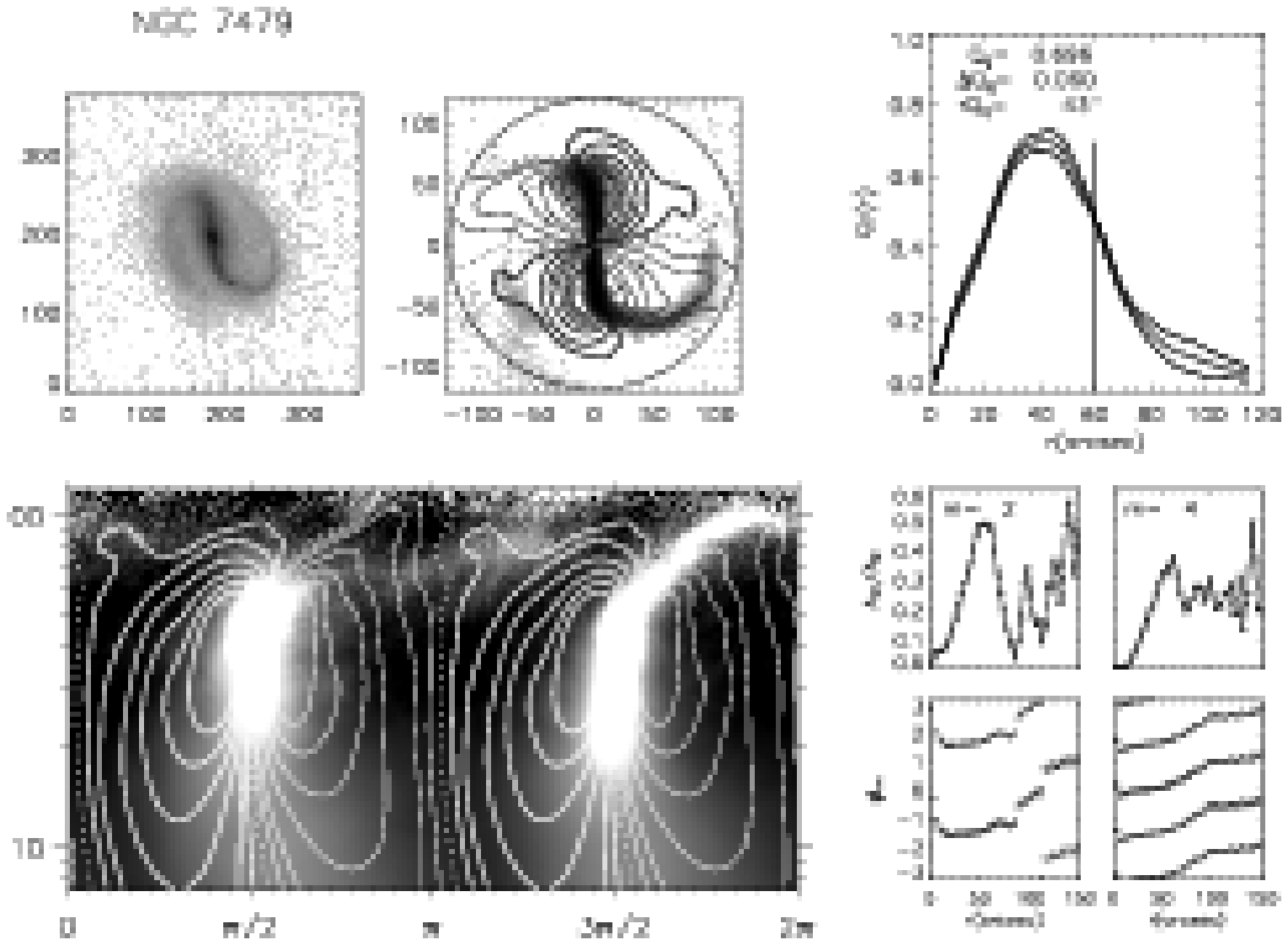,width=16cm}
Fig. 8h
\vfill
\eject

\psfig{file=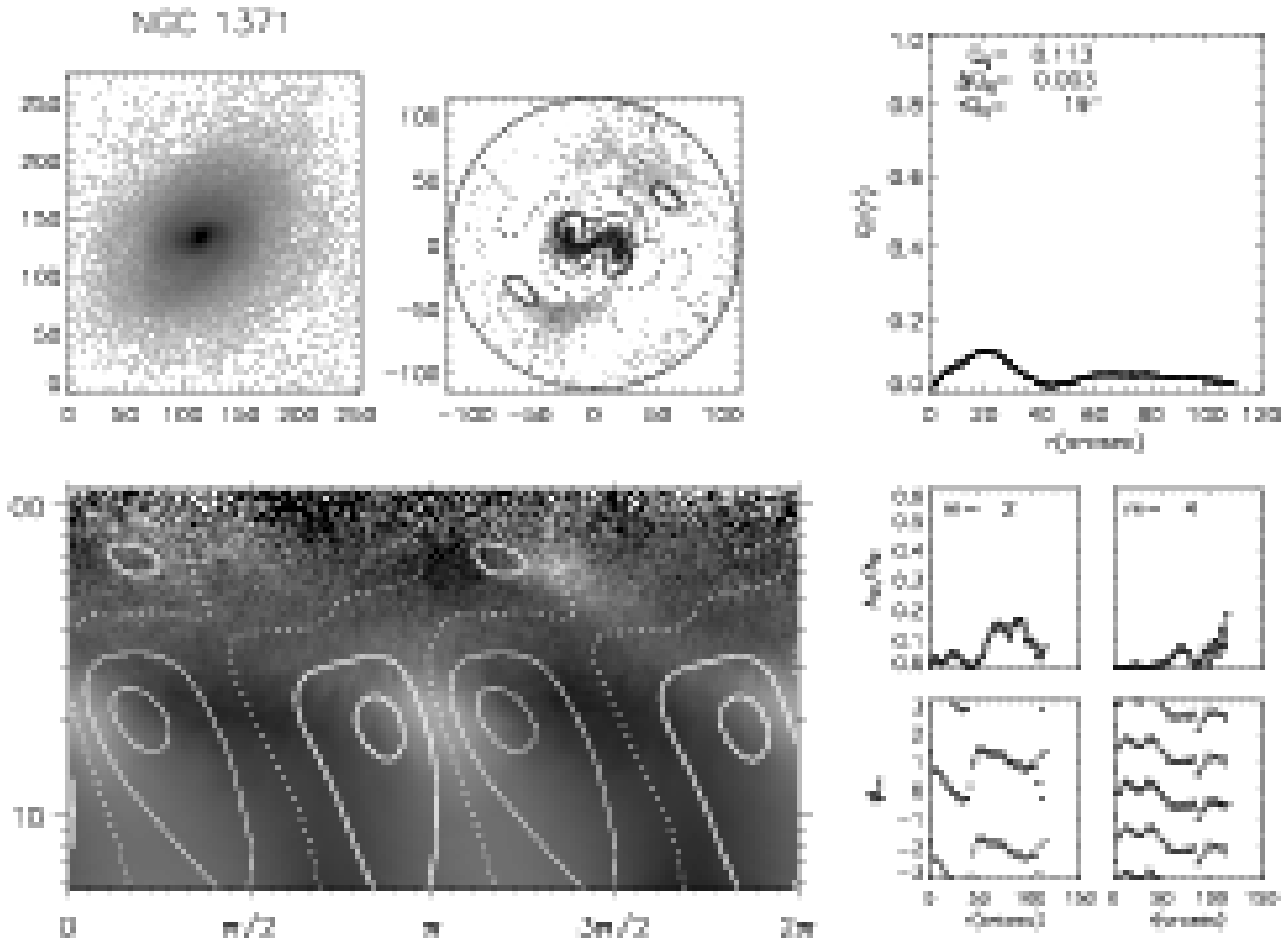,width=16cm}
Fig. 8i
\vfill
\eject

\psfig{file=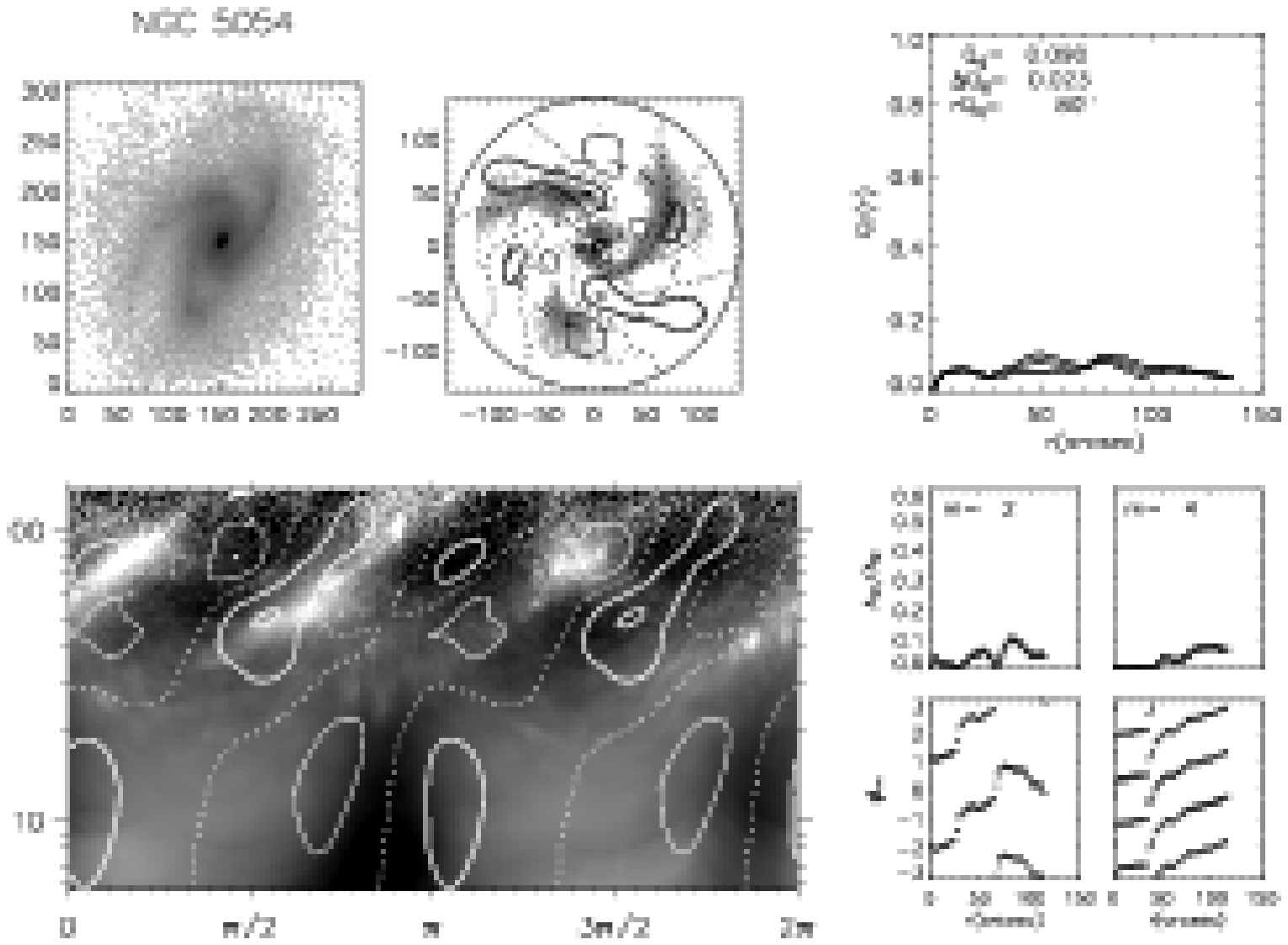,width=16cm}
Fig. 8j
\vfill
\eject


\section*{References}

\beginrefs

\bibitem Block, D., Buta, R., Knapen, J., Elmegreen, D., Elmegreen, B., Puerari, I. 2004, AJ, 128, 183
\bibitem Buta, R. 1995, ApJS, 96, 39
\bibitem Buta, R. 1996, in Barred Galaxies, R. Buta.,
D. A. Crocker, \& B. G. Elmegreen, eds., ASP Conf. Ser. 91, p. 11
\bibitem Buta, R., Laurikainen, E., Salo, H. 2004 (BLS), AJ, 127, 279
\bibitem Buta, R., Block, D., Knapen, D. 2003, AJ, 126, 1148
\bibitem Buta, R., Block, D. 2001 (BB), AJ, 550, 243
\bibitem Byun, Y. I., Freeman, K. C. 1995, ApJ, 448, 563
\bibitem Combes, F., Sanders, R. 1981, A\&A, 96, 164
\bibitem Crocker, D., Baugus, P., Buta, R. 1996, ApJS, 105
\bibitem de Grijs, R. 1998, MNRAS, 299, 595 
\bibitem de Jong, R. 1996, A\&A, 313, 45 
\bibitem de Souza, R., Gadotti, R., and dos Anjos, S. 2004, ApJ, in press
\bibitem de Vaucouleurs, G., and Freeman, K. C. 1972, Vistas in Astronomy, 14, 163
\bibitem de Vaucouleurs, G., de Vaucouleurs, A., Corwin, H.G. Jr., Buta, R., Paturel, G.,  and Fouque, P., 1991, Third Reference Cataloque of Bright Galaxies. Springer-Verlag, New York (RC3)
\bibitem Eskridge, P. B., Frogel, J. A., Pogge, R. W., Quillen, A. C., Berlind, A. A., Davies, R. L., DePoy, D. L., Gilbert, K. M., Houdashelt, M. L., Kuchinski, L.  2002 (EFP) ApJS, 143, 73
\bibitem Garcia-Gomez, C., Barber\'a, C., Athanassoula, E., Bosma, A., \& Whyte, L. 2004, A\&A, in press
\bibitem Jogee, S., Knapen, J. H., Laine, S., Shlosman, I., Scoville, N. Z., \& Englmaier, P. 2002, ApJ, 570, L55
\bibitem Knapen, J. H., de Jong, R. S., Stedman, S., Bramich, D. M. 2003, MNRAS, 344, 527 
\bibitem Kormendy, J. 2004, in Penetrating Bars Through Masks of Cosmic Dust: The Hubble Turning Fork Strikes a New Note, eds. D.L. Block, K.C. Freeman, I. Puerari, R. Groess, E.K. Block, Dordrecht, Kluwer, in press
\bibitem Laurikainen, E., Salo H., Buta, R. 2004 (LSB), ApJ, 697, 103 
\bibitem Laurikainen, E., Salo H. 2002 (LS), MNRAS, 337, 1118 
\bibitem Laurikainen, E., Salo H. 2000, A\&A, 141, 103
\bibitem Mac Arthur, A. A., Courteau, S., Holtzman, J. 2003, ApJ, 582, 689 
\bibitem M\"ollenhoff, C., Heidt, J. 2001 (MH), A\&A, 368, 16 
\bibitem Peng, C. Y., Ho, L., Impey, C. D., Rix, H. 2002, AJ, 124, 266
\bibitem Peng, C. Y. 2002, AJ, 124, 294
\bibitem Quillen, A., Frogel, J., Gonzalez, R. 1994, ApJ, 437, 162
\bibitem Salo, H., Rautiainen, P., Buta, R., Purcell, G. B., Cobb, M. L., Crocker, D. A., \& Laurikainen, E. 1999, AJ, 117, 792
\bibitem Salo, H., Buta, R., Laurikainen, E. 2004, in preparation
\bibitem Sanders, R., Tubbs, A., 1980, ApJ, 235, 803
\bibitem Sersic, J. L. 1968, Atlas de Galaxias Australes (Gordoba: Obs. Astron. Univ. Nac. Cordoba) 
\bibitem Shaw, M. A., Gilmore, G. 1989, MNRAS, 237, 903 
\bibitem Simard, L., Willmer, N. A., Vogt, N. P., Sarahedidi V. L., Phillips, A. C., Weiner, B. J., Koo, D., Im, M., Illingworth, G. D., Faber, S. M. 2002, ApJS, 142, 1 
\bibitem Skrutzkie et al. 1997, in The Impact of Large-Scale Near-IR Surveys, F. Grazon et al., eds., Dordrecht, Kluwer, p. 25
\bibitem Tully, B. 1988, Nearby Galaxies Catalogue, Cambridge University Press 
\bibitem Wadadekar, Y., Robbason, B., Kembavi, A. 1999, AJ, 117, 1219 

\endrefs
\bye